\documentclass[a4paper,fleqn]{cas-sc}

\pdfoutput=1

\usepackage[numbers]{natbib}

\usepackage{subfig}

\usepackage{tcolorbox}

\newtcolorbox{mybox}{
    left=3pt, 
    right=3pt, 
    top=1pt, 
    bottom=1pt, 
}

\def\tsc#1{\csdef{#1}{\textsc{\lowercase{#1}}\xspace}}
\tsc{WGM}
\tsc{QE}

\begin{document}
\let\WriteBookmarks\relax
\def\floatpagepagefraction{1}
\def\textpagefraction{.001}

\shorttitle{Bibliometric Analysis of Scientific Publications on Blockchain Research and Applications}    

\shortauthors{Bao et al. }  

\title [mode=title]{Bibliometric Analysis of Scientific Publications on Blockchain Research and Applications}



%

\author[1,2]{Lingfeng Bao}[orcid=0000-0003-1846-0921]
\cormark[1]
\ead{lingfengbao@zju.edu.cn}



\affiliation[1]{organization={The State Key Laboratory of Blockchain and Data Security, Zhejiang University},
            country={China}}
\affiliation[2]{organization={Hangzhou High-Tech Zone (Binjiang) Blockchain and Data Security Research Institute},
            country={China}}
\affiliation[3]{organization={Zhejiang University Press},
            country={China}}
\affiliation[4]{organization={University of Stavanger, Norway}}

\author[3]{Jiameng Yang}
\author[1,2]{Xiaohu Yang}
\author[4]{Chunming Rong}






\cortext[1]{Corresponding author}



\begin{abstract}
Since the introduction of Bitcoin in 2008, blockchain technology has garnered widespread attention. Scholars from various research fields, countries, and institutions have published a significant number of papers on this subject. However, there is currently a lack of comprehensive analysis specifically focusing on the scientific publications in the field of blockchain.

To conduct a comprehensive analysis, we compiled a corpus of 41,497 publications in blockchain research from 2008 to 2023 using the Clarivate databases\footnote{Clarivate is a global analytics company that provides insights and analytics based on scientific research, intellectual property, and business data. It provides data and analytic services in this paper.}. Through bibliometric and citation analyses, we gained valuable insights into the field. Our study offers an overview of the blockchain research landscape, including country, institution, authorship, and subject categories. Additionally, we identified Emerging Research Areas (ERA) using the co-citation clustering approach, examining factors such as recency, growth, and contributions from different countries/regions. Furthermore, we identified influential publications based on citation velocity and analyzed five representative Research Fronts in detail. This analysis provides a fine-grained examination of specific areas within blockchain research. Our findings contribute to understanding evolving trends, emerging applications, and potential directions for future research in the multidisciplinary field of blockchain.
\end{abstract}



\maketitle

\section{Introduction}

Blockchain is a decentralized and distributed digital ledger technology that records transactions across multiple computers or nodes. The concept of blockchain can be traced back to 1991 when Stuart Haber and W. Scott Stornetta~\cite{haber1991time} proposed the idea of a chain of hashes to create a total order of commitments to a dynamically growing set of documents. However, the technology gained widespread recognition with the advent of Bitcoin~\cite{nakamoto2008bitcoin} in 2009, which is the first and most well-known application of blockchain. 
Then, Vitalik Buterin proposed Ethereum~\cite{wood2014ethereum} in 2013 as a blockchain platform that introduced the groundbreaking concept of smart contracts. Ethereum revolutionized blockchain technology by extending its capabilities beyond digital currency and facilitating the creation of decentralized applications (DApps).



The rapid growth of blockchain research has attracted scholars from diverse disciplines, including computer science, economics, finance, and engineering. Researchers have explored various aspects of blockchain technology, ranging from theoretical foundations and system architecture to hardware implementations and software development. Moreover, the integration of blockchain with domains such as finance, supply chain management, healthcare, and Internet of Things (IoT) has led to a proliferation of interdisciplinary studies.

Since the introduction of Bitcoin in 2008, the field of blockchain has witnessed a significant number of research papers published by scholars, researchers, institutions, and organizations from various disciplines, countries, and backgrounds.
Therefore, in this paper, we want to to conduct a comprehensive analysis of scientific publications in the field of blockchain research and application.
The multidisciplinary nature of blockchain research necessitates a comprehensive analysis of scientific publications to gain insights into the evolving trends, emerging applications, and collaborations within the field. By analyzing the collective body of knowledge, we can identify key research areas, influential authors, and potential directions for future research.

In this paper, we employ the Clarivate databases to compile a corpus of 41,497 publications focused on blockchain research, encompassing the years 2008 to 2023. Leveraging this extensive collection, we undertake bibliometric and citation analyses to extract valuable insights. Our study begins by presenting a comprehensive overview of the blockchain research landscape. We delve into various aspects of blockchain research, including country, institutions, authorship, and subject categories. Additionally, we employ the co-citation clustering approach to identify Emerging Research Areas (ERA). By examining factors such as recency, growth, and contributions from different countries/regions, we discern different groups of ERA. Furthermore, we identify the most influential publications in blockchain research based on citation velocity. Lastly, we select five representative Research Fronts to conduct a detailed analysis, providing a fine-grained examination of blockchain research within these specific areas.

\smallskip \noindent \textbf{Paper Structure}: 
The data source and methodology employed are detailed in Section~\ref{sec:data}. The search string utilized to query the Clarivate databases and the corresponding extracted results are presented in Section~\ref{sec:publications}. The analysis of the obtained publications is outlined in Section~\ref{sec:results}. Section~\ref{sec:era} delves into the findings regarding emerging research areas. The analysis of top publications in blockchain research is discussed in Section~\ref{sec:top}. Section~\ref{sec:key} provides an examination of five representative research fronts in the blockchain field. Finally, Section~\ref{sec:conclusion} offers concluding remarks for this paper.




\section{Data Source and Methodology}\label{sec:data}
In this section, we describe the data source and the analysis methodology used in the paper.

\subsection{Terminologies}

\vspace{1mm} \noindent \textbf{Publications/Papers}: Clarivate abstracts publications including editorials, meeting abstracts, conference proceedings and book reviews, as well as research journal articles. This analysis covers publications, which includes papers (i.e., substantive peer-reviewed research journal articles and reviews) and conference proceedings.

\vspace{1mm} \noindent \textbf{Citations}: The citation count is the number of times that a citation has been recorded for a given publication since it was published. Not all citations are necessarily recorded since not all publications are indexed. However, the material indexed by Clarivate is estimated to attract about 95\% of global citations.

\vspace{1mm} \noindent \textbf{Highly-cited publications}: Highly cited work is recognized as having a greater impact and Clarivate has shown that high citation rates are correlated with other qualitative evaluations of research performance, such as peer review. In this project, publications that are in the top 10\% in terms of citation frequency are considered to be highly cited, taking into account the year of publication and journal subject category (JSC). Conference proceedings and journal papers were treated separately as citation rates are different for different modes of publication.

\vspace{1mm} \noindent \textbf{Citation velocity}: Citation Velocity is a measure of the rate of citation accumulation based off certain frequency within a set period of time. In this report, the citation velocity for each publication is a measure of the rate of citation accumulation per month starting from the month it is published until December 2023. Note that publications that are only cited in the same month when published will have citation velocity of 0. 

\vspace{1mm} \noindent \textbf{Consistently-cited publications}: In this paper, publications that are in the top 10\% in terms of citation velocity are considered to be consistently-cited, taking into account the year of publication and JSC.

\vspace{1mm} \noindent \textbf{Interdisciplinarity}: In this paper, interdisciplinarity corresponds to the average number of distinct journal subject categories from the core publications. 

\subsection{Data Source}

We rely on the following Clarivate\footnote{https://clarivate.com/} databases to extract bibliometric data:  Web of Science Core Collection$^{TM}$, Essential Science Indicators$^{TM}$, InCites$^{TM}$.

For the publications analyzed in this paper, bibliometric data was sourced from databases underlying \textit{the Web of Science}, which gives access to conference proceedings, patents, websites, and chemical structures, compounds, and reactions in addition to journals. It has a unified structure that integrates all data and search terms together and therefore provides a level of comparability not found in other databases. It is widely acknowledged to be the world’s leading source of citation and bibliometric data. \textit{The Web of Science Core Collection} is part of the Web of Science and focuses on research published in journals and conferences in science, medicine, arts, humanities, and social sciences. The authoritative, multidisciplinary content covers over 27,000 of the highest impact journals worldwide, including Open Access journals and over 190,000 conference proceedings. Coverage is both current and retrospective in the sciences, social sciences, arts, and humanities, in some cases back to 1900. Within the research community, these data are often still referred to by the acronym `ISI'. Clarivate has extensive experience with databases on research inputs, activity, and outputs and has developed innovative analytical approaches for benchmarking and interpreting international, national, and institutional research impact.

InCites$^{TM}$ is a customized, citation-based research evaluation tool enabling analysis of productivity and benchmarking of output against peers worldwide, drawing on data from the Web of Science. Essential Science Indicators is a tool that is part of InCites$^{TM}$.

\subsection{Bibliometrics and Citation Analysis}

In this paper, we conduct bibliometric and citation analysis based on the extracted bibliometric data.

\subsubsection{Bibliometrics}
Research evaluation is increasingly making wider use of bibliometric data and analyses. Bibliometrics is the analysis of data derived from publications and their citations. Publication of research outcomes is an integral part of the research process and is a universal activity. Consequently, bibliometric data have a currency across subjects, time, and location that are found in few other sources of research-relevant data. The use of bibliometric analysis, allied to informed review by experts, increases the objectivity of, and confidence in, evaluation.

Bibliometric indicators have been found to be more informative for core natural sciences, especially for basic science, than they are for applied and professional areas and for social sciences. In professional areas the range of publication modes used by leading researchers is likely to be diverse as they target a diverse, non-academic audience. In social sciences there is also a diversity of publication modes and citation rates are typically much lower than in natural sciences.

\subsubsection{Citation Analysis}
Research publications accumulate citation counts when they are referred to by more recent publications. Citations to prior work are a normal part of publication and reflect the value placed on a work by later researchers. Some publications get cited frequently and many remain uncited. Highly cited work is recognized as having a greater impact and Clarivate has shown that high citation rates are correlated with other qualitative evaluations of research performance, such as peer review. This relationship holds across most science and technology areas and, to a limited extent, in social sciences and even in some humanities subjects.

Indicators derived from publication and citation data should always be used with caution. Some fields publish at faster rates than others and citation rates also vary. Citation counts must be carefully normalized to account for such variations by field. Because citation counts naturally grow over time, it is essential to account for growth by year. Normalization is usually done by reference to the relevant global average for the field and for the year of publication.

Therefore, to measure the impact of publications, we use \textit{Category Normalized Citation Impact (CNCI)}, which is a bibliometric indicator used to assess the impact of scientific publications within specific subject categories. It is a normalized measure that takes into account the average citation impact of articles in a particular subject category and compares it to the global average.
The CNCI is calculated by dividing the average number of citations received by articles in a specific subject category by the average number of citations received by articles across all subject categories. The result is a ratio that indicates whether the publications within a particular subject category receive more or fewer citations compared to the global average. A CNCI value greater than 1 indicates that the publications in the subject category receive more citations on average compared to the global average, suggesting a higher impact within that field. Conversely, a value less than 1 suggests a lower impact compared to the global average.


\subsection{Co-citation Clustering}\label{sec:cocitation}

In order to identify the prominent publications and researchers in the field of blockchain research, we conduct co-citation analysis. The methodology of co-citation analysis was initially introduced by Henry Small in 1973, who served as the Chief Scientist of Clarivate at that time~\citep{small1973co, small1974structure, small2003paradigms, small2006tracking}.


\textit{Co-citation} occurs when two publications are cited by a third, later publication. The greater the frequency of co-citation for a given pair of publications, the greater the likelihood that it defines an established or emerging topic or subspecialty. The citation pair can be used in a citation index search to retrieve related publications. One pair can usually identify a small research front, but active research fronts generally involve several interrelated co-citation pairs~\cite{Garfield1988NewTF}. The larger the number of pairs included in a cluster, the broader the scope.

Single-link clustering, for which the algorithm selects a single document and searches for all the other items that are linked to it, is used to form clusters of co-cited publications. With the publications that cite them, \textit{research fronts} are identified. Frequency thresholds are used to modulate clustering by controlling the relative number of pairs selected. With co-citation analysis, the scope can be adjusted by increasing or decreasing the threshold. In other words, the larger the number of pairs included in the cluster, the broader the scope. This concept is important when trying to create maps of the literature at different levels of detail. Threshold strength refers to the degree of association between co-cited pairs in terms of the proportion of their total citations that are co-citations~\cite{October1998ABCSOC}.

\begin{figure}
    \centering
    \includegraphics[width=0.5\textwidth]{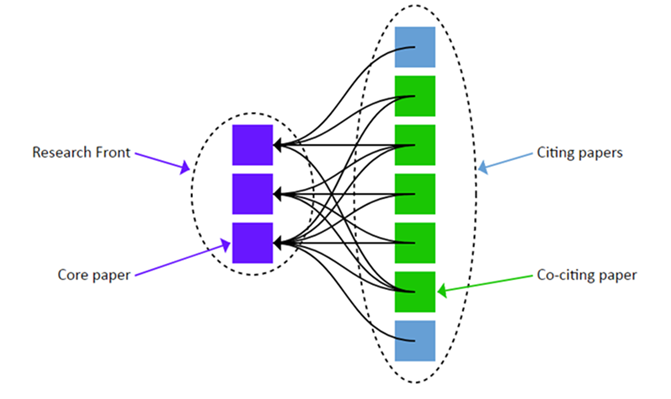}
    \caption{Research Front Explanation}
    \label{fig:5-1}
\end{figure}

A \textit{Research Front} consists of two or more highly cited papers. These highly cited papers are referred to as ``core papers'' in this presentation and are shown in purple in Figure~\ref{fig:5-1}. These highly cited papers have been frequently co-cited indicating that these core papers have a strong intellectual connection. The papers that co-cite the core papers are shown in green and are referred to as ``co-citing'' papers. In order to provide a fuller description of the research fields developing from the Research Fronts, the Research Fronts will be extended into Emerging Research Areas – defined as the core Research Fronts papers and all co-citing publications to these core papers.

In the original Research Front method, the cluster creation algorithm links all publications that are co-cited with a cosine similarity above a given threshold as a single cluster. Such a cluster is a Research Front if its size is 50 or fewer. If the cluster is too big, then the process is repeated with a higher cosine similarity threshold. This all repeats until no co-citation pairs exceed the cosine similarity threshold. The breadth of the subspecialty within each cluster can vary widely; its size depends upon the frequency thresholds used.

In this paper, the co-citation clustering method described in the original Research Front is applied to the publication set collected.


\section{Identification of Research Publications}\label{sec:publications}


In order to gather a comprehensive selection of relevant publications on the topic of blockchain, we begin by defining a set of keywords that are relevant to the subject, including terms such as `blockchain', `cryptocurrency', and `decentralized application'. Since blockchain technology emerged in 2008, we narrow down our search to the time period from 2008 to 2023. Additionally, to refine our results, we exclude papers that do not fall into the categories of \emph{Article}, \emph{Review}, or \emph{Proceedings}. The full query is shown in Table~\ref{tab:search}

\begin{table*}[]
    \centering
    \caption{Search String Used for Gathering Dataset}
    \resizebox{\textwidth}{!}{%
    \begin{tabular}{p{0.9\textwidth}} 
    \toprule
        ((TS = (blockchain* or cryptocurrenc* or ``decentralized application*'' or ``decentralised application*'' or ``decentralized financ*`` or ``decentralised financ*'' or ``decentralized autonomous organization*'' or ``decentralised autonomous organisation*'' or ``digit asset'' or ``distributed ledger'' or hyperledger or ``smart contract*'' or ``proof of stake'' or ``decentralized ledger*'' or ``decentralised ledger*'' or ``crypto-currenc*'' or ``Non-Fungible Token\$'' or ``Layer 2 Scaling Solution\$'' or ``zk-Rollups'' or ``Optimistic Rollup\$'' or ``zk-SNARK\$'' or ``Zero-Knowledge Succinct Non-Interactive Arguments of Knowledge''
	)) 
	OR TMIC==(``4.187.12014 Blockchain'' or ``4.187.2134 Blockchain'')
	OR ((TS = (bitcoin or ethereum or cryptocurrency)) NOT (TMIC==(``6.10.80 Option Pricing'') OR TS =(``futures'' or forex or forecasting)))
                OR (UT = (WOS:000886932500017 or WOS:000899554100022 or WOS:000862763100002)) 
)
AND (PY==(``2023'' OR ``2022'' OR ``2021'' OR "2020" OR "2019" OR "2018" OR "2017" OR "2016" OR "2015" OR "2014" OR "2013" OR "2012" OR "2011" OR "2010" OR "2009" OR "2008") 
AND DT==("ARTICLE" OR "REVIEW" OR "PROCEEDINGS PAPER")) \\ 
 \bottomrule
 \multicolumn{1}{l}{\footnotesize TS: Topic; TMIC: Micro topics; UT: Accession Number; PY: Publication Year; DT: Document Type.}
    \end{tabular}
    }
    \label{tab:search}
\end{table*}

Finally, we retrieved a total of 42,053 publications from the Web of Science Collection$^{TM}$. Among these, 41,497 publications are accessible within InCites. The analysis presented in the paper is conducted based on this subset of 41,497 publications.

\section{The Landscape of Blockchain Research}\label{sec:results}

In this section, we provide an overview of the analysis results obtained from the gathered publications. Initially, we present the comprehensive landscape of blockchain research. Subsequently, we delve into a detailed examination of various aspects of blockchain research, encompassing \emph{country}, \emph{institution}, \emph{authorship}, and \emph{subject categories}.

\subsection{Overall Scientific Output Analysis}

\begin{table*}[]
\centering
\caption{Trend in Blockchain Research}
\resizebox{0.95\textwidth}{!}{%
\begin{tabular}{@{}crrrrr@{}}
\toprule
\textbf{\begin{tabular}[c]{@{}c@{}}Publication \\ Year\end{tabular}} & \multicolumn{1}{c}{\textbf{\begin{tabular}[c]{@{}c@{}}Number of \\ Publications\end{tabular}}} & \multicolumn{1}{c}{\textbf{\begin{tabular}[c]{@{}c@{}}Category Normalized \\ Citation Impact\end{tabular}}} & \multicolumn{1}{c}{\textbf{Citation Impact}} & \multicolumn{1}{c}{\textbf{\begin{tabular}[c]{@{}c@{}}\% of Documents in \\ Top 10\%\end{tabular}}} & \multicolumn{1}{c}{\textbf{\begin{tabular}[c]{@{}c@{}}\% of International \\ Collaboration\end{tabular}}} \\ \midrule
2008                                                                 & 32                                                                                             & 1.04                                                                                                        & 14.19                                        & 15.62\%                                                                                             & 15.62\%                                                                                                   \\
2009                                                                 & 29                                                                                             & 0.64                                                                                                        & 6.07                                         & 10.34\%                                                                                             & 3.45\%                                                                                                    \\
2010                                                                 & 33                                                                                             & 0.87                                                                                                        & 18.88                                        & 6.06\%                                                                                              & 21.21\%                                                                                                   \\
2011                                                                 & 29                                                                                             & 0.67                                                                                                        & 6.10                                         & 6.90\%                                                                                              & 10.34\%                                                                                                   \\
2012                                                                 & 34                                                                                             & 1.28                                                                                                        & 7.56                                         & 14.71\%                                                                                             & 5.88\%                                                                                                    \\
2013                                                                 & 61                                                                                             & 5.69                                                                                                        & 38.52                                        & 19.67\%                                                                                             & 14.75\%                                                                                                   \\
2014                                                                 & 174                                                                                            & 5.74                                                                                                        & 39.25                                        & 33.33\%                                                                                             & 15.52\%                                                                                                   \\
2015                                                                 & 223                                                                                            & 6.57                                                                                                        & 43.48                                        & 34.98\%                                                                                             & 22.42\%                                                                                                   \\
2016                                                                 & 410                                                                                            & 8.15                                                                                                        & 61.63                                        & 43.17\%                                                                                             & 20.24\%                                                                                                   \\
2017                                                                 & 1,002                                                                                          & 5.38                                                                                                        & 43.98                                        & 48.80\%                                                                                             & 18.56\%                                                                                                   \\
2018                                                                 & 2,768                                                                                          & 3.80                                                                                                        & 31.01                                        & 40.17\%                                                                                             & 23.70\%                                                                                                   \\
2019                                                                 & 4,974                                                                                          & 2.43                                                                                                        & 23.30                                        & 31.87\%                                                                                             & 25.01\%                                                                                                   \\
2020                                                                 & 6,265                                                                                          & 2.05                                                                                                        & 19.13                                        & 26.37\%                                                                                             & 31.22\%                                                                                                   \\
2021                                                                 & 7,449                                                                                          & 1.77                                                                                                        & 11.91                                        & 22.89\%                                                                                             & 34.70\%                                                                                                   \\
2022                                                                 & 10,708                                                                                         & 1.58                                                                                                        & 4.56                                         & 16.74\%                                                                                             & 33.97\%                                                                                                   \\
2023                                                                 & 7,306                                                                                          & 1.78                                                                                                        & 1.15                                         & 14.15\%                                                                                             & 34.22\%                                                                                                   \\ \midrule
Aggregated  &  41,497 & 2.18 & 13.43 & 23.4\% & 31\% \\   
\bottomrule
\end{tabular}
}
\label{tab:overall}
\end{table*}

\begin{figure}
  \centering
  \subfloat[Category Normalized Citation Impact]{\includegraphics[width=0.45\textwidth]{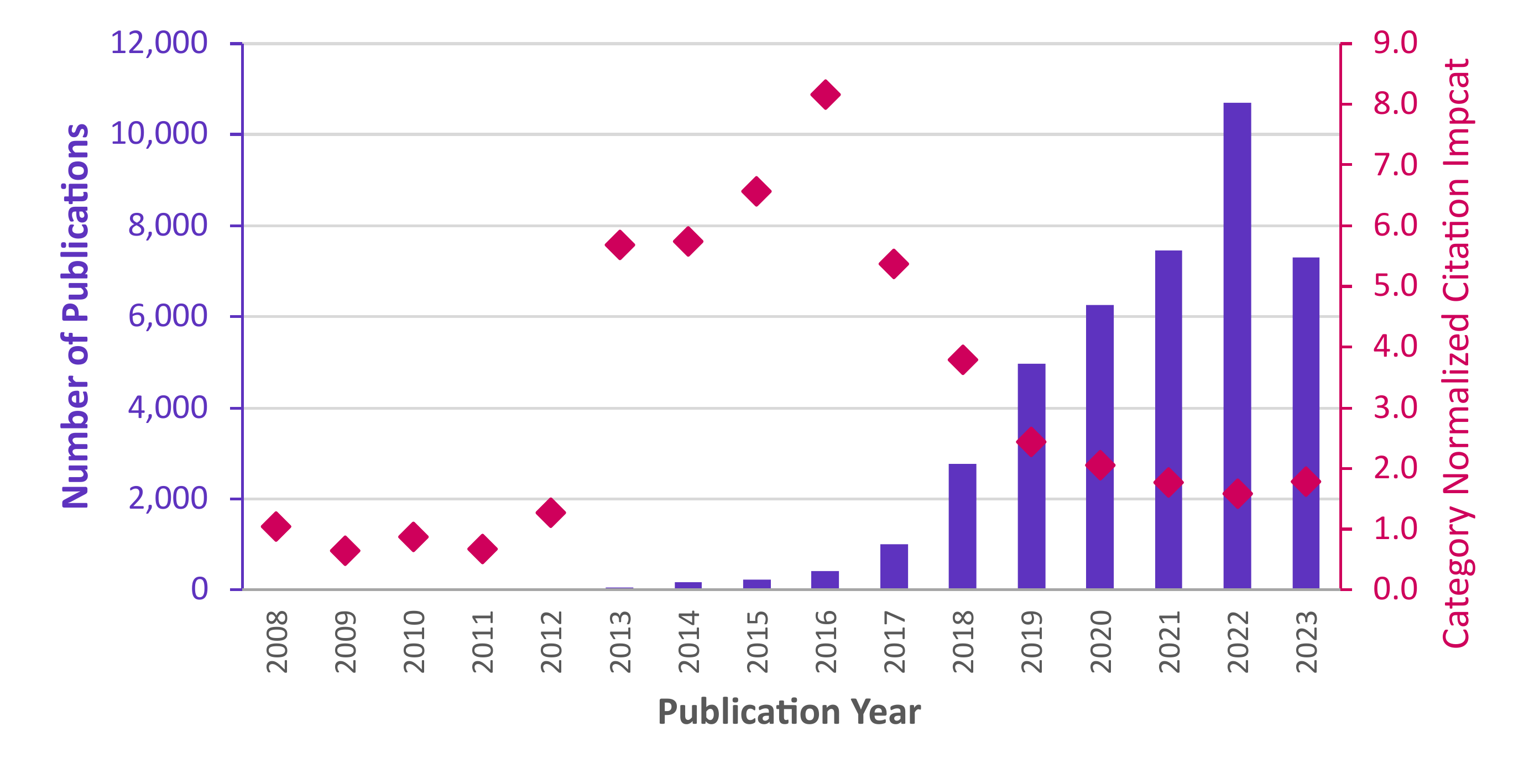}}
  \hfill
  \subfloat[\% of Documents in Top 10\%]{\includegraphics[width=0.45\textwidth]{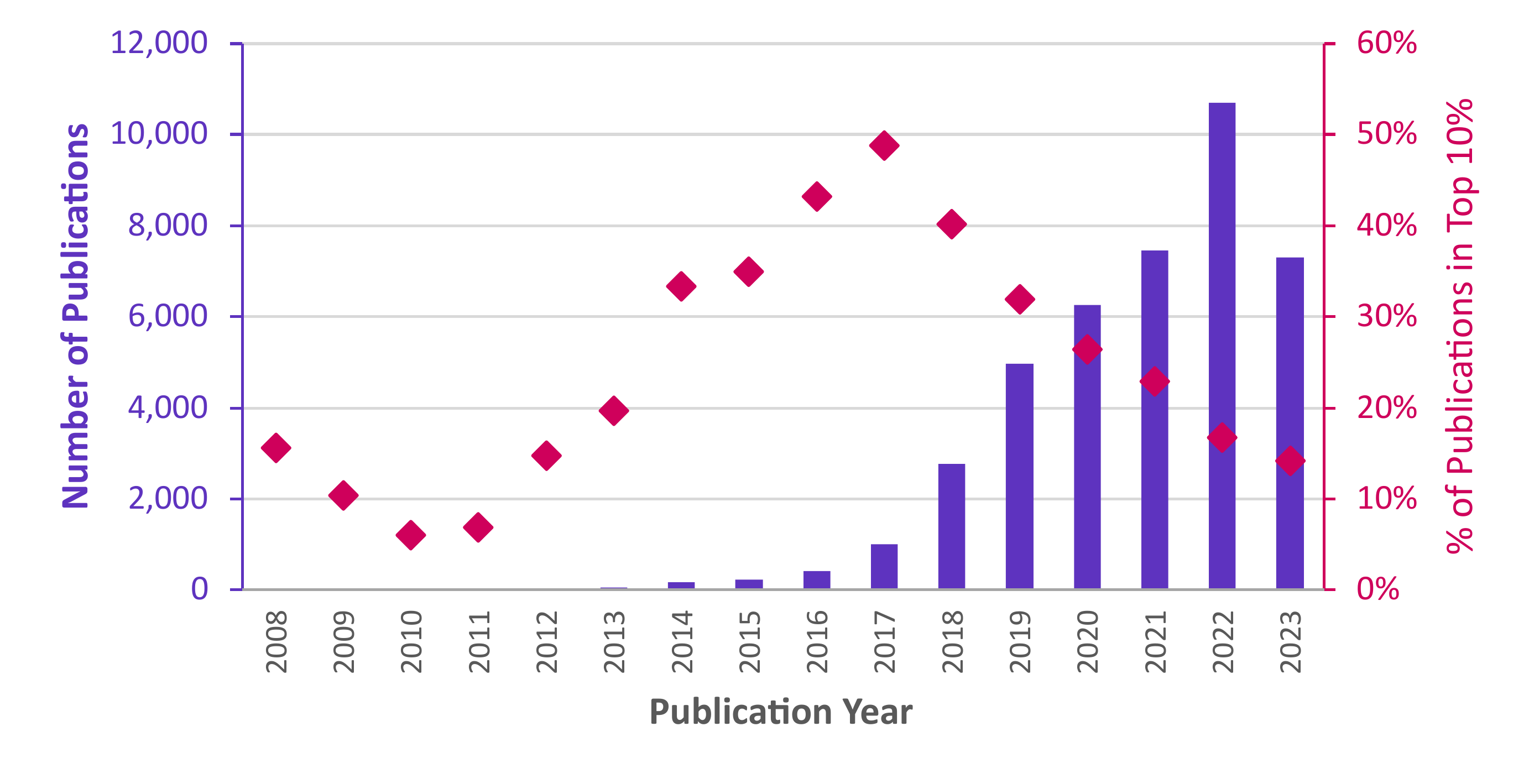}}
  \vfill
  \subfloat[\% of International Collaboration]{\includegraphics[width=0.45\textwidth]{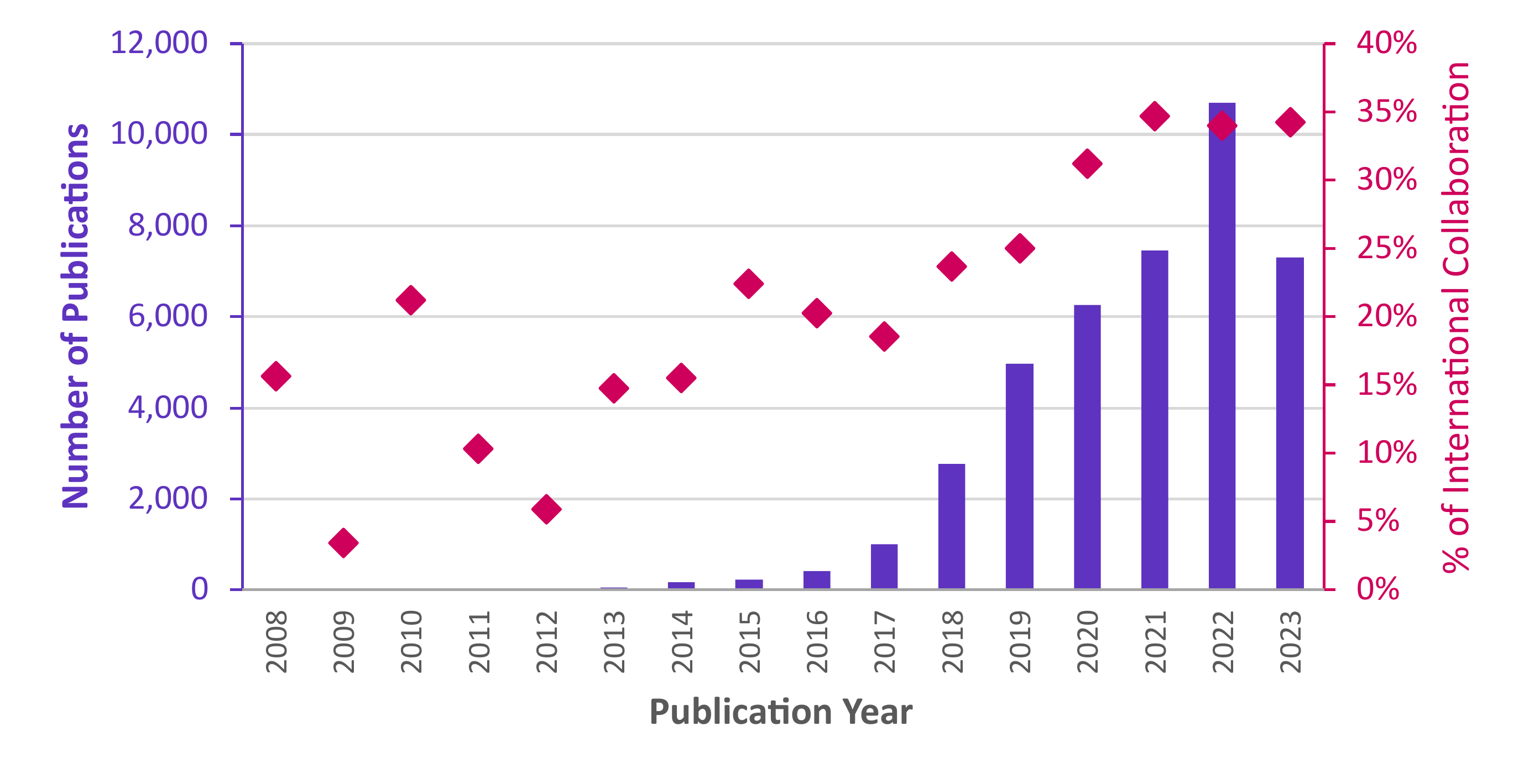}}
  \caption{Number of Publications}
  \label{fig:overall}
\end{figure}

Table~\ref{tab:overall} shows the trend in blockchain research. The last row of this table shows the aggregated data for the whole period from 2008 to 2023. 
In this table, we also show the percentage of publications in the top 10\% based on citations (\textit{\% of Documents in Top 10\%}) and the percentage of publications that contain one or more international co-authors (\textit{\% of International Collaboration}).
Figure~\ref{fig:overall} represents the trend in the volume of Blockchain research with Category Normalized Citation Impact, Percentage of Documents in the Top 10\%, and Percentage of International Collaboration in the period from 2008 to 2023, resepctively.

In terms of the number of publications, there has been a steady growth in blockchain-related papers between 2008 and 2022. 
But more than 60\% of blockchain research was published within the last three years (2021-2023).
Prior to 2021, there were several years of rapid growth, including 2014, 2017 and 2018 where publication outputs more than doubled in each year. 
By 2022, the number of papers had reached tens of thousands.

In terms of Category Normalized Citation Impact (CNCI), blockchain research was below the global average (with a value of 1) between 2009 and 2011. This could be attributed to the early stage of blockchain technology when it had not yet garnered widespread attention from researchers. However, from 2012 onwards, the CNCI values consistently exceeded 1, reaching its peak in 2016 at 8.15.
CNCI has been lower in more recent years. However, since very few papers were published in earlier years this could instead be due to the fact that the CNCI could be inflated by the lower number of publications.

The percentage of publications in the top 10\% followed a similar pattern as the CNCI where it peaked in 2017 and has since declined and in recent years has ranged between 14 and 26\%, above the world average of 10\%. 

The percentage of international collaboration has steadily increased since 2017 and around a third of publications has been internationally collaborative since 2021. 

\begin{mybox}
\textbf{Summary}: Since its emergence, blockchain-related research papers have experienced significant growth. Moreover, both the CNCI values and the proportion of top papers in the top 10\% indicate that the field surpasses the global average. International collaboration in blockchain research has also steadily increased, with over one-third of papers being the result of international cooperation by 2023.
\end{mybox}





\begin{table*}[]
\centering
\caption{Top 10 Countries/Regions Based on Number of Publications in Blockchain Research Field, 2008-2023}
\resizebox{0.95\textwidth}{!}{%
\begin{tabular}{lccccc}
\toprule
\textbf{Country / Region} & \textbf{\begin{tabular}[c]{@{}c@{}}Number of \\ Publications\end{tabular}} & \textbf{\begin{tabular}[c]{@{}c@{}}Category Normalized \\ Citation Impact\end{tabular}} & \textbf{\begin{tabular}[c]{@{}c@{}}Citation \\ Impact\end{tabular}} & \textbf{\begin{tabular}[c]{@{}c@{}}\% of Documents in \\ Top 10\%\end{tabular}} & \multicolumn{1}{l}{\textbf{\begin{tabular}[c]{@{}l@{}}\% of International \\ Collaboration\end{tabular}}} \\
\midrule
CHINA MAINLAND            & 11,419                                                                     & 2.12                                                                                    & 13.20                                                               & 23.300                                                                          & 35.34                                                                                                     \\
USA                       & 6,587                                                                      & 3.21                                                                                    & 21.68                                                               & 31.230                                                                          & 51.24                                                                                                     \\
INDIA                     & 4,328                                                                      & 2.10                                                                                    & 11.09                                                               & 24.190                                                                          & 42.35                                                                                                     \\
UNITED KINGDOM            & 3,363                                                                      & 3.12                                                                                    & 21.42                                                               & 34.340                                                                          & 68.33                                                                                                     \\
AUSTRALIA                 & 2,171                                                                      & 3.16                                                                                    & 20.94                                                               & 33.670                                                                          & 64.39                                                                                                     \\
SOUTH KOREA               & 1,803                                                                      & 2.47                                                                                    & 15.53                                                               & 26.460                                                                          & 42.32                                                                                                     \\
ITALY                     & 1,776                                                                      & 2.42                                                                                    & 14.31                                                               & 28.380                                                                          & 44.93                                                                                                     \\
CANADA                    & 1,747                                                                      & 3.17                                                                                    & 19.20                                                               & 33.430                                                                          & 66.11                                                                                                     \\
GERMANY (FED REP GER)     & 1,693                                                                      & 2.33                                                                                    & 17.04                                                               & 28.290                                                                          & 46.31                                                                                                     \\
SAUDI ARABIA              & 1,609                                                                      & 2.18                                                                                    & 12.50                                                               & 27.280                                                                          & 78.87                 \\ \bottomrule                                                                                  
\end{tabular}
}
\label{tab:country_publication}
\end{table*}

\begin{table*}[]
\centering
\caption{Top 10 Countries/Regions Based on Category Normalized Citation Impact in Blockchain Research Field, 2008-2023 (With Least 10 Papers)}
\resizebox{0.95\textwidth}{!}{%
\begin{tabular}{@{}lccccc@{}}
\toprule
\textbf{Country / Region} & \textbf{\begin{tabular}[c]{@{}c@{}}Number of \\ Publications\end{tabular}} & \textbf{\begin{tabular}[c]{@{}c@{}}Category Normalized \\ Citation Impact\end{tabular}} & \textbf{\begin{tabular}[c]{@{}c@{}}Citation \\ Impact\end{tabular}} & \textbf{\begin{tabular}[c]{@{}c@{}}\% of Documents in \\ Top 10\%\end{tabular}} & \textbf{\begin{tabular}[c]{@{}c@{}}\% of International \\ Collaboration\end{tabular}} \\ \midrule
ISRAEL                    & 247                                                                        & 5.52                                                                                    & 30.41                                                               & 29.55                                                                           & 59.51                                                                                 \\
LEBANON                   & 248                                                                        & 4.34                                                                                    & 19.56                                                               & 38.31                                                                           & 90.73                                                                                 \\
IRELAND                   & 497                                                                        & 4.26                                                                                    & 24.86                                                               & 40.44                                                                           & 70.62                                                                                 \\
SINGAPORE                 & 706                                                                        & 4.23                                                                                    & 27.11                                                               & 36.12                                                                           & 75.92                                                                                 \\
MACAU, CHINA              & 160                                                                        & 4.01                                                                                    & 35.06                                                               & 32.50                                                                           & 33.12                                                                                 \\
VIETNAM                   & 395                                                                        & 4.00                                                                                    & 17.21                                                               & 40.51                                                                           & 69.11                                                                                 \\
HONG KONG, CHINA          & 903                                                                        & 3.69                                                                                    & 20.95                                                               & 40.20                                                                           & 44.08                                                                                 \\
LUXEMBOURG                & 172                                                                        & 3.48                                                                                    & 15.37                                                               & 30.81                                                                           & 65.70                                                                                 \\
YEMEN                     & 39                                                                         & 3.37                                                                                    & 11.90                                                               & 46.15                                                                           & 100.00                                                                                \\
FINLAND                   & 449                                                                        & 3.37                                                                                    & 21.92                                                               & 35.63                                                                           & 72.83                                                                                 \\ \bottomrule
\end{tabular}
}
\label{tab:country_cnci}
\end{table*}

\begin{figure}
  \centering
  \subfloat[Category Normalized Citation Impact]{\includegraphics[width=0.45\textwidth]{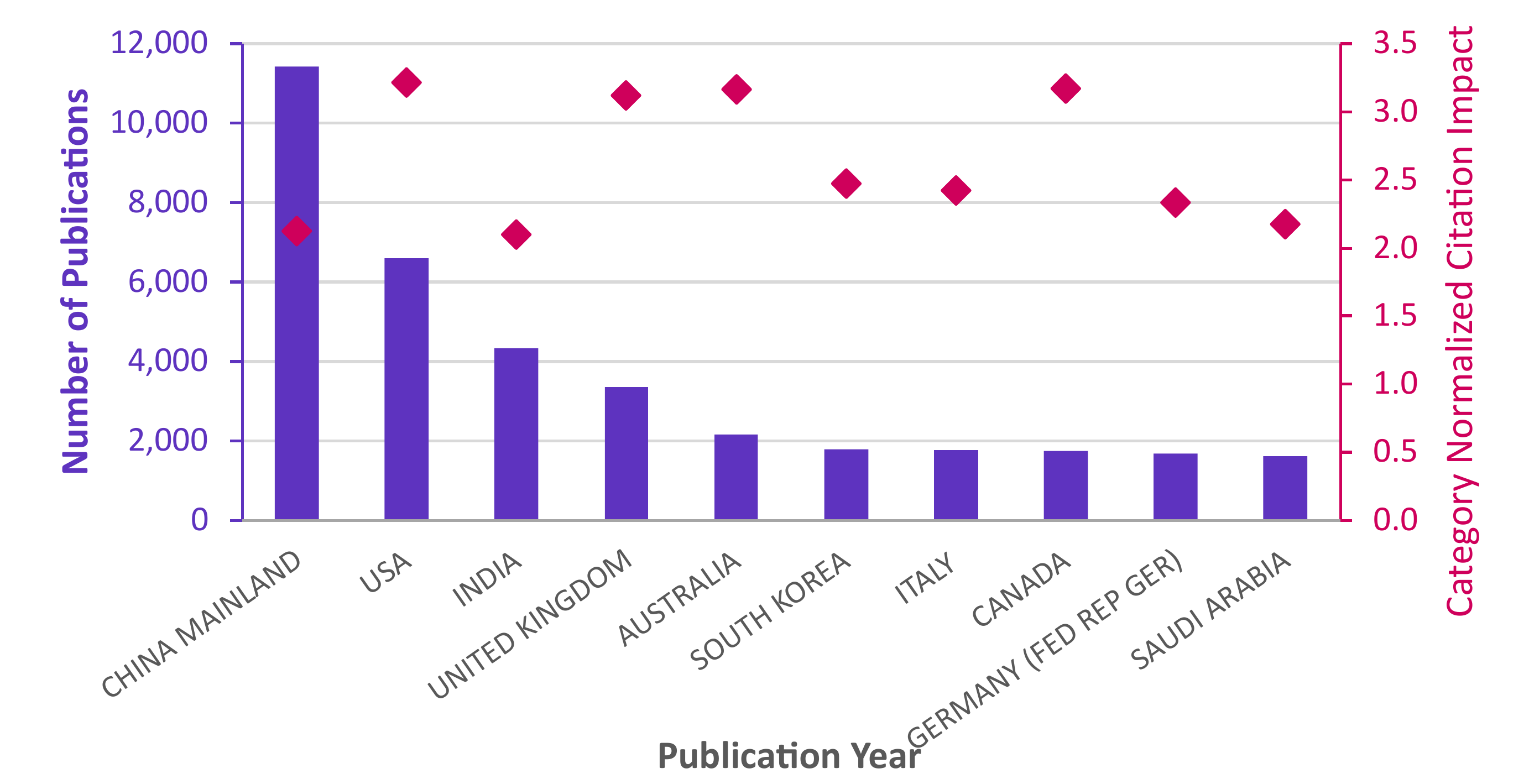}}
  \hfill
  \subfloat[\% of Documents in Top 10\%]{\includegraphics[width=0.45\textwidth]{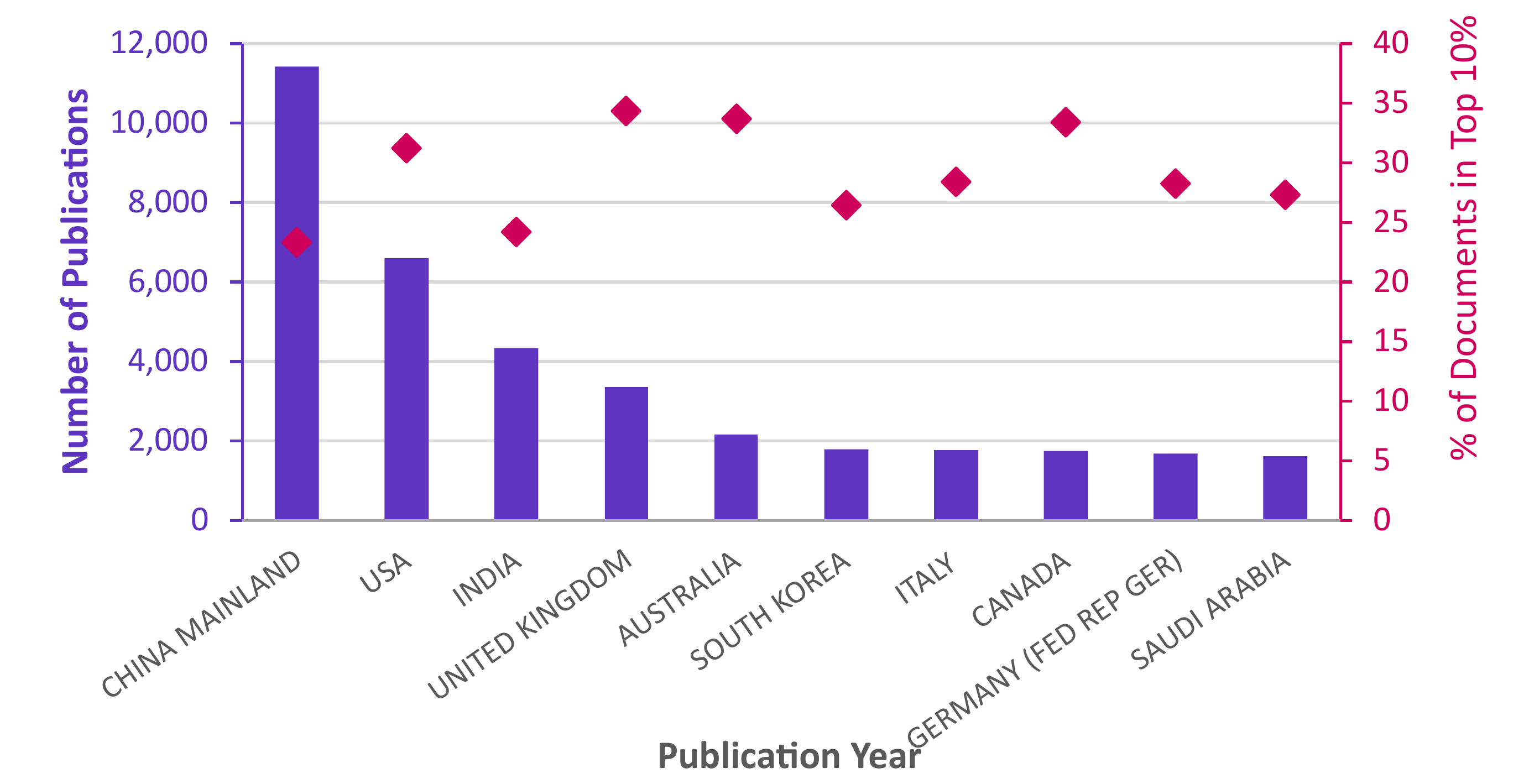}}
  \vfill
  \subfloat[\% of International Collaboration]{\includegraphics[width=0.45\textwidth]{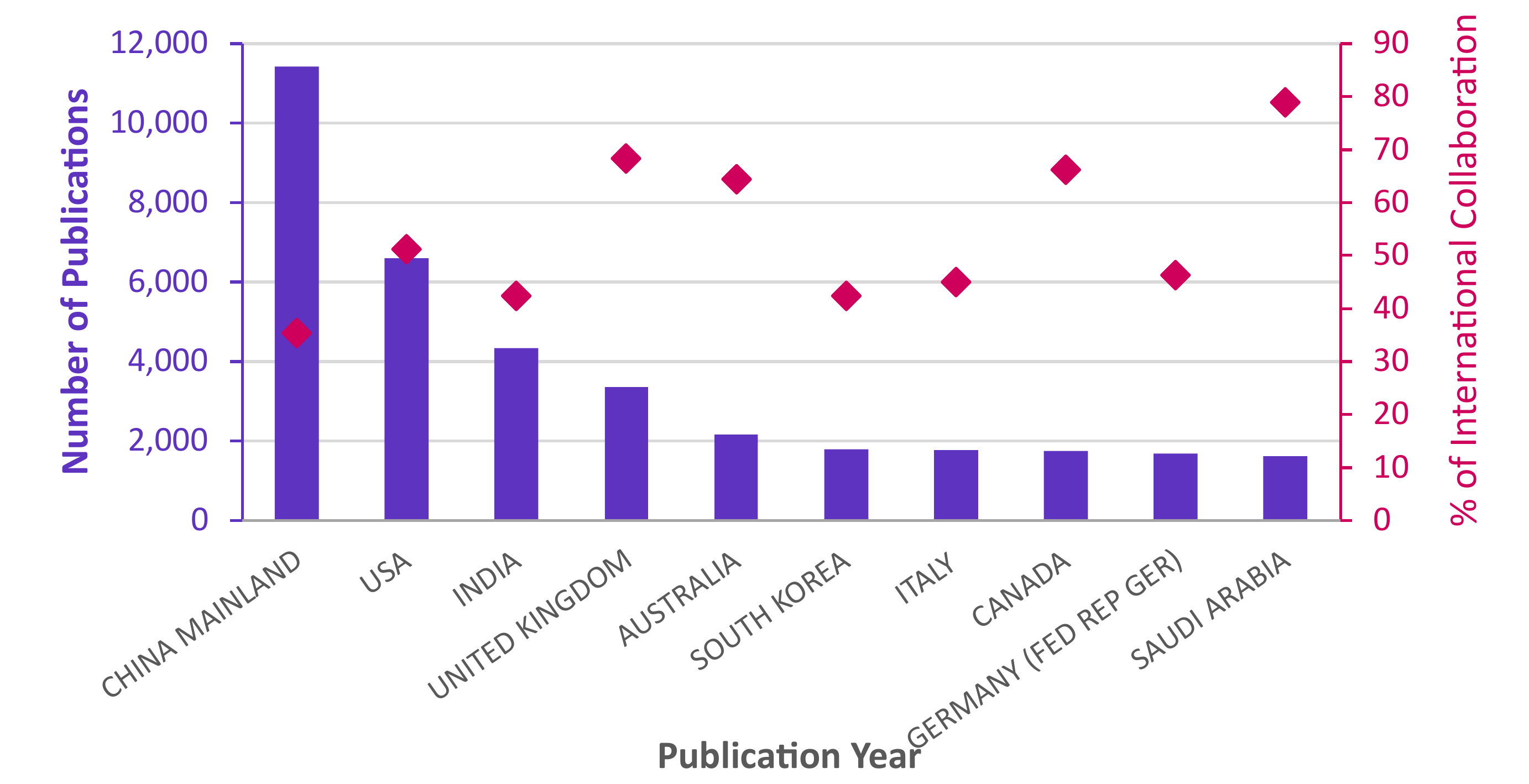}}
  \caption{Top 10 Countries/Regions Based on Number of Publications}
  \label{fig:country_publication}
\end{figure}

\begin{figure}
  \centering
  \subfloat[Category Normalized Citation Impact]{\includegraphics[width=0.45\textwidth]{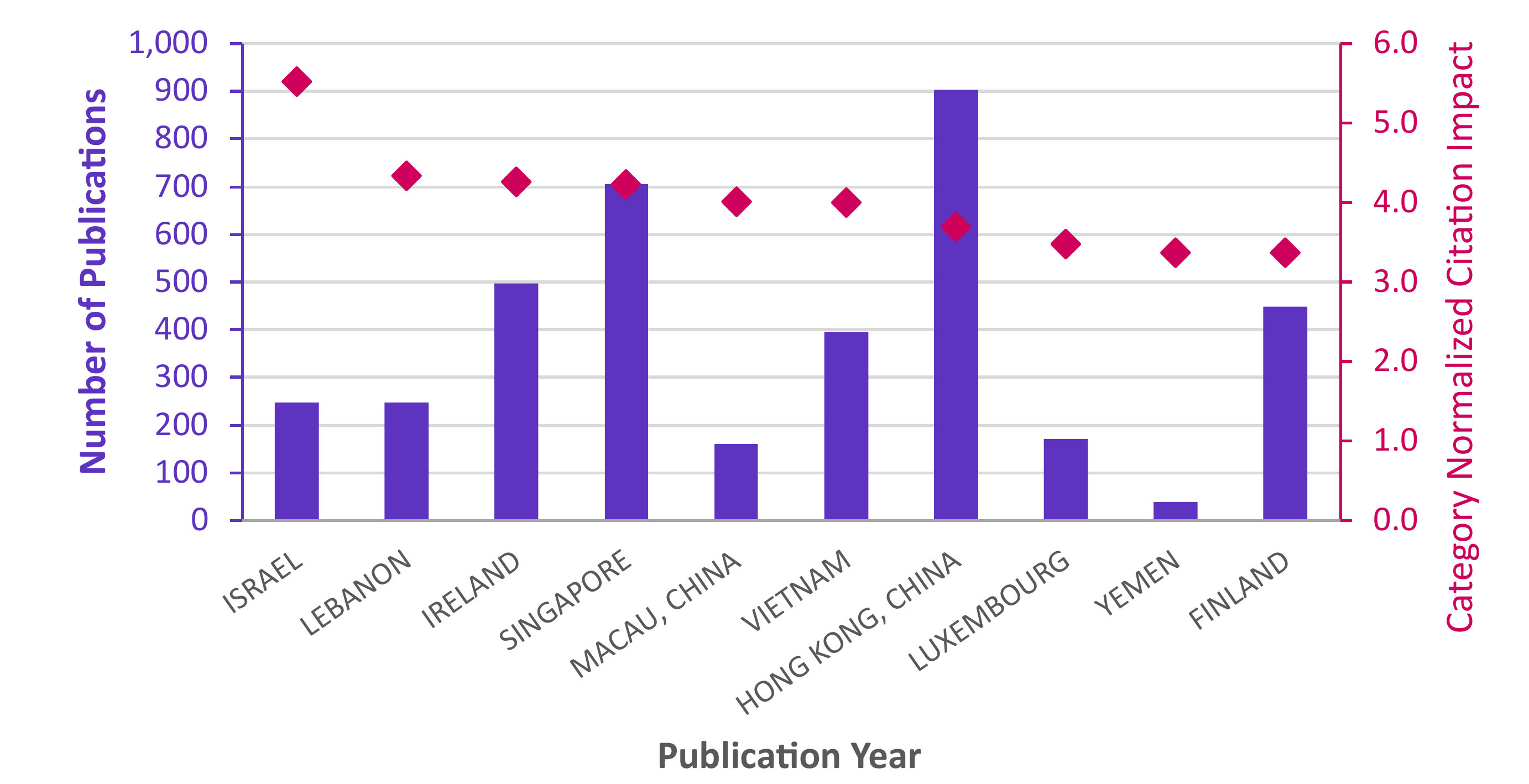}}
  \hfill
  \subfloat[\% of Documents in Top 10\%]{\includegraphics[width=0.45\textwidth]{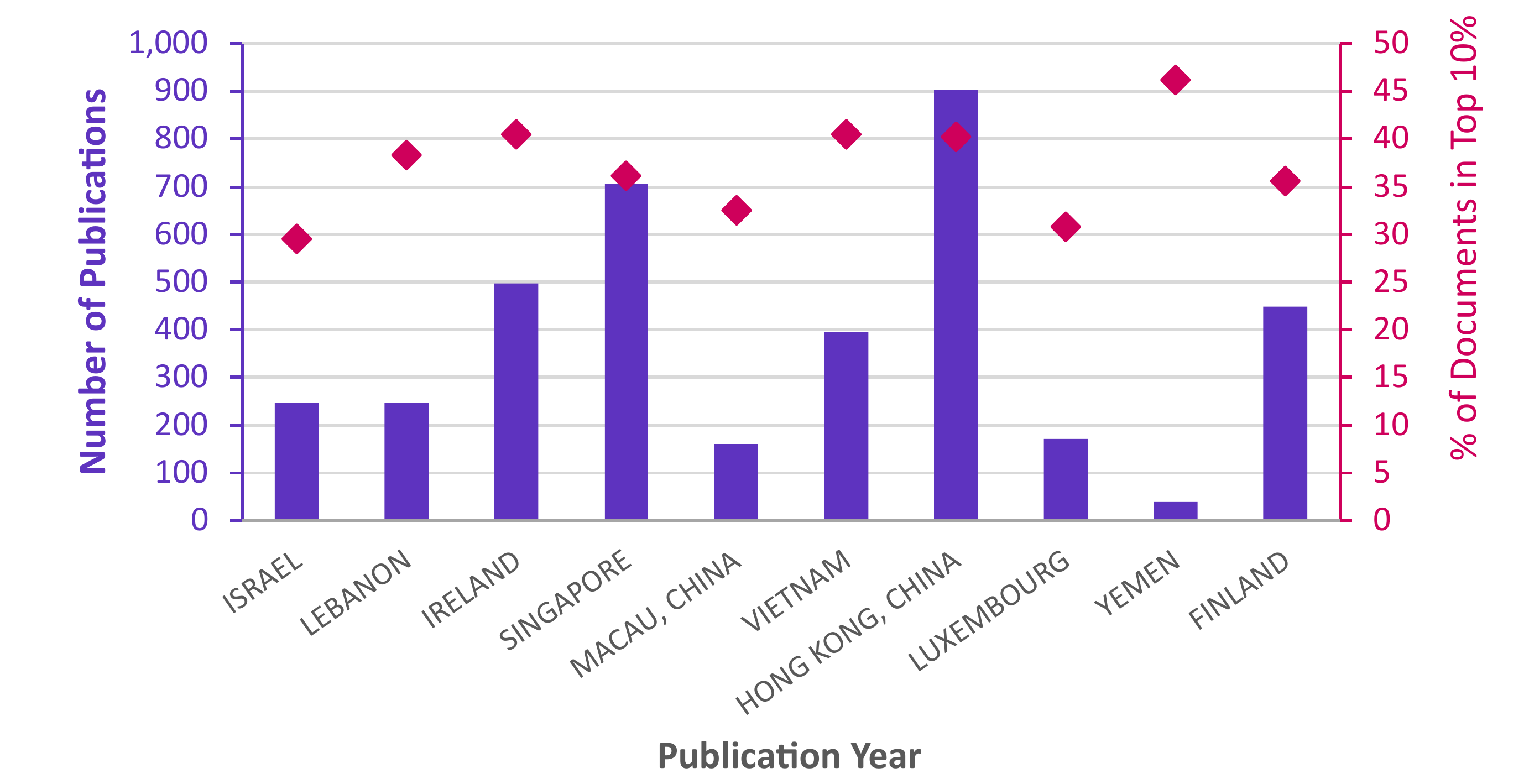}}
  \vfill
  \subfloat[\% of International Collaboration]{\includegraphics[width=0.45\textwidth]{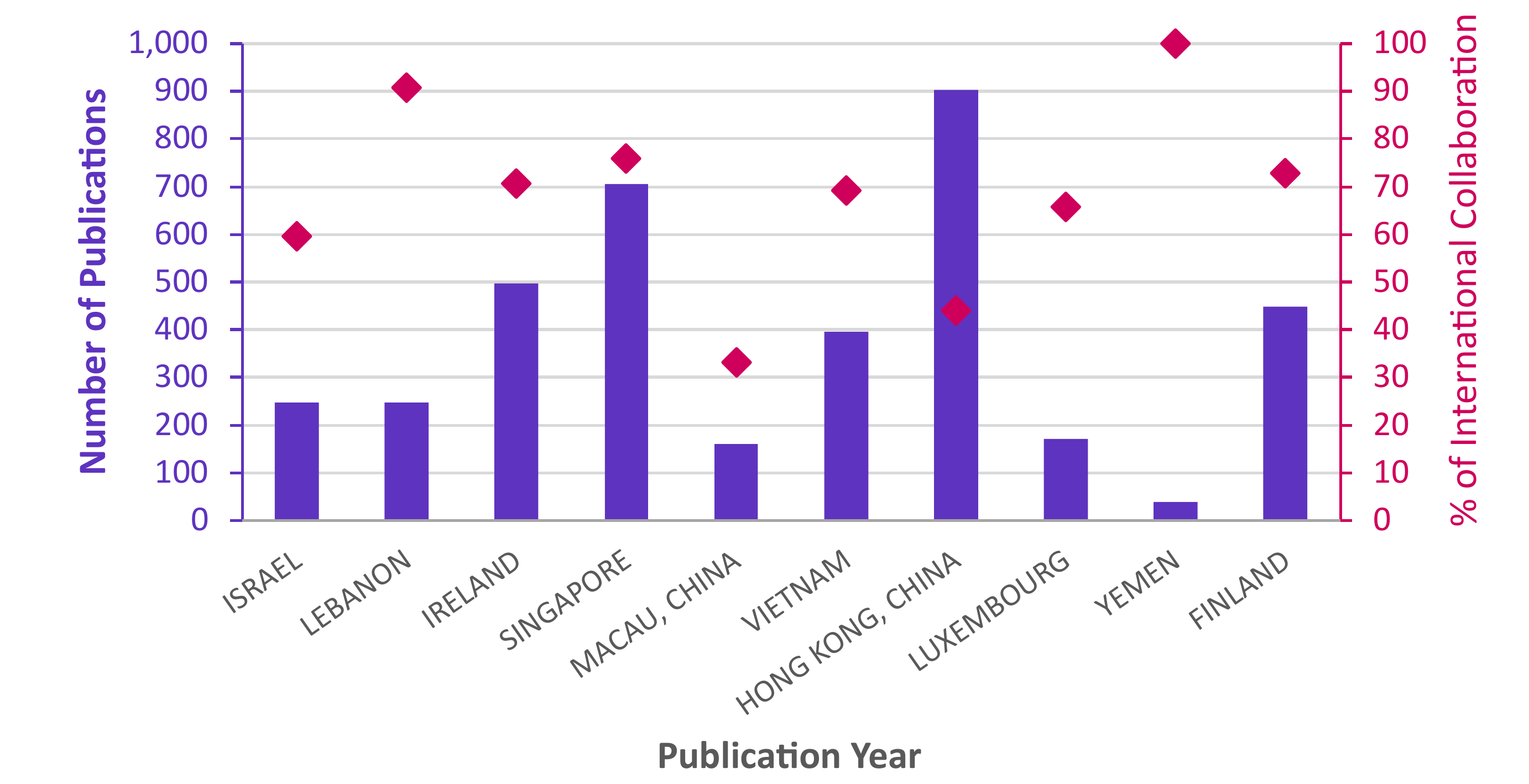}}
  \caption{Top 10 Countries/Regions Based on Category Normalized Citation Impact}
  \label{fig:country_cnci}
\end{figure}

\subsection{Country Analysis}

We identified the top 10 countries based on the number of publications and Category Normalized Citation Impact in the blockchain research field respectively. Table~\ref{tab:country_publication}
and~\ref{tab:country_cnci} present the selected metrics for each country based on number of publications and CNCI respectively, while Figure~\ref{fig:country_publication} and~\ref{fig:country_cnci}  present the visualization of number of publications for each country, as well as Category Normalized Citation Impact (CNCI), percentage of documents in the top 10\%, and percentage of international collaboration per country for the period 2008 to 2023.


\subsubsection{Based on Number of Publications}

China leads in terms of the number of papers published in the field of blockchain research with 11,419 articles. However, when compared to the top 10 countries, China's CNCI of 2.12 is only 0.02 higher than Indiass, placing it second to last. Additionally, the proportion of papers published within the top 10\% is the lowest at 23.30\%. These findings suggest that while China has the highest publication count, some of the papers may lack quality. China also had the least amount of internationally collaborative research with just over a third (35.3\%) involving international collaboration. 

In terms of the number of published papers, the United States follows China in second place with 6,587 papers. Furthermore, it has achieved the highest CNCI of 3.21, surpassing the world average by three times. This indicates that the quality of research papers from the United States is relatively high.

The country ranking third is India, with 4,328 published papers. However, it has the lowest CNCI and a relatively lower percentage of international collaboration. The remaining countries in the top 10 are economically affluent, including some Western developed countries such as the United Kingdom, Australia, and Germany. The United Kingdom had the highest percentage of the documents in the top 10\% - more than a third of research is in the top 10\%. The most internationally collaborative research in the field was from Saudia Arabia, with more than three-quarters of documents being internationally collaborative.

\subsubsection{Based on Category Normalized Citation Impact}

Compared to the total number of publications, the top 10 countries based on CNCI are predominantly smaller countries or regions. Israel has the highest category normalized citation impact in the field, with a citation impact of 5.52, which is five times the world average. Lebanon follows closely with a citation impact of 4.34. Remarkably, four out of the top 10 countries are from Asia, three from the Middle East, and three from Europe.

In terms of international collaboration, eight out of the ten countries have achieved a collaboration level close to or exceeding 60\%, and all of Yemen’s publication were internationally collaborative. It is worth noting that several of these countries are developing nations, including Lebanon, Vietnam, and Yemen.


\begin{mybox}
\textbf{Summary}: China has published the highest number of papers in the field of blockchain, followed by the United States. However, the United States has a higher CNCI than China (3.21 vs. 2.12). When ranked by CNCI, many smaller countries are in the top 10, and these countries tend to have a relatively high proportion of international collaboration in their papers. 
\end{mybox}




\begin{table*}[]
\centering
\caption{Top 10 Institutions Based on Number of Publications in Blockchain Research Field, 2008-2023}
\resizebox{0.95\textwidth}{!}{%
\begin{tabular}{@{}lccccc@{}}
\toprule
\textbf{Institution} & \textbf{\begin{tabular}[c]{@{}c@{}}Number of \\ Publications\end{tabular}} & \textbf{\begin{tabular}[c]{@{}c@{}}Category \\ Normalized \\ Citation Impact\end{tabular}} & \textbf{\begin{tabular}[c]{@{}c@{}}Citation \\ Impact\end{tabular}} & \textbf{\begin{tabular}[c]{@{}c@{}}\% of \\ Documents in \\ Top 10\%\end{tabular}} & \textbf{\begin{tabular}[c]{@{}c@{}}\% of \\ International \\ Collaboration\end{tabular}} \\ \midrule
Chinese Academy of Sciences & 697 & 2.82 & 18.11 & 27.40 & 32.28 \\
\rowcolor[HTML]{EFEFEF} 
Beijing University of Posts \& Telecommunications & 502 & 2.67 & 21.42 & 28.69 & 37.25 \\
University of London & 448 & 3.17 & 19.53 & 34.82 & 52.01 \\
\rowcolor[HTML]{FFFFFF} 
\hspace{0.6cm}University College London & 247 & 3.43 & 21.21 & 37.65 & 53.04 \\
\rowcolor[HTML]{EFEFEF} 
National Institute of Technology (NIT System) & 405 & 2.61 & 13.64 & 28.40 & 30.86 \\
\rowcolor[HTML]{EFEFEF} 
\hspace{0.6cm}National Institute of Technology Raipur & 42 & 5.81 & 32.43 & 64.29 & 47.62 \\
\rowcolor[HTML]{FFFFFF} 
University of Electronic Science \& Technology of China & 370 & 3.30 & 25.91 & 33.24 & 54.05 \\
\rowcolor[HTML]{EFEFEF} 
Indian Institute of Technology System (IIT System) & 367 & 2.40 & 11.34 & 27.25 & 38.69 \\
\rowcolor[HTML]{EFEFEF} 
\hspace{0.6cm}Indian Institute of Technology (IIT) - Kharagpur & 71 & 1.82 & 7.08 & 16.90 & 42.25 \\
\rowcolor[HTML]{FFFFFF} 
Xidian University & 366 & 2.29 & 15.89 & 29.78 & 49.45 \\
\rowcolor[HTML]{EFEFEF} 
University of California System & 341 & 3.53 & 24.62 & 35.78 & 48.68 \\
\rowcolor[HTML]{EFEFEF} 
\hspace{0.6cm}University of California Berkeley & 82 & 5.24 & 30.99 & 42.68 & 48.78 \\
\rowcolor[HTML]{FFFFFF} 
University of Texas System & 337 & 3.09 & 25.60 & 37.09 & 64.09 \\
\rowcolor[HTML]{FFFFFF} 
\hspace{0.6cm}University of Texas at San Antonio (UTSA) & 172 & 3.77 & 35.92 & 44.19 & 83.14 \\
\rowcolor[HTML]{EFEFEF} 
Hong Kong Polytechnic University & 329 & 4.40 & 26.53 & 47.11 & 45.29 \\ \bottomrule
\end{tabular}
}
\label{tab:Institution_publication}
\end{table*}

\begin{table*}[]
\centering
\caption{Top 10 Institutions Based on Category Normalized Citation Impact in Blockchain Research Field, 2008-2023}
\resizebox{0.95\textwidth}{!}{%
\begin{tabular}{@{}lccccc@{}}
\toprule
\textbf{Institution}                & \textbf{\begin{tabular}[c]{@{}c@{}}Number of \\ Publications\end{tabular}} & \textbf{\begin{tabular}[c]{@{}c@{}}Category \\ Normalized Citation \\ Impact\end{tabular}} & \textbf{\begin{tabular}[c]{@{}c@{}}Citation \\ Impact\end{tabular}} & \textbf{\begin{tabular}[c]{@{}c@{}}\% of \\ Documents in \\ Top 10\%\end{tabular}} & \textbf{\begin{tabular}[c]{@{}c@{}}\% of \\ International \\ Collaboration\end{tabular}} \\ \midrule
Tel Aviv University                 & 32                                                                         & 20.36                                                                                      & 108.09                                                              & 56.3                                                                               & 87.5                                                                                     \\
University of Maryland College Park & 47                                                                         & 13.95                                                                                      & 79.28                                                               & 55.3                                                                               & 51.1                                                                                     \\
Cornell University                  & 98                                                                         & 12.75                                                                                      & 73.13                                                               & 54.1                                                                               & 38.8                                                                                     \\
Tennessee State University          & 13                                                                         & 12.16                                                                                      & 66.38                                                               & 53.9                                                                               & 61.5                                                                                     \\
Memorial University Newfoundland    & 20                                                                         & 12.09                                                                                      & 20.45                                                               & 30.0                                                                               & 70.0                                                                                     \\
Inje University                     & 24                                                                         & 11.93                                                                                      & 14.38                                                               & 45.8                                                                               & 16.7                                                                                     \\
University of Rochester             & 11                                                                         & 10.61                                                                                      & 84.91                                                               & 54.6                                                                               & 72.7                                                                                     \\
Anglia Ruskin University            & 12                                                                         & 10.45                                                                                      & 131.50                                                              & 50.0                                                                               & 75.0                                                                                     \\
University of Akron                 & 39                                                                         & 9.91                                                                                       & 23.74                                                               & 64.1                                                                               & 92.3                                                                                     \\
Istanbul Medeniyet University       & 24                                                                         & 9.78                                                                                       & 50.83                                                               & 66.7                                                                               & 87.5                                                                                     \\ \bottomrule
\end{tabular}
}
\label{tab:Institution_cnci}
\end{table*}

\subsection{Institution Analysis}
\begin{figure}
  \centering
  \subfloat[Category Normalized Citation Impact]{\includegraphics[width=0.45\textwidth]{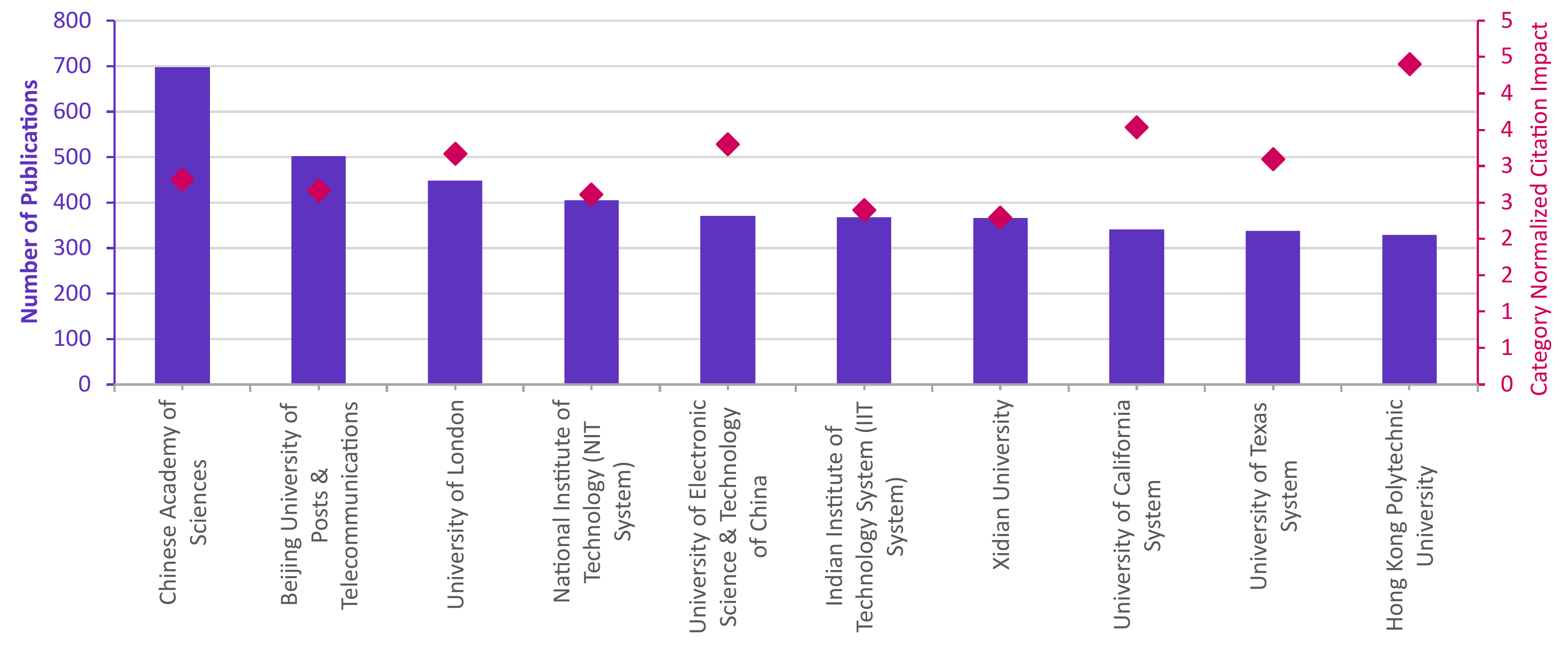}}
  \hfill
  \subfloat[\% of Documents in Top 10\%]{\includegraphics[width=0.45\textwidth]{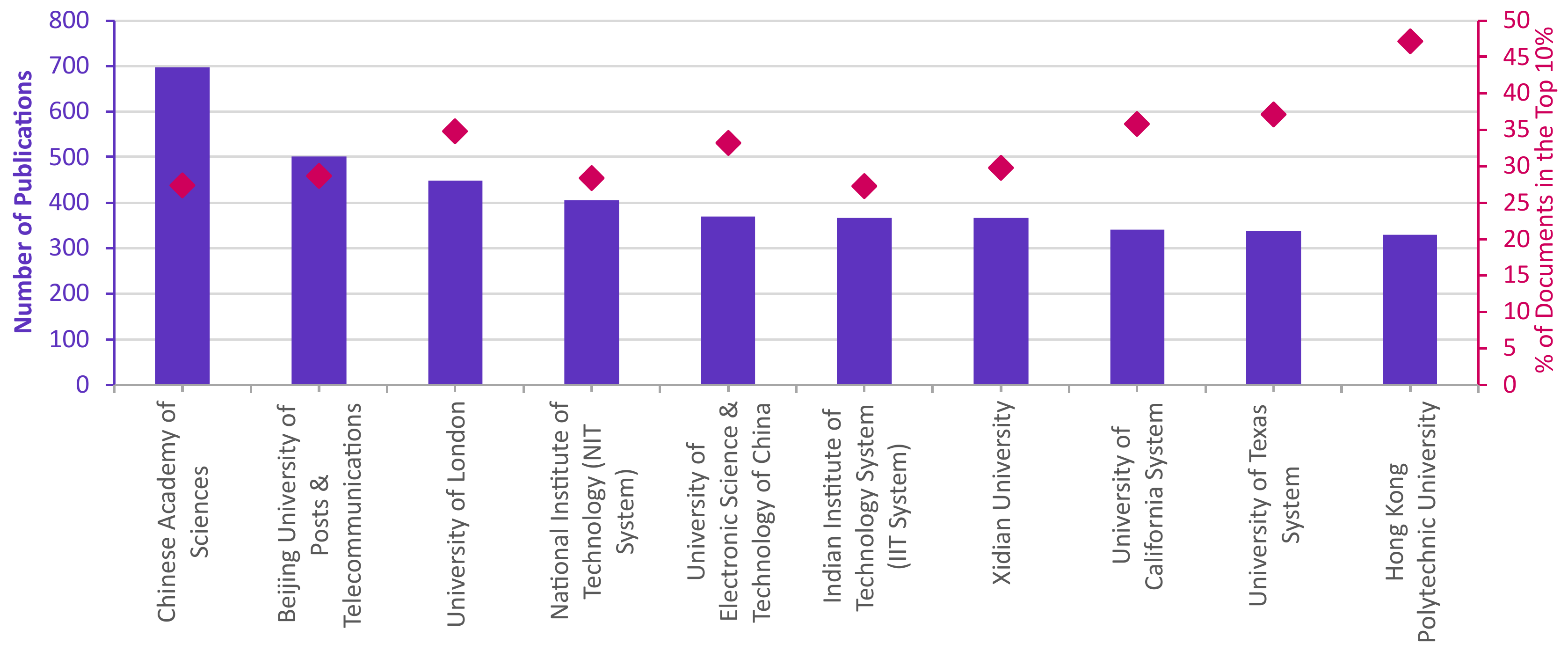}}
  \vfill
  \subfloat[\% of International Collaboration]{\includegraphics[width=0.45\textwidth]{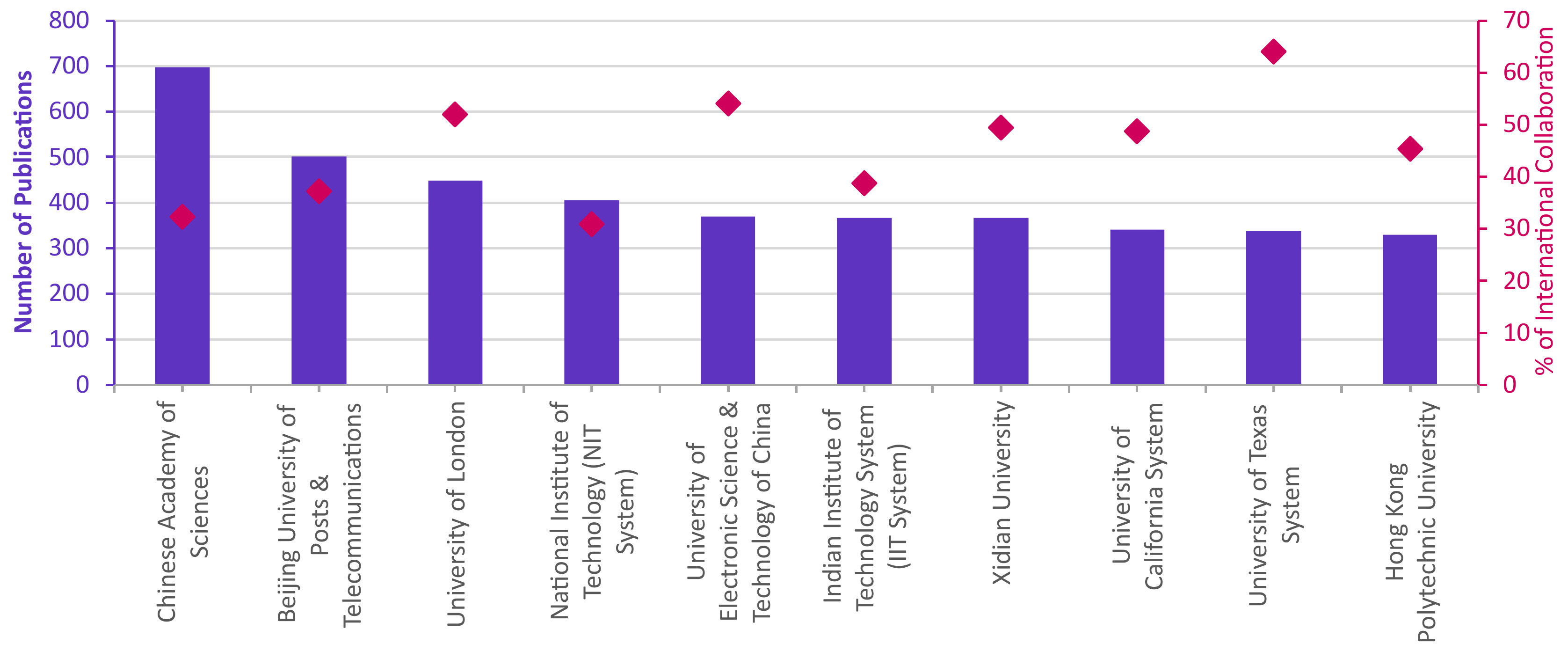}}
  \caption{Top 10 Institutions Based on Number of Publications}
  \label{fig:institution_publication}
\end{figure}

\begin{figure}
  \centering
  \subfloat[Category Normalized Citation Impact]{\includegraphics[width=0.45\textwidth]{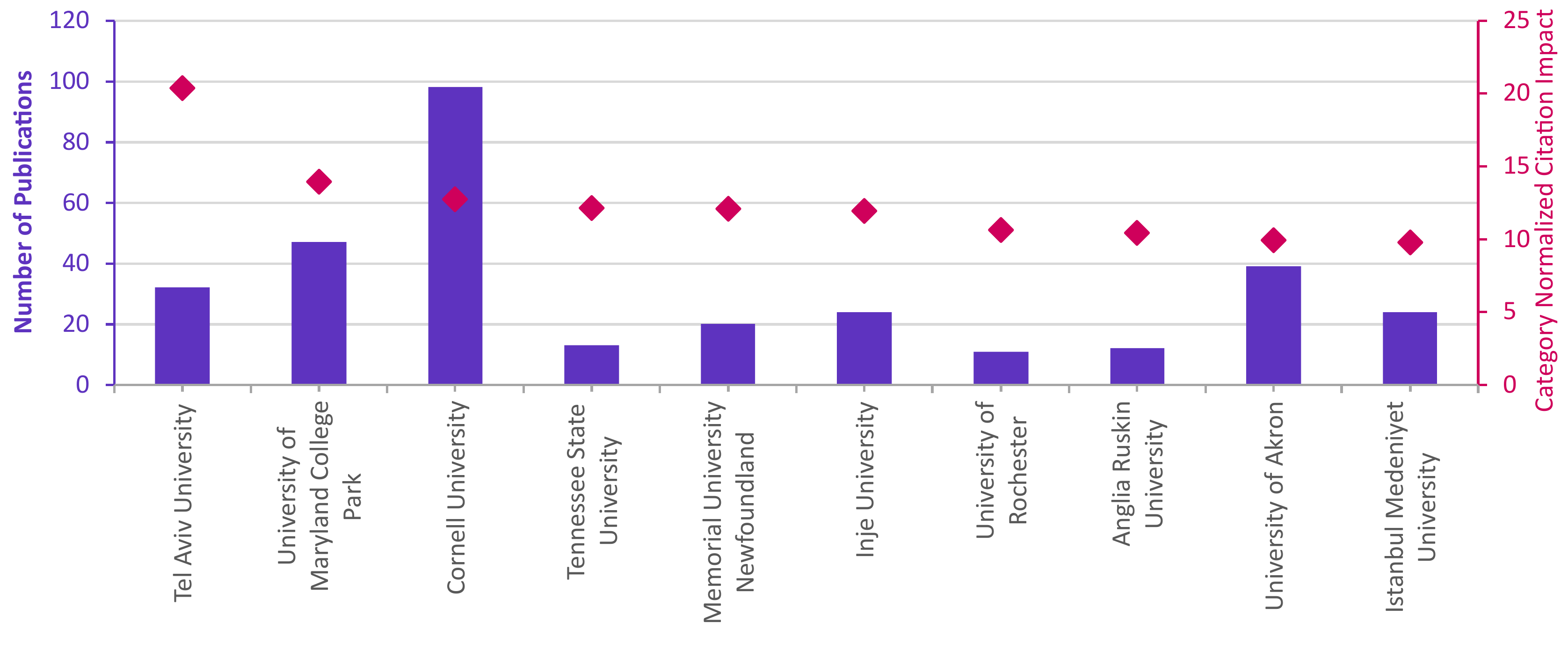}}
  \hfill
  \subfloat[\% of Documents in Top 10\%]{\includegraphics[width=0.45\textwidth]{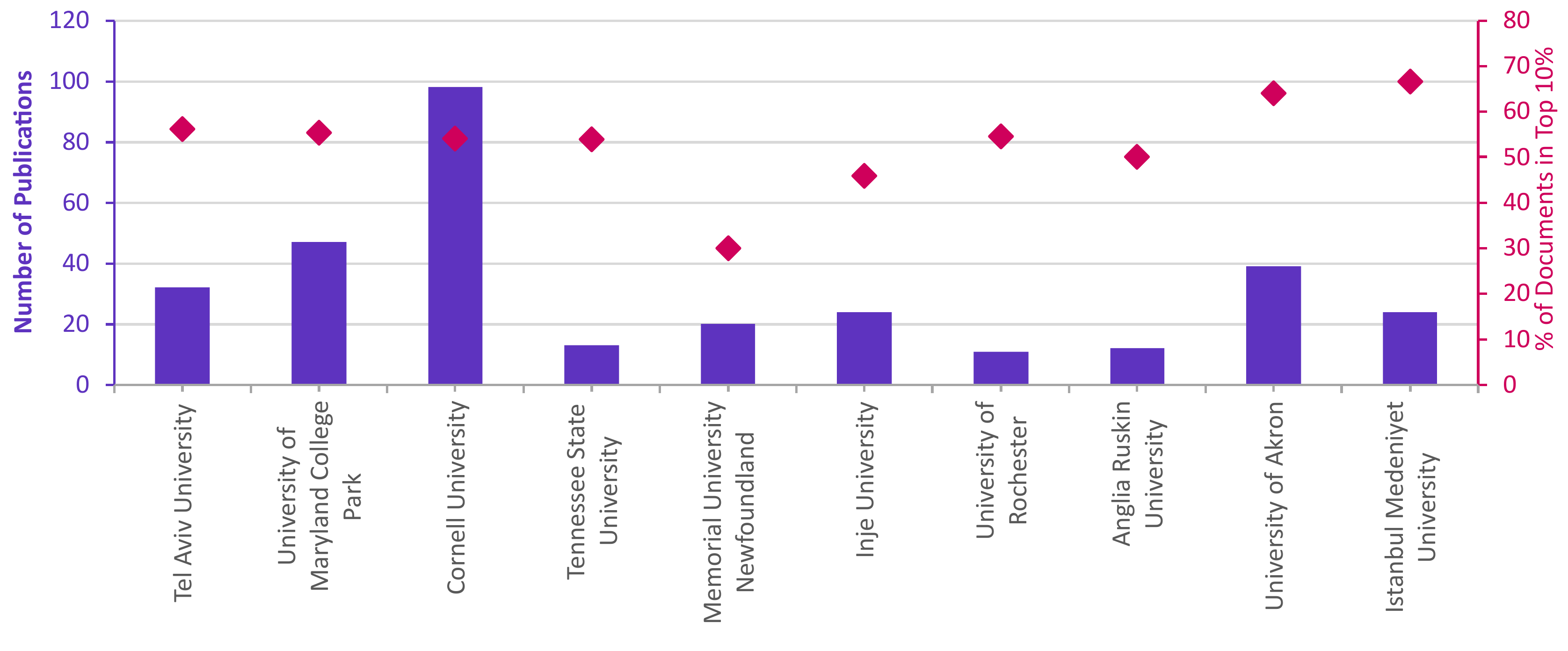}}
  \vfill
  \subfloat[\% of International Collaboration]{\includegraphics[width=0.45\textwidth]{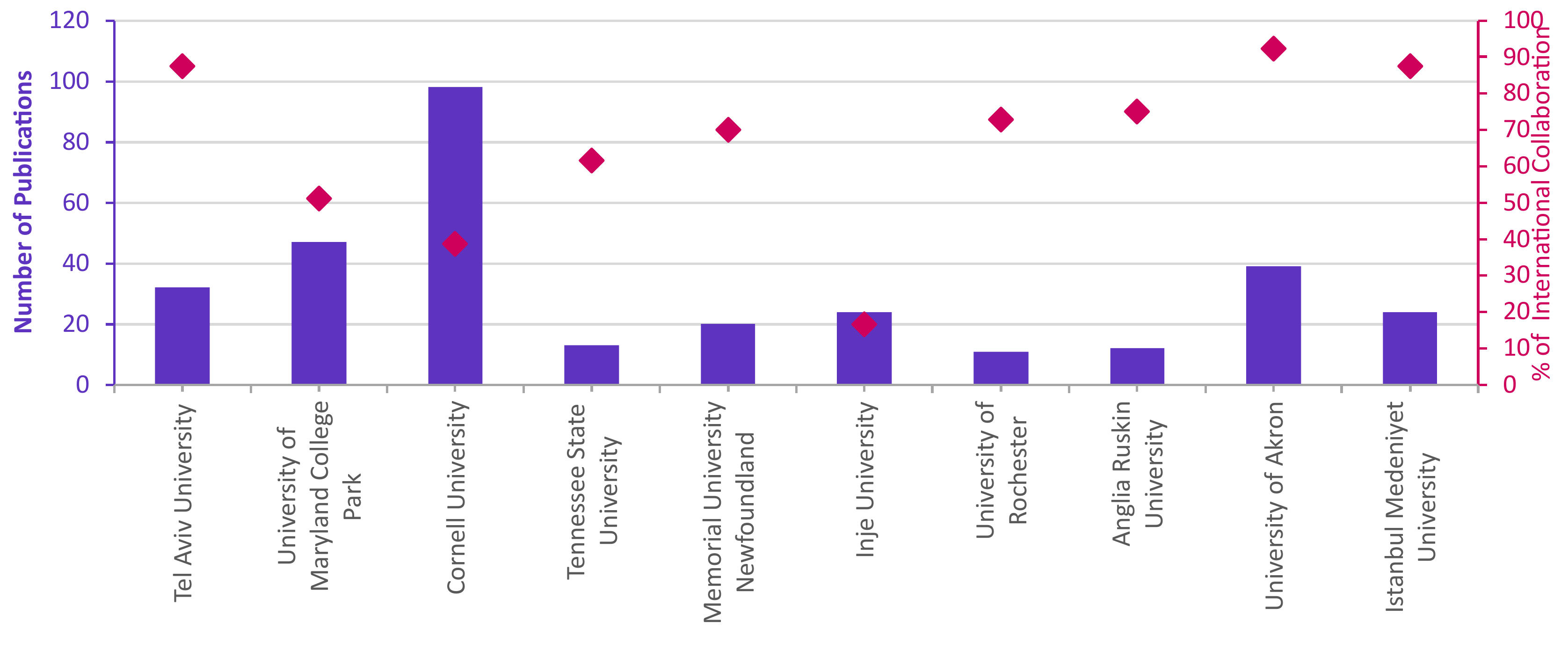}}
  \caption{Top 10 Institutions Based on Category Normalized Citation Impact}
  \label{fig:institution_cnci}
\end{figure}

The top 10 institutions were identified based on the number of publications and Category Normalized Citation Impact in blockchain research, respectively.
Table~\ref{tab:Institution_publication} and ~\ref{tab:Institution_cnci} shows the selected metrics for each institution. 
Figure~\ref{fig:institution_publication} and~\ref{fig:institution_cnci} provide the visualization of number of publications for each institution, as well as Category Normalized Citation Impact (CNCI), percentage of documents in the top 10\%, and percentage of international collaboration per institution for the period 2008 to 2023.  

\subsubsection{Based on Number of Publications}

Among the top 10 institutions, mainland China is represented by four institutions: the Chinese Academy of Sciences, Beijing University of Posts and Telecommunications, University of Electronic Science and Technology of China, and Xidian University. 
Chinese Academy of Sciences had the highest output (697 publications) of Blockchain-related research in the period from 2008 to 2023. Beijing University of Posts \& Telecommunications – China had the second highest output, with 502 publications. Moreover, the percentage of internationally collaborative publications from these two institutions are relatively modest, with percentages of 32.28\% and 37.25\% respectively.


Hong Kong Polytechnic University - Hong Kong, China had the highest CNCI of 4.40, which is more than four times higher than the world’s average CNCI. The University of California System - USA had the second highest CNCI at 3.53, three times higher than the world’s average. All top ten institutions had average CNCIs, in the period from 2008 to 2023, that were at least two times higher than the world average; with Xidian University - China having the lowest at 2.29. 

Hong Kong Polytechnic – Hong Kong, China also had the highest percentage of publications in the top 10\% (47.11\% of publications). Followed by the University of Texas System – USA (37.09\%). While most of the top institutions had at least a quarter of their publications in the top 10\%.

University of Texas System – USA had the highest percentage of internationally collaborative publications (64.09\%), followed by the University of Electronic Science \& Technology of China with 54.05\%.  The National Institute of Technology (NIT System) – India had the lowest percentage of internationally collaborative publications (30.86\%).

\subsubsection{Based on Category Normalized Citation Impact}

Based on CNCI, the United States is represented by five institutions out of the top 10, namely University of Maryland College Park, Cornell University, Tennessee State University, University of Rochester, and University of Akron.
All the Top 10 institutions by normalized citation impact had citation impacts of nearly 10 times higher than the world average. 
Tel Aviv had the highest with a citation impact of 20 times higher than the world average. It is worth noting that the number of publications for most of these institutions are quite low and therefore the category normalized citation impact and percentage of documents in top 10\% should be interpreted with caution. 
Apart from Memorial University Newfoundland and Inje University, all of the institutions had at least half of their publications in the top 10\%, with Istanbul Medeniyet University having the highest with 66.7\% of its documents in the top 10\%. University of Akron had the highest percentage of internationally collaborative papers (92.3\%). While Inje had the lowest (16.7\%)

\begin{mybox}
\textbf{Summary}: In terms of the number of publications, China Mainland has the highest number of institutions among the top 10 (4 institutions). However, based on CNCI, the United States has the highest number of institutions among the top 10 (5 institutions).
\end{mybox}




\subsection{Author Analysis}

\begin{table*}[]
\centering
\caption{Top 10 Authors Based on Number of Publications in Blockchain Research Field, 2008-2023}
\resizebox{0.95\textwidth}{!}{%
\begin{tabular}{@{}lccccc@{}}
\toprule
\textbf{Author (Surname, First Name)}                                                                                     & \textbf{\begin{tabular}[c]{@{}c@{}}Number of \\ Publications\end{tabular}} & \textbf{\begin{tabular}[c]{@{}c@{}}Category \\ Normalized Citation    \\ Impact\end{tabular}} & \textbf{\begin{tabular}[c]{@{}c@{}}Citation \\ Impact\end{tabular}} & \textbf{\begin{tabular}[c]{@{}c@{}}\% of \\ Documents in \\ Top 10\%\end{tabular}} & \textbf{\begin{tabular}[c]{@{}c@{}}\% of \\ International \\ Collaboration\end{tabular}} \\ \midrule
Tanwar,   Sudeep (Nirma   University)                                                                                     & 139                                                                        & 3.48                                                                                          & 26.0                                                                & 46.04                                                                              & 66.91                                                                                    \\
Kumar, Neeraj (Thapar   Institute of Engineering and Technology)                                                          & 129                                                                        & 3.44                                                                                          & 29.6                                                                & 52.71                                                                              & 76.74                                                                                    \\
Choo,   Kim-Kwang Raymond (University of Texas San Antonio)                                                               & 110                                                                        & 4.22                                                                                          & 49.2                                                                & 53.64                                                                              & 92.73                                                                                    \\
Salah, Khaled (Khalifa   University of Science and Technology)                                                            & 99                                                                         & 3.91                                                                                          & 49.9                                                                & 48.48                                                                              & 33.33                                                                                    \\
Zhu, Liehuang   (Beijing   Institute of Technology)                                                                       & 88                                                                         & 2.49                                                                                          & 28.9                                                                & 32.95                                                                              & 48.86                                                                                    \\
Guizani, Mohsen (Mohamed bin Zayed University of Artificial Intelligence) & 87                                                                         & 3.51                                                                                          & 37.4                                                                & 52.87                                                                              & 97.7                                                                                     \\
Gupta,   Dr. Rajesh (Nirma   University)                                                                                  & 72                                                                         & 3.41                                                                                          & 19.9                                                                & 50.00                                                                              & 65.28                                                                                    \\
Kanhere, Salil (University   of New South Wales, Sydney)                                                                  & 65                                                                         & 3.54                                                                                          & 19.5                                                                & 43.08                                                                              & 55.38                                                                                    \\
Jayaraman,   Raja (Khalifa   University of Science and Technology)                                                        & 64                                                                         & 2.80                                                                                          & 26.7                                                                & 40.62                                                                              & 29.69                                                                                    \\
Du, Xiaojiang (Stevens   Institute of Technology)                                                                         & 63                                                                         & 3.14                                                                                          & 32.0                                                                & 49.21                                                                              & 100                                                                                      \\ \bottomrule
\end{tabular}
}
\label{tab:author_publication}
\end{table*}

\begin{figure}
  \centering
  \subfloat[Category Normalized Citation Impact]{\includegraphics[width=0.45\textwidth]{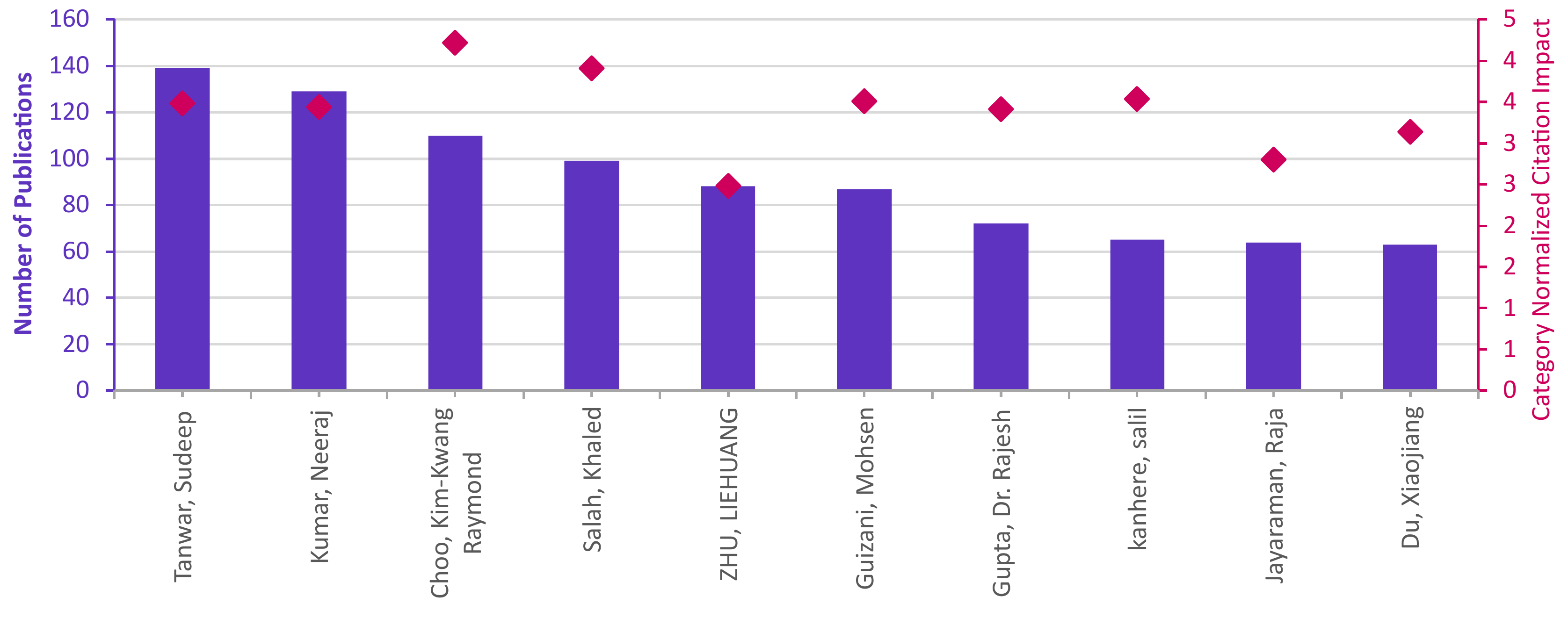}}
  \hfill
  \subfloat[\% of Documents in Top 10\%]{\includegraphics[width=0.45\textwidth]{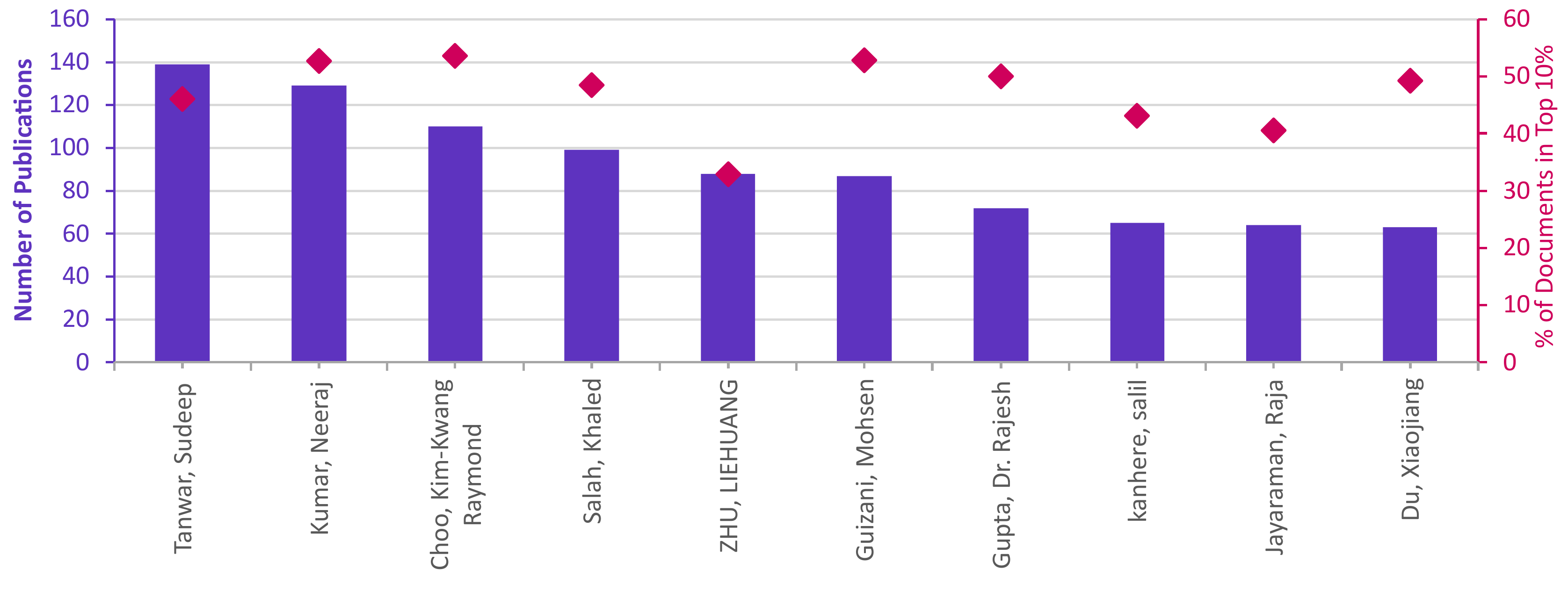}}
  \vfill
  \subfloat[\% of International Collaboration]{\includegraphics[width=0.45\textwidth]{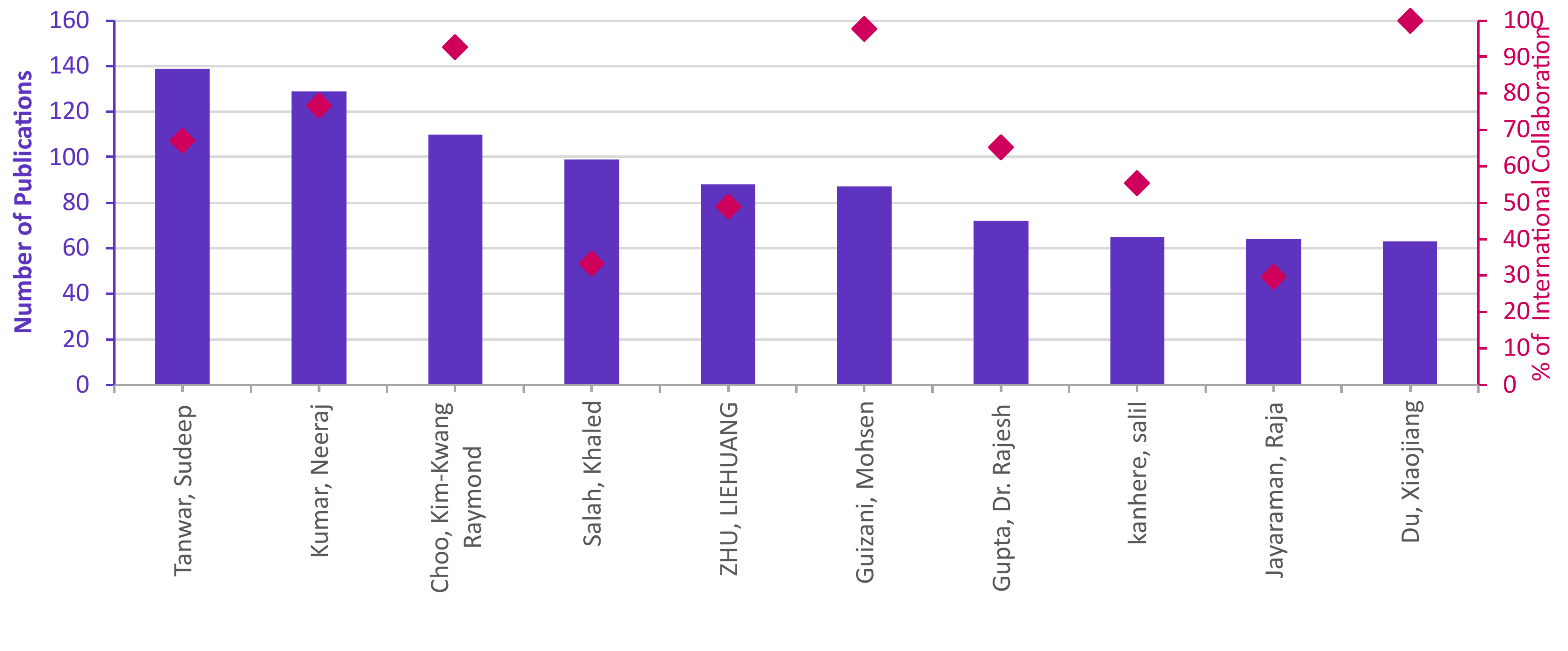}}
  \caption{Top 10 Authors Based on Number of Publications}
  \label{fig:author_publication}
\end{figure}

The top 10 authors by number of publications in Blockchain research field were identified. 
Table~\ref{tab:author_publication} shows the selected metrics for each author and Figures~\ref{fig:author_publication} provides the visualization of the number of publications for each author, as well as CNCI, percentage of documents in the top 10\%, and percentage of international collaboration per author for the period of time from 2008 to 2023.

Sundeep Tanwar from Nirma University published the highest number of publications (139 publications) in the field from 2008 to 2023, followed by Neeraj Kumar from Thapar Institute of Engineering and Technology with 129 publications. Notably, there were two researchers each affiliated with Nirma University and Khalifa University of Science and Technology in the top 10. India and the United Arab Emirates have the highest number of authors among the top 10 authors, with three authors each. 

Kim-Kwang Raymond Choo from University of Texas San Antonio published the most impactful publications of the top 10 authors by number of publications with a CNCI of 4.22, which is more than 4 times higher than the world average CNCI of 1.0. Likewise, they also had the largest percentage of documents in the top 10\% (53.64\%). All other authors in the top 10 by publication had CNCIs of at least 2 times higher than the world average and at least a third of their publications in the top 10\% of publications. 

All (100\%) of Xiaojiang Du’s from Stevens Institute of Technology publications were internationally collaborative research in the field. While a third (33.3\%) of Khaled Saled’s from Khalifa University of Science and Technology publications were internationally collaborative.




\begin{mybox}
\textbf{Summary}: The author (Tanwar Sudeep from Nirma University, India) with the highest number of publications in the field of blockchain has already published 139 articles. India and the United Arab Emirates have the highest number of authors among the top 10 authors, with three authors each.
\end{mybox}

\begin{table*}[]
\centering
\caption{Top 10 Subject Categories, Based on Number of Publications in Blockchain Research Field, 2008-2023}
\resizebox{0.95\textwidth}{!}{%
\begin{tabular}{@{}lccccc@{}}
\toprule
\textbf{Subject Category}                        & \textbf{\begin{tabular}[c]{@{}c@{}}Number of \\ Publications\end{tabular}} & \textbf{\begin{tabular}[c]{@{}c@{}}Category \\ Normalized Citation \\ Impact\end{tabular}} & \textbf{\begin{tabular}[c]{@{}c@{}}Citation \\ Impact\end{tabular}} & \textbf{\begin{tabular}[c]{@{}c@{}}\% of \\ Documents in \\ Top 10\%\end{tabular}} & \textbf{\begin{tabular}[c]{@{}c@{}}\% of \\ International \\ Collaboration\end{tabular}} \\ \midrule
COMPUTER SCIENCE, INFORMATION SYSTEMS            & 13,754                                                                     & 2.09                                                                                       & 13.53                                                               & 20.1                                                                               & 32.4                                                                                     \\
COMPUTER SCIENCE, THEORY \& METHODS              & 9,714                                                                      & 1.86                                                                                       & 10.98                                                               & 18.7                                                                               & 27.0                                                                                     \\
ENGINEERING, ELECTRICAL \& ELECTRONIC            & 9,645                                                                      & 2.44                                                                                       & 14.81                                                               & 26.0                                                                               & 34.6                                                                                     \\
TELECOMMUNICATIONS                               & 8,292                                                                      & 2.62                                                                                       & 15.12                                                               & 25.3                                                                               & 35.3                                                                                     \\
COMPUTER SCIENCE, INTERDISCIPLINARY APPLICATIONS & 4,780                                                                      & 1.94                                                                                       & 11.56                                                               & 19.6                                                                               & 27.6                                                                                     \\
COMPUTER SCIENCE, SOFTWARE ENGINEERING           & 3,298                                                                      & 1.99                                                                                       & 12.17                                                               & 19.3                                                                               & 31.2                                                                                     \\
COMPUTER SCIENCE, ARTIFICIAL INTELLIGENCE        & 3,245                                                                      & 0.89                                                                                       & 6.69                                                                & 10.7                                                                               & 25.4                                                                                     \\
COMPUTER SCIENCE, HARDWARE \& ARCHITECTURE       & 2,634                                                                      & 2.07                                                                                       & 13.52                                                               & 21.8                                                                               & 36.4                                                                                     \\
BUSINESS, FINANCE                                & 2,453                                                                      & 3.09                                                                                       & 16.67                                                               & 30.4                                                                               & 37.6                                                                                     \\
ECONOMICS                                        & 2,152                                                                      & 2.61                                                                                       & 14.29                                                               & 25.5                                                                               & 29.3                                                                                     \\ \bottomrule
\end{tabular}
}
\label{tab:category}
\end{table*}

\begin{figure}
  \centering
  \subfloat[Category Normalized Citation Impact]{\includegraphics[width=0.45\textwidth]{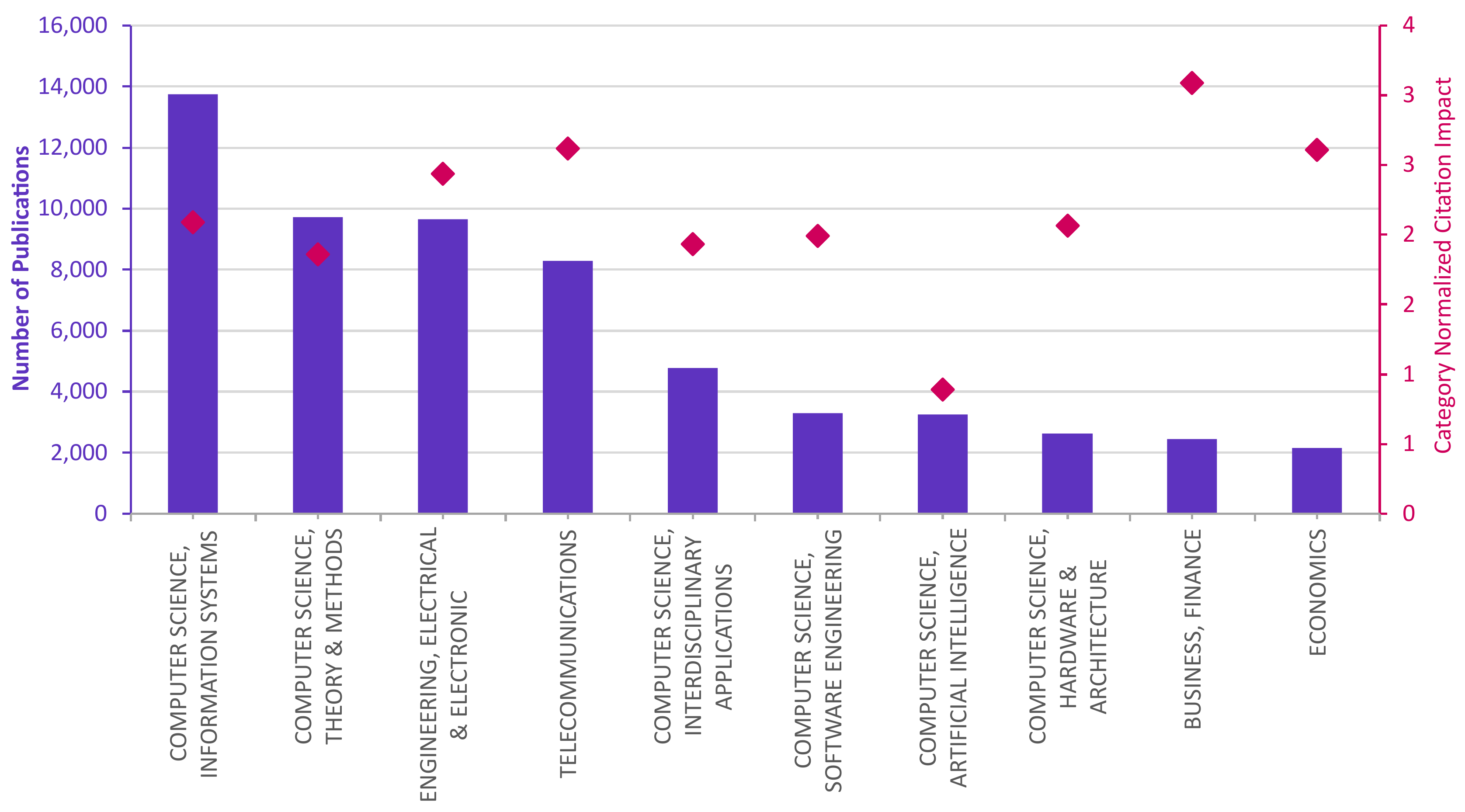}}
  \hfill
  \subfloat[\% of Documents in Top 10\%]{\includegraphics[width=0.45\textwidth]{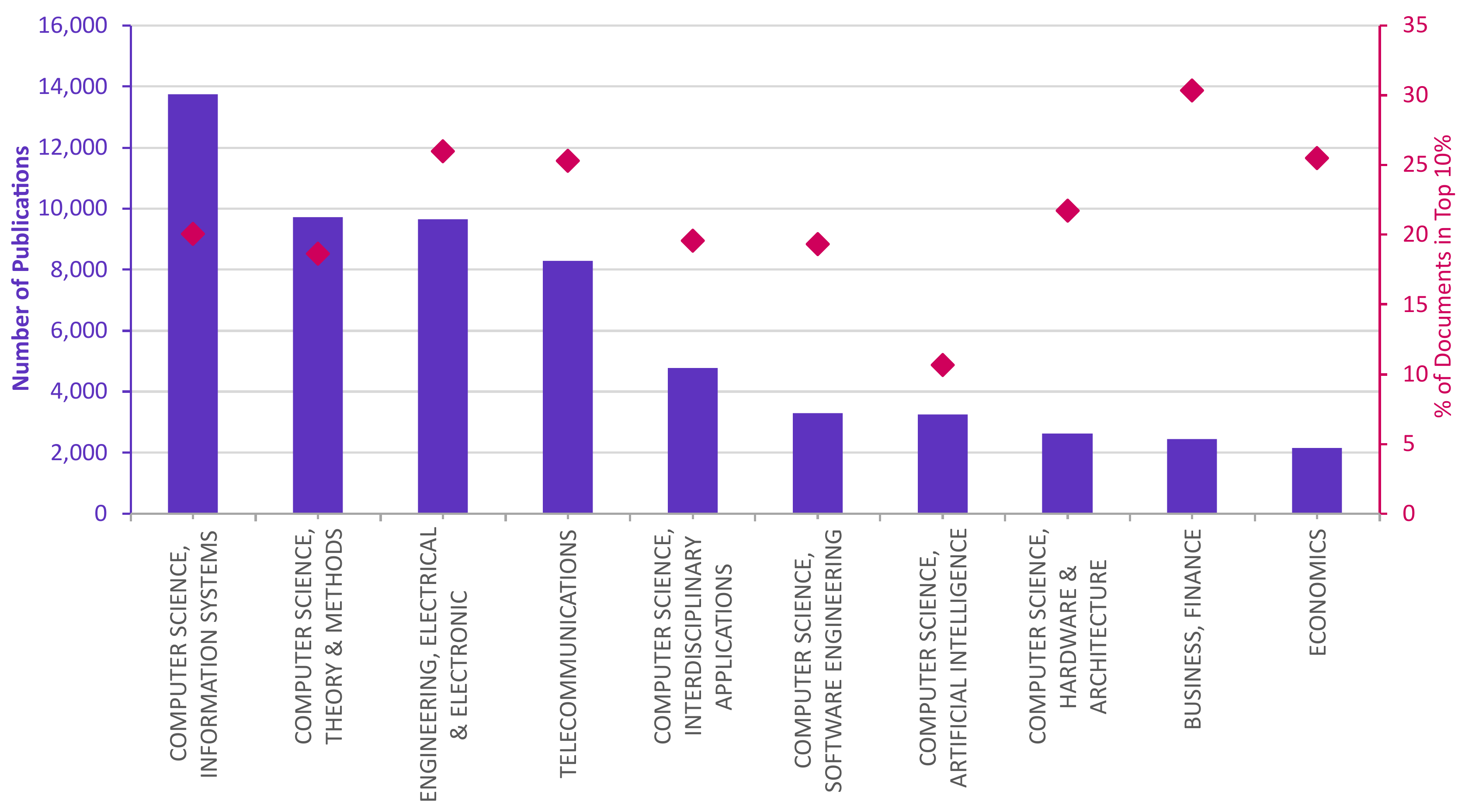}}
  \vfill
  \subfloat[\% of International Collaboration]{\includegraphics[width=0.45\textwidth]{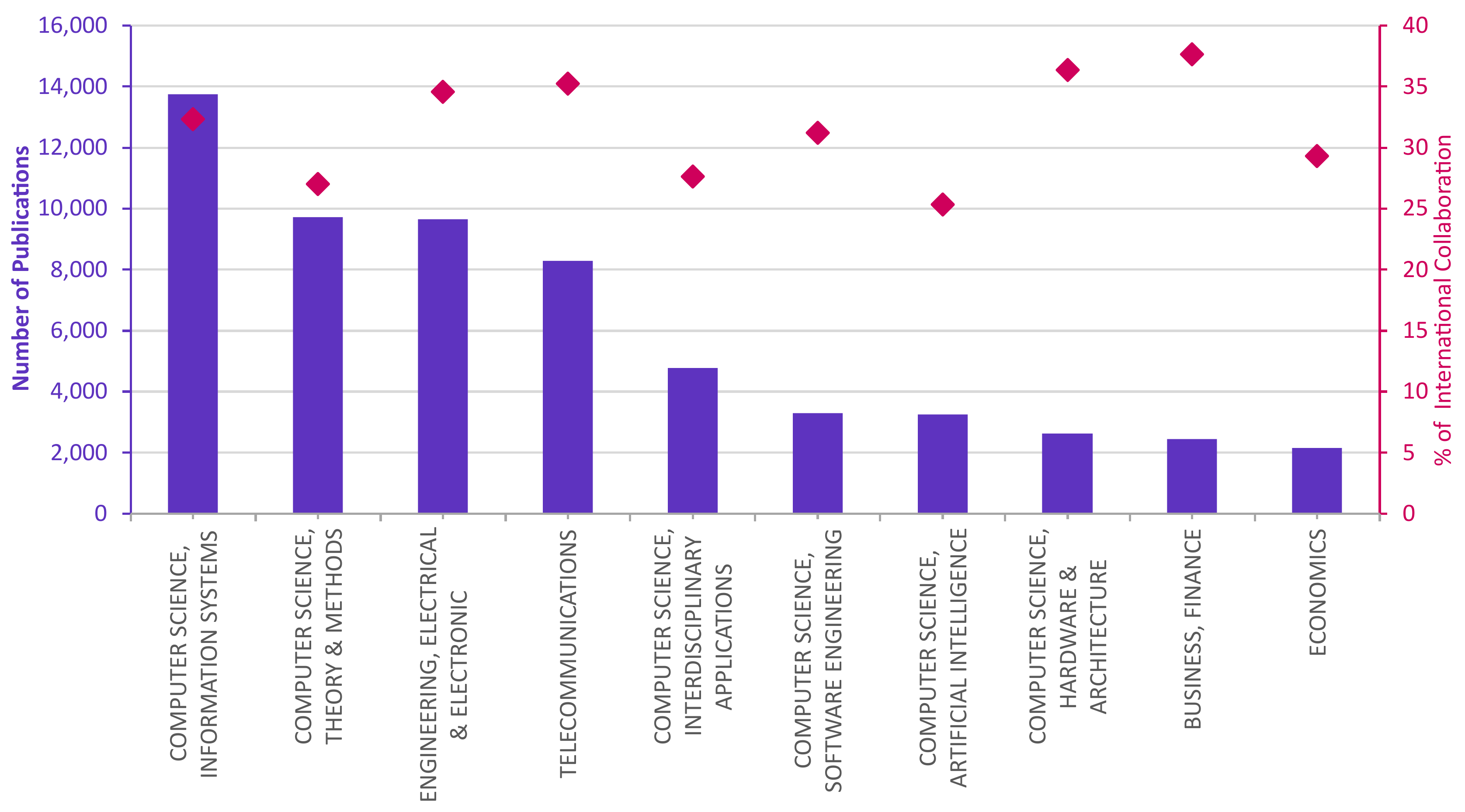}}
  \caption{Top 10 Subject Categories Based on Number of Publications}
  \label{fig:category_publication}
\end{figure}

\begin{figure}
    \centering
    \includegraphics[width=0.7\textwidth]{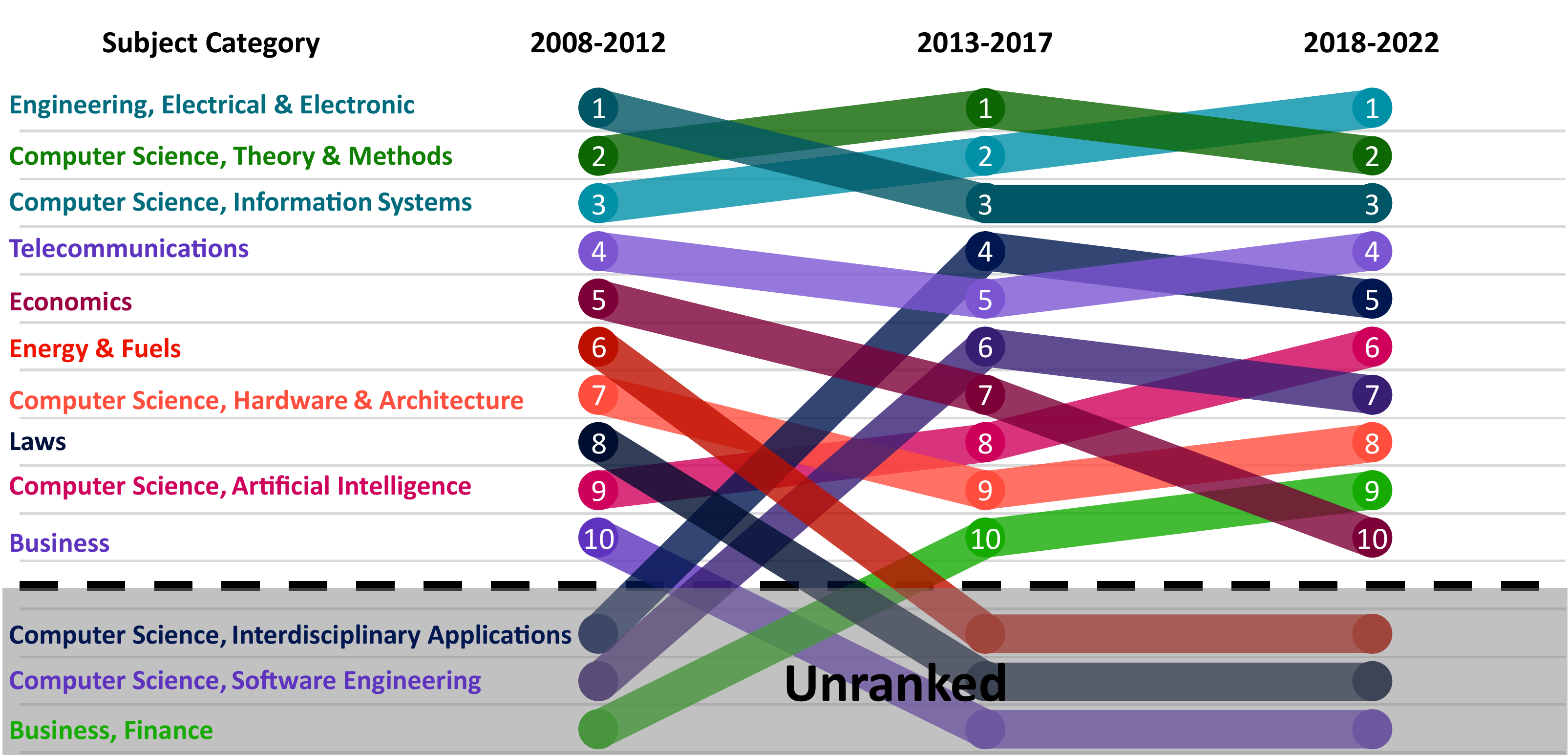}
    \caption{Top 10 Journal Subject Category By Publication Count In Each 5 Year Range, (2008-2012, 2013-2017, 2018-2022)}
    \label{fig:category_time}
\end{figure}

\subsection{Subject Category Analysis}
The top ten subject categories based on the number of publications in blockchain research were identified. 
Table~\ref{tab:category} shows the selected metrics for each category and Figures~\ref{fig:category_publication} provides the visualization of the number of publications for each category, as well as CNCI, percentage of documents in the top 10\%, and percentage of international collaboration per category for the period of time from 2008 to 2023.

Of the top 10 subject categories based on number of publications, six were related to \textit{Computer Science} with \textit{Computer Science, Information Systems} having the largest number of publications with 13,754 publications. 
This is because blockchain is a system that integrates technological achievements from various fields of computer science, including computer architecture, cryptography, distributed consensus, and smart contracts. It also relies on interdisciplinary research in areas such as game theory and economics.
Two of the categories (i.e., \textit{Business, Finance} and \textit{Economics}) were finance/economics related likely because of the applications of this technology to the area specifically its use in cryptocurrencies. 

The \textit{Business, Finance} category was the most impactful category in the list with a CNCI of 3.09, three times the world average of 1. This was followed by the \textit{Telecommunications} category with a CNCI of 2.62. \textit{Computer Science, Artificial Intelligence} had the lowest CNCI (0.89) and was the only category in the list that had a CNCI below the world average. 
Despite this, this category also had around the world average (10\%) percentage of publications in the top 10\% with 10.7\% of the publications being in the top 10\%. \textit{Business, Finance} also had the highest percentage of publications in the top 10\% with a little less than a third (30.4\%) of publications being in the top 10\%. 

All the categories had less than 40\% of publications being internationally collaborative. With the \textit{Business, Finance} category having the largest amount 37.6\% and \textit{Computer Science, Artificial Intelligence} having the least amount with around a quarter (25.4\%) being internationally collaborative research.




Figure~\ref{fig:category_time} presents the top 10 subject categories by publication count in each 5 year range. 
The top four journal subject categories have remained relatively the same over the years.
While publications in \textit{Computer Science, Interdisciplinary Applications}, and \textit{Computer Science, Software Engineering} Journals rapidly increased from 2008-2012 and 2013-2017. 
The growth in the number of papers in the field of \textit{Computer Science, Interdisciplinary Applications} is primarily due to the exploration of blockchain applications in various domains such as economics, supply chain, agriculture, and the Internet of Things (IoT) following the development of blockchain technology.
On the other hand, the increase in papers in the field of \textit{Computer Science, Software Engineering} is mainly attributed to the emergence of smart contracts. Blockchain, as a programmable platform, has demonstrated more potential in this area, leading to numerous research questions such as smart contract vulnerability detection and the development and maintenance of smart contracts.

\begin{mybox}
\textbf{Summary}: Blockchain is an interdisciplinary field, with the main published papers covering various domains within computer science, including theory, hardware, system architecture, software, and applications. Additionally, it also extends to areas such as economics and finance.
\end{mybox}


\section{The Analysis of Emerging Research Area (ERA) of Blockchain Research}\label{sec:era}

In this section, we use the co-citation clustering approach (see Section~\ref{sec:cocitation}) to identify Emerging Research Areas (ERA). 
An ERA consists of a Research Front (RF) and all co-citing publications that cite core papers. The citing publications represent the most recent work and therefore represent the leading edge of the research front.

For each ERA, we also calculate several metrics in the group, which are as follows:
\begin{itemize}
    \item \textbf{Size of foundation}: The overall number of core papers. The number of core Research Front papers would range from 2 to 50 by original definition in the Research Front methodology. 
    \item \textbf{Recency}: The average publication year of core papers, this would indicate when the Research Front emerged.
    \item \textbf{Growth}: The average annual growth rate of co-citing papers, this would measure how fast the Research Front is growing. Growth is not shown for recent ERAs.
    \item \textbf{Country/Region contribution to ERA}: The number of ERA publications (core + co-citing) that have at least one affiliation in selected country/region. 
\end{itemize}

We identified the following ERA groups: 
\begin{itemize}
    \item Top ERAs by recency
    \item Top ERAs by growth
    \item Top ERAs by China contribution in ERA
    \item Top ERAs by United States contribution in ERA
    \item Top ERAs by United Kingdom contribution in ERA
    \item Top ERAs by Asia-Pacific Region (excluding China)
\end{itemize}
By analyzing the top ERAs (Emerging Research Areas) in terms of recency and growth, we can identify the youngest research areas and determine which fields are experiencing the most rapid development. These insights are valuable for tracking the latest research trends in blockchain, inspiring further investigations for researchers, understanding the evolving landscape of blockchain development for the industry, and providing a foundation for government policy formulation.
Additionally, we categorize the top ERAs into four regions, namely China, the United States, the United Kingdom, and the Asia-Pacific Region (excluding China). This can help us understand the focus areas and distinctions in blockchain research among several major countries or regions.






\begin{table}[]
\centering
\caption{Top 10 Youngest ERAs in Blockchain Research}
\resizebox{\textwidth}{!}{%
\begin{tabular}{@{}lccccccc@{}}
\toprule
\textbf{Field}                                                                                                        & \textbf{\begin{tabular}[c]{@{}c@{}}Size of \\ foundation\end{tabular}} & \textbf{Recency} & \textbf{Growth} & \textbf{\begin{tabular}[c]{@{}c@{}}United \\ States \\ Contribution \\ to ERA\end{tabular}} & \textbf{\begin{tabular}[c]{@{}c@{}}United \\ Kingdom \\ Contribution \\ to ERA\end{tabular}} & \textbf{\begin{tabular}[c]{@{}c@{}}China \\ Contribution \\ to ERA\end{tabular}} & \textbf{\begin{tabular}[c]{@{}c@{}}Asia Pacific \\ Region \\ Contribution \\ to ERA\end{tabular}} \\ \midrule
BLOCKCHAIN ENABLED IOMT SYSTEMS                                                                                       & 2                                                                      & 2023             & 4.00            & 13\%                                                                                        & 0\%                                                                                          & 38\%                                                                             & 75\%                                                                                              \\
\rowcolor[HTML]{EFEFEF} 
\begin{tabular}[c]{@{}l@{}}BLOCKCHAIN-BASED AUTHENTICATION \\ FOR IoT SYSTEMS/APPLICATIONS\end{tabular}               & 2                                                                      & 2023             & -0.86           & 0\%                                                                                         & 0\%                                                                                          & 70\%                                                                             & 50\%                                                                                              \\
\begin{tabular}[c]{@{}l@{}}BLOCKCHAIN'S ROLE IN ACHIEVING \\ CARBON NEUTRALITY\end{tabular}                           & 2                                                                      & 2023             &                 & 0\%                                                                                         & 0\%                                                                                          & 75\%                                                                             & 13\%                                                                                              \\
\rowcolor[HTML]{EFEFEF} 
BLOCKCHAIN APPLICATIONS IN IoT DRONES                                                                                 & 3                                                                      & 2022.7           & 3.85            & 13\%                                                                                        & 9\%                                                                                          & 41\%                                                                             & 36\%                                                                                              \\
BLOCKCHAIN-BASED PRIVACY-PRESERVING                                                                                   & 2                                                                      & 2022.5           & 0.38            & 12\%                                                                                        & 9\%                                                                                          & 58\%                                                                             & 48\%                                                                                              \\
\rowcolor[HTML]{EFEFEF} 
BLOCKCHAIN-BASED IOT HEALTHCARE SYSTEMS                                                                               & 2                                                                      & 2022.5           & -0.33           & 0\%                                                                                         & 0\%                                                                                          & 14\%                                                                             & 86\%                                                                                              \\
BLOCKCHAIN APPLICATIONS IN FAULT DETECTION                                                                            & 2                                                                      & 2022.5           &                 & 0\%                                                                                         & 0\%                                                                                          & 56\%                                                                             & 28\%                                                                                              \\
\rowcolor[HTML]{EFEFEF} 
FOOD INDUSTRY 4.0 TECHNOLOGIES                                                                                        & 4                                                                      & 2022.3           & 1.14            & 28\%                                                                                        & 28\%                                                                                         & 28\%                                                                             & 52\%                                                                                              \\
BLOCKCHAIN-BASED ACCESS CONTROL SYSTEMS                                                                               & 3                                                                      & 2022.3           & -0.43           & 0\%                                                                                         & 8\%                                                                                          & 100\%                                                                            & 50\%                                                                                              \\
\rowcolor[HTML]{EFEFEF} 
\begin{tabular}[c]{@{}l@{}}INTERRELATEDNESS OF CRYPTOCURRIENCIES, \\ NFTS CLEAN ENERGY AND GREEN MARKETS\end{tabular} & 23                                                                     & 2022.3           & 0.40            & 9\%                                                                                         & 15\%                                                                                         & 36\%                                                                             & 55\%                                                                                              \\ \bottomrule
\end{tabular}
}
\label{tab:5-2}
\end{table}

\begin{table}[]
\centering
\caption{Top 10 Youngest ERAs in Blockchain Research, Foundation Size >= 5}
\resizebox{\textwidth}{!}{%
\begin{tabular}{@{}lccccccc@{}}
\toprule
\textbf{Field}                                                                                                        & \textbf{\begin{tabular}[c]{@{}c@{}}Size of \\ Foundation\end{tabular}} & \textbf{Recency} & \textbf{Growth} & \textbf{\begin{tabular}[c]{@{}c@{}}United\\ States \\ Contribution \\ to ERA\end{tabular}} & \textbf{\begin{tabular}[c]{@{}c@{}}United \\ Kingdom \\ Contribution \\ to ERA\end{tabular}} & \textbf{\begin{tabular}[c]{@{}c@{}}China \\ Contribution \\ to ERA\end{tabular}} & \textbf{\begin{tabular}[c]{@{}c@{}}Asia Pacific \\ Region \\ Contribution \\ to ERA\end{tabular}} \\ \midrule
\begin{tabular}[c]{@{}l@{}}INTERRELATEDNESS OF CRYPTOCURRIENCIES, \\ NFTS CLEAN ENERGY AND GREEN MARKETS\end{tabular} & 23                                                                     & 2022.3           & 0.40            & 9\%                                                                                        & 15\%                                                                                         & 36\%                                                                             & 55\%                                                                                              \\
\rowcolor[HTML]{EFEFEF} 
CENTRAL BANK DIGITAL CURRENCIES                                                                                       & 5                                                                      & 2022             & 0.40            & 20\%                                                                                       & 23\%                                                                                         & 33\%                                                                             & 40\%                                                                                              \\
\begin{tabular}[c]{@{}l@{}}BLOCKCHAIN BASED SECURE DATA \\ STORAGE AND DATA SHARING\end{tabular}                      & 6                                                                      & 2021             & 1.44            & 23\%                                                                                       & 11\%                                                                                         & 93\%                                                                             & 23\%                                                                                              \\
\rowcolor[HTML]{EFEFEF} 
BLOCKCHAIN-BASED DATA AUDITING SCHEMES;                                                                               & 5                                                                      & 2020.8           & 5.50            & 8\%                                                                                        & 5\%                                                                                          & 94\%                                                                             & 23\%                                                                                              \\
BLOCKCHAIN PRIVACY \& SECURITY                                                                                        & 27                                                                     & 2020.6           & 0.50            & 2\%                                                                                        & 4\%                                                                                          & 86\%                                                                             & 23\%                                                                                              \\
\rowcolor[HTML]{EFEFEF} 
\begin{tabular}[c]{@{}l@{}}BLOCKCHAIN-BASED INFORMATION SYSTEMS \\ MANAGEMENT AND SECURITY\end{tabular}               & 14                                                                     & 2020.5           & 9.22            & 15\%                                                                                       & 11\%                                                                                         & 40\%                                                                             & 47\%                                                                                              \\
BLOCKCHAINED FEDERATED LEARNING                                                                                       & 10                                                                     & 2020.5           & 1.36            & 14\%                                                                                       & 7\%                                                                                          & 54\%                                                                             & 38\%                                                                                              \\
\rowcolor[HTML]{EFEFEF} 
BLOCKCHAIN APPLICATIONS IN CONSTRUCTION INDUSTRY                                                                      & 7                                                                      & 2020.4           & 1.67            & 11\%                                                                                       & 17\%                                                                                         & 47\%                                                                             & 28\%                                                                                              \\
SECURE BLOCKCHAIN-BASED DATA TRANSMISSION                                                                             & 5                                                                      & 2020.4           & -0.25           & 10\%                                                                                       & 3\%                                                                                          & 91\%                                                                             & 55\%                                                                                              \\
\rowcolor[HTML]{EFEFEF} 
BLOCKCHAIN AND FEDERATED LEARNING                                                                                     & 6                                                                      & 2020.2           & 1.65            & 19\%                                                                                       & 5\%                                                                                          & 69\%                                                                             & 41\%                                                                                              \\ \bottomrule
\end{tabular}
}
\label{tab:5-3}
\end{table}

\smallskip \noindent \textbf{Top ERAs by Recency}
Table~\ref{tab:5-2} (\ref{tab:5-3}) presents the top 10 youngest ERAs in blockchain research (with a foundation size of 5 or more). As shown in the tables, the majority of these ERAs primarily revolve around the applications of blockchain technology, encompassing areas like IOMT (Internet-of-Medical-Things) systems, IoT systems, IoT drones, Industry 4.0, and more. 
Among the top 10 youngest ERAs with a foundation size of 5 or more, blockchain researchers also exhibit interest in the domains of privacy and security, as evidenced by research areas such as \textit{Blockchain-Based Secure Data Storage and Data Share} and \textit{Blockchain Privacy \& Security}. Additionally, there is a focus on the intersection of blockchain and federated learning, as seen in research topics like \textit{Blckchained Federated Learning} and \textit{Blockchain and Federated Learning}.
In terms of contributions to the top 10 youngest ERAs, China and the Asia Pacific Region exhibit a higher level of involvement compared to the United States and the United Kingdom. However, for the top 10 youngest ERAs with a foundation size of 5 or more, the contribution from the United States and the United Kingdom shows an increase. 

\begin{mybox}
\textbf{Summary}: The most recent ERAs tend to surround the topic of Blockchain technology applications; China and the Asia Pacific Region make more contributions than the United States and the United Kingdom in the top 10 youngest ERAs. 
\end{mybox}

\begin{table}[]
\centering
\caption{Top 10 ERAs By Greatest Growth in Blockchain Research}
\resizebox{\textwidth}{!}{%
\begin{tabular}{@{}lccccccc@{}}
\toprule
\textbf{Field}                                                                                                 & \textbf{\begin{tabular}[c]{@{}c@{}}Size of \\ Foundation\end{tabular}} & \textbf{Recency} & \textbf{Growth} & \textbf{\begin{tabular}[c]{@{}c@{}}United \\ States \\ Contribution \\ to ERA\end{tabular}} & \textbf{\begin{tabular}[c]{@{}c@{}}United \\ Kingdom \\ Contribution \\ to ERA\end{tabular}} & \textbf{\begin{tabular}[c]{@{}c@{}}China \\ Contribution \\ to ERA\end{tabular}} & \textbf{\begin{tabular}[c]{@{}c@{}}Asia Pacific \\ Region \\ Contribution \\ to ERA\end{tabular}} \\ \midrule
CRYPTOCURRENCY TRADING AND BLOCKCHAIN                                                                       & 2                                                                      & 2020.5           & 9.50            & 16\%                                                                                        & 4\%                                                                                          & 24\%                                                                             & 40\%                                                                                              \\
\rowcolor[HTML]{EFEFEF} 
\begin{tabular}[c]{@{}l@{}}BLOCKCHAIN-BASED INFORMATION SYSTEMS \\ MANAGEMENT AND SECURITY\end{tabular}     & 14                                                                     & 2020.5           & 9.22            & 15\%                                                                                        & 11\%                                                                                         & 40\%                                                                             & 47\%                                                                                              \\
BLOCKCHAIN-BASED ECDSA;BLOCKCHAIN-ENABLED IOMT                                                              & 2                                                                      & 2022             & 9.13            & 0\%                                                                                         & 14\%                                                                                         & 54\%                                                                             & 46\%                                                                                              \\
\rowcolor[HTML]{EFEFEF} 
BLOCKCHAIN APPLICATIONS TO 6G TECHNOLOGY                                                                    & 4                                                                      & 2022             & 8.69            & 19\%                                                                                        & 6\%                                                                                          & 38\%                                                                             & 54\%                                                                                              \\
FEDERATED LEARNING AND BLOCKCHAIN                                                                           & 2                                                                      & 2021             & 5.59            & 13\%                                                                                        & 13\%                                                                                         & 29\%                                                                             & 39\%                                                                                              \\
\rowcolor[HTML]{EFEFEF} 
BLOCKCHAIN-BASED DATA AUDITING SCHEMES;                                                                     & 5                                                                      & 2020.8           & 5.50            & 8\%                                                                                         & 5\%                                                                                          & 94\%                                                                             & 23\%                                                                                              \\
\begin{tabular}[c]{@{}l@{}}BLOCKCHAIN TECHNOLOGY APPLICATIONS \\ IN SUPPLY CHAIN CHAINS\end{tabular}        & 2                                                                      & 2019             & 5.05            & 16\%                                                                                        & 30\%                                                                                         & 19\%                                                                             & 37\%                                                                                              \\
\rowcolor[HTML]{EFEFEF} 
BLOCKCHAIN AND IOT INTEGRATION AND SECURITY                                                                 & 4                                                                      & 2018             & 4.39            & 13\%                                                                                        & 11\%                                                                                         & 17\%                                                                             & 49\%                                                                                              \\
\begin{tabular}[c]{@{}l@{}}BLOCKCHAIN-BASED AGRI-FOOD \\ SUPPLY CHAIN AND TRACEABILITY SYSTEMS\end{tabular} & 3                                                                      & 2019.3           & 4.38            & 5\%                                                                                         & 12\%                                                                                         & 29\%                                                                             & 49\%                                                                                              \\
\rowcolor[HTML]{EFEFEF} 
BLOCKCHAIN-BASED VEHICULAR NETWORKS                                                                         & 4                                                                      & 2018.8           & 3.61            & 16\%                                                                                        & 10\%                                                                                         & 64\%                                                                             & 34\%                                                                                              \\ \bottomrule
\end{tabular}
}
\label{tab:5-4}
\end{table}
\begin{table}[]
\centering
\caption{Top 10 ERAs By Greatest Growth in Blockchain Research, Foundation Size >= 5}
\resizebox{\textwidth}{!}{%
\begin{tabular}{@{}lccccccc@{}}
\toprule
\textbf{Field}                                                                                          & \textbf{\begin{tabular}[c]{@{}c@{}}Size of \\ Foundation\end{tabular}} & \textbf{Recency} & \textbf{Growth} & \textbf{\begin{tabular}[c]{@{}c@{}}United \\ States \\ Contribution \\ to ERA\end{tabular}} & \textbf{\begin{tabular}[c]{@{}c@{}}United \\ Kingdom \\ Contribution \\ to ERA\end{tabular}} & \textbf{\begin{tabular}[c]{@{}c@{}}China \\ Contribution \\ to ERA\end{tabular}} & \textbf{\begin{tabular}[c]{@{}c@{}}Asia Pacific \\ Region \\ Contribution \\ to ERA\end{tabular}} \\ \midrule
\begin{tabular}[c]{@{}l@{}}BLOCKCHAIN-BASED INFORMATION \\ SYSTEMS MANAGEMENT AND SECURITY\end{tabular} & 14                                                                     & 2020.5           & 9.22            & 15\%                                                                                        & 11\%                                                                                         & 40\%                                                                             & 47\%                                                                                              \\
\rowcolor[HTML]{EFEFEF} 
BLOCKCHAIN-BASED   DATA AUDITING SCHEMES;                                                               & 5                                                                      & 2020.8           & 5.50            & 8\%                                                                                         & 5\%                                                                                          & 94\%                                                                             & 23\%                                                                                              \\
BLOCKCHAIN IN HEALTHCARE SECTOR                                                                         & 14                                                                     & 2018.6           & 3.32            & 17\%                                                                                        & 9\%                                                                                          & 19\%                                                                             & 48\%                                                                                              \\
\rowcolor[HTML]{EFEFEF} 
CRYPTOCURRENCIES AS A SAFE HAVEN FOR EQUITY MARKETS                                                     & 14                                                                     & 2019.4           & 2.67            & 9\%                                                                                         & 14\%                                                                                         & 19\%                                                                             & 38\%                                                                                              \\
CRYPTOCURRENCY MARKET   EFFIECIENCY                                                                     & 5                                                                      & 2018.2           & 2.62            & 9\%                                                                                         & 15\%                                                                                         & 15\%                                                                             & 31\%                                                                                              \\
\rowcolor[HTML]{EFEFEF} 
BLOCKCHAIN   APPLICATIONS TO ENERGY TRADING (PEER-TO-PEER)                                              & 29                                                                     & 2019.2           & 2.56            & 12\%                                                                                        & 12\%                                                                                         & 29\%                                                                             & 28\%                                                                                              \\
BLOCKCHAIN APPLICATIONS IN   CONSTRUCTION INDUSTRY                                                      & 7                                                                      & 2020.4           & 1.67            & 11\%                                                                                        & 17\%                                                                                         & 47\%                                                                             & 28\%                                                                                              \\
\rowcolor[HTML]{EFEFEF} 
BLOCKCHAIN   AND FEDERATED LEARNING                                                                     & 6                                                                      & 2020.2           & 1.65            & 19\%                                                                                        & 5\%                                                                                          & 69\%                                                                             & 41\%                                                                                              \\
BLOCKCHAIN BASED SECURE DATA  STORAGE AND DATA SHARING                                                  & 6                                                                      & 2021             & 1.44            & 23\%                                                                                        & 11\%                                                                                         & 93\%                                                                             & 23\%                                                                                              \\
\rowcolor[HTML]{EFEFEF} 
BLOCKCHAINED   FEDERATED LEARNING                                                                       & 10                                                                     & 2020.5           & 1.36            & 14\%                                                                                        & 7\%                                                                                          & 54\%                                                                             & 38\%                                                                                              \\ \bottomrule
\end{tabular}
}
\label{tab:5-5}
\end{table}

\smallskip \noindent \textbf{Top ERAs by Growth}
Table~\ref{tab:5-4} (~\ref{tab:5-5}) presents the top 10 ERAs by greatest growth in blockchain research (with a foundation size of 5 or more).
The ERA with the greatest growth was \textit{Cryptocurrency Trading}, followed by \textit{Blockchain enabled Information Systems and Security}. 
Of the top ERAs with the greatest growth, the most recent are \textit{Blockchain Applications to 6G Technology} and \textit{Blockchain-based ECDSA; Blockchain-enabled IOMT}. 
If we restrict the ERA's foundation size to be greater than or equal to 5, the ERA that experienced the highest growth was \textit{Blockchain-based Information Systems Management and Security}. 

\begin{mybox}
\textbf{Summary}: The area that experienced the most significant growth was \textit{Cryptocurrency Trading}, and many of the rapidly growing areas are also associated with the application of Blockchain technology.
\end{mybox}



\begin{table}[]
\centering
\caption{Top 10 ERAs Where China Has the Highest Contribution in Blockchain Research}
\resizebox{\textwidth}{!}{%
\begin{tabular}{@{}lccccccc@{}}
\toprule
\textbf{Field}                                           & \textbf{\begin{tabular}[c]{@{}l@{}}Size of \\ Foundation\end{tabular}} & \textbf{Recency} & \textbf{Growth} & \textbf{\begin{tabular}[c]{@{}l@{}}United \\ States \\ Contribution \\ to ERA\end{tabular}} & \textbf{\begin{tabular}[c]{@{}l@{}}United \\ Kingdom \\ Contribution \\ to ERA\end{tabular}} & \textbf{\begin{tabular}[c]{@{}l@{}}China \\ Contribution \\ to ERA\end{tabular}} & \textbf{\begin{tabular}[c]{@{}l@{}}Asia Pacific \\ Region \\ Contribution \\ to ERA\end{tabular}} \\ \midrule
BLOCKCHAIN-BASED ACCESS CONTROL SYSTEMS               & 3                                                                      & 2022.3           & -0.43           & 0\%                                                                                         & 8\%                                                                                          & 100\%                                                                            & 50\%                                                                                              \\
\rowcolor[HTML]{EFEFEF} 
BLOCKCHAIN AND GREEN INNOVATION                       & 3                                                                      & 2021.7           & -0.75           & 0\%                                                                                         & 0\%                                                                                          & 100\%                                                                            & 11\%                                                                                              \\
BLOCKCHAIN-ENABLED MOBILE-EDGE COMPUTING              & 2                                                                      & 2020.5           & -0.18           & 17\%                                                                                        & 6\%                                                                                          & 100\%                                                                            & 29\%                                                                                              \\
\rowcolor[HTML]{EFEFEF} 
MOBILE EDGE CROWDSENSING AND BLOCKCHAIN               & 3                                                                      & 2021             & 1.64            & 16\%                                                                                        & 2\%                                                                                          & 98\%                                                                             & 4\%                                                                                               \\
EDGE INTELLIGENCE AND BLOCKCHAIN                      & 2                                                                      & 2019             & -0.46           & 2\%                                                                                         & 16\%                                                                                         & 98\%                                                                             & 31\%                                                                                              \\
\rowcolor[HTML]{EFEFEF} 
RISK ASSESSMENT OF CRYPTOCURRENCIES                   & 2                                                                      & 2021             & -0.41           & 0\%                                                                                         & 4\%                                                                                          & 96\%                                                                             & 67\%                                                                                              \\
BLOCKCHAIN-ENABLED COMPUTATION OFFLOADING             & 5                                                                      & 2019.2           & 0.44            & 11\%                                                                                        & 5\%                                                                                          & 96\%                                                                             & 23\%                                                                                              \\
\rowcolor[HTML]{EFEFEF} 
BLOCKCHAIN-BASED DATA AUDITING SCHEMES;               & 5                                                                      & 2020.8           & 5.50            & 8\%                                                                                         & 5\%                                                                                          & 94\%                                                                             & 23\%                                                                                              \\
BLOCKCHAIN BASED SECURE DATA STORAGE AND DATA SHARING & 6                                                                      & 2021             & 1.44            & 23\%                                                                                        & 11\%                                                                                         & 93\%                                                                             & 23\%                                                                                              \\
\rowcolor[HTML]{EFEFEF} 
BLOCKCHAIN-BASED ON-DEMAND COMPUTING RESOURCE TRADING & 2                                                                      & 2020.5           & -0.94           & 3\%                                                                                         & 0\%                                                                                          & 92\%                                                                             & 15\%                                                                                              \\ \bottomrule
\end{tabular}
}
\label{tab:5-6}
\end{table}
\begin{table}[]
\centering
\caption{Top 10 ERAs Where China Has the Highest Contribution in Blockchain Research, Foundation Size >=5}
\resizebox{\textwidth}{!}{%
\begin{tabular}{@{}llllllll@{}}
\toprule
\textbf{Field}                                               & \textbf{\begin{tabular}[c]{@{}l@{}}Size of \\ Foundation\end{tabular}} & \textbf{Recency} & \textbf{Growth} & \textbf{\begin{tabular}[c]{@{}l@{}}United \\ States \\ Contribution \\ to ERA\end{tabular}} & \textbf{\begin{tabular}[c]{@{}l@{}}United \\ Kingdom \\ Contribution \\ to ERA\end{tabular}} & \textbf{\begin{tabular}[c]{@{}l@{}}China \\ Contribution \\ to ERA\end{tabular}} & \textbf{\begin{tabular}[c]{@{}l@{}}Asia Pacific \\ Region \\ Contribution \\ to ERA\end{tabular}} \\ \midrule
BLOCKCHAIN-ENABLED COMPUTATION OFFLOADING                    & 5                                                                      & 2019.2           & 0.44            & 11\%                                                                                        & 5\%                                                                                          & 96\%                                                                             & 23\%                                                                                              \\
BLOCKCHAIN-BASED DATA AUDITING SCHEMES;                      & 5                                                                      & 2020.8           & 5.50            & 8\%                                                                                         & 5\%                                                                                          & 94\%                                                                             & 23\%                                                                                              \\
BLOCKCHAIN BASED SECURE DATA STORAGE AND DATA SHARING        & 6                                                                      & 2021             & 1.44            & 23\%                                                                                        & 11\%                                                                                         & 93\%                                                                             & 23\%                                                                                              \\
SECURE BLOCKCHAIN-BASED DATA TRANSMISSION                    & 5                                                                      & 2020.4           & -0.25           & 10\%                                                                                        & 3\%                                                                                          & 91\%                                                                             & 55\%                                                                                              \\
BLOCKCHAIN PRIVACY \& SECURITY                               & 27                                                                     & 2020.6           & 0.50            & 2\%                                                                                         & 4\%                                                                                          & 86\%                                                                             & 23\%                                                                                              \\
BLOCKCHAIN SUPPLY CHAIN TRACEABILITY                         & 11                                                                     & 2020.1           & 0.75            & 16\%                                                                                        & 15\%                                                                                         & 70\%                                                                             & 19\%                                                                                              \\
BLOCKCHAIN AND FEDERATED LEARNING                            & 6                                                                      & 2020.2           & 1.65            & 19\%                                                                                        & 5\%                                                                                          & 69\%                                                                             & 41\%                                                                                              \\
BLOCKCHAINED FEDERATED LEARNING                              & 10                                                                     & 2020.5           & 1.36            & 14\%                                                                                        & 7\%                                                                                          & 54\%                                                                             & 38\%                                                                                              \\
BLOCKCHAIN APPLICATIONS IN CONSTRUCTION INDUSTRY             & 7                                                                      & 2020.4           & 1.67            & 11\%                                                                                        & 17\%                                                                                         & 47\%                                                                             & 28\%                                                                                              \\
BLOCKCHAIN-BASED INFORMATION SYSTEMS MANAGEMENT AND SECURITY & 14                                                                     & 2020.5           & 9.22            & 15\%                                                                                        & 11\%                                                                                         & 40\%                                                                             & 47\%                                                                                              \\ \bottomrule
\end{tabular}
}
\label{tab:5-7}
\end{table}

\begin{table}[]
\centering
\caption{Top 10 ERAs Where APAC, Excluding China, Has the Highest Contribution in Blockchain Research}
\resizebox{\textwidth}{!}{%
\begin{tabular}{@{}lccccccc@{}}
\toprule
\textbf{Field}                                          & \textbf{\begin{tabular}[c]{@{}c@{}}Size of \\ Foundation\end{tabular}} & \textbf{Recency} & \textbf{Growth} & \textbf{\begin{tabular}[c]{@{}c@{}}United \\ States \\ Contribution \\ to ERA\end{tabular}} & \textbf{\begin{tabular}[c]{@{}c@{}}United \\ Kingdom \\ Contribution \\ to ERA\end{tabular}} & \textbf{\begin{tabular}[c]{@{}c@{}}China \\ Contribution \\ to ERA\end{tabular}} & \textbf{\begin{tabular}[c]{@{}c@{}}Asia Pacific \\ Region \\ Contribution \\ to ERA\end{tabular}} \\ \midrule
BLOCKCHAIN-BASED TRUST MODELS                           & 2                                                                      & 2020.5           & -0.25           & 22\%                                                                                        & 33\%                                                                                         & 44\%                                                                             & 89\%                                                                                              \\
\rowcolor[HTML]{EFEFEF} 
BLOCKCHAIN-BASED SMART APPLICATIONS                     & 2                                                                      & 2020.5           & -0.34           & 13\%                                                                                        & 0\%                                                                                          & 17\%                                                                             & 88\%                                                                                              \\
BLOCKCHAIN-BASED IOT HEALTHCARE SYSTEMS                 & 2                                                                      & 2022.5           & -0.33           & 0\%                                                                                         & 0\%                                                                                          & 14\%                                                                             & 86\%                                                                                              \\
\rowcolor[HTML]{EFEFEF} 
BLOCKCHAIN-BASED HEALTHCARE                             & 2                                                                      & 2022             & 1.50            & 33\%                                                                                        & 0\%                                                                                          & 0\%                                                                              & 78\%                                                                                              \\
FEDERATED LEARNING AND BLOCKCHAIN                       & 2                                                                      & 2022             & -0.43           & 8\%                                                                                         & 0\%                                                                                          & 46\%                                                                             & 77\%                                                                                              \\
\rowcolor[HTML]{EFEFEF} 
BLOCKCHAIN-DRIVEN SUPPLY CHAIN                          & 3                                                                      & 2022             & 0.25            & 0\%                                                                                         & 58\%                                                                                         & 25\%                                                                             & 75\%                                                                                              \\
BLOCKCHAIN ENABLED IOMT SYSTEMS                         & 2                                                                      & 2023             & 4.00            & 13\%                                                                                        & 0\%                                                                                          & 38\%                                                                             & 75\%                                                                                              \\
\rowcolor[HTML]{EFEFEF} 
BLOCKCHAIN-ENHANCED DATA SHARING AND PRIVACY PRESERVING & 2                                                                      & 2021             & -0.73           & 7\%                                                                                         & 11\%                                                                                         & 59\%                                                                             & 68\%                                                                                              \\
SECURE BLOCKCHAIN ENABLED CYBER-PHYSICAL SYSTEMS        & 2                                                                      & 2021             & 0.99            & 12\%                                                                                        & 9\%                                                                                          & 32\%                                                                             & 68\%                                                                                              \\
\rowcolor[HTML]{EFEFEF} 
RISK ASSESSMENT OF CRYPTOCURRENCIES                     & 2                                                                      & 2021             & -0.41           & 0\%                                                                                         & 4\%                                                                                          & 96\%                                                                             & 67\%                                                                                              \\ \bottomrule
\end{tabular}
}
\label{tab:5-10}
\end{table}
\begin{table}[]
\centering
\caption{Top 10 ERAs Where APAC, Excluding China, Has the Highest Contribution in Blockchain Research, Foundation Size >= 5}
\resizebox{\textwidth}{!}{%
\begin{tabular}{@{}lccccccc@{}}
\toprule
\textbf{Field}                                                                                                        & \textbf{\begin{tabular}[c]{@{}c@{}}Size of \\ Foundation\end{tabular}} & \textbf{Recency} & \textbf{Growth} & \textbf{\begin{tabular}[c]{@{}c@{}}United \\ States \\ Contribution \\ to ERA\end{tabular}} & \textbf{\begin{tabular}[c]{@{}c@{}}United \\ Kingdom \\ Contribution \\ to ERA\end{tabular}} & \textbf{\begin{tabular}[c]{@{}c@{}}China \\ Contribution \\ to ERA\end{tabular}} & \textbf{\begin{tabular}[c]{@{}c@{}}Asia Pacific \\ Region \\ Contribution \\ to ERA\end{tabular}} \\ \midrule
\begin{tabular}[c]{@{}l@{}}INTERRELATEDNESS OF CRYPTOCURRIENCIES, \\ NFTS CLEAN ENERGY AND GREEN MARKETS\end{tabular} & 23                                                                     & 2022.3           & 0.40            & 9\%                                                                                         & 15\%                                                                                         & 36\%                                                                             & 55\%                                                                                              \\
\rowcolor[HTML]{EFEFEF} 
SECURE BLOCKCHAIN-BASED DATA TRANSMISSION                                                                             & 5                                                                      & 2020.4           & -0.25           & 10\%                                                                                        & 3\%                                                                                          & 91\%                                                                             & 55\%                                                                                              \\
BLOCKCHAIN IN HEALTHCARE SECTOR                                                                                       & 14                                                                     & 2018.6           & 3.32            & 17\%                                                                                        & 9\%                                                                                          & 19\%                                                                             & 48\%                                                                                              \\
\rowcolor[HTML]{EFEFEF} 
\begin{tabular}[c]{@{}l@{}}BLOCKCHAIN-BASED INFORMATION SYSTEMS \\ MANAGEMENT AND SECURITY\end{tabular}               & 14                                                                     & 2020.5           & 9.22            & 15\%                                                                                        & 11\%                                                                                         & 40\%                                                                             & 47\%                                                                                              \\
BLOCKCHAIN AND FEDERATED LEARNING                                                                                     & 6                                                                      & 2020.2           & 1.65            & 19\%                                                                                        & 5\%                                                                                          & 69\%                                                                             & 41\%                                                                                              \\
\rowcolor[HTML]{EFEFEF} 
CENTRAL BANK DIGITAL CURRENCIES                                                                                       & 5                                                                      & 2022             & 0.40            & 20\%                                                                                        & 23\%                                                                                         & 33\%                                                                             & 40\%                                                                                              \\
CRYPTOCURRENCIES AS A SAFE HAVEN FOR EQUITY MARKETS                                                                   & 14                                                                     & 2019.4           & 2.67            & 9\%                                                                                         & 14\%                                                                                         & 19\%                                                                             & 38\%                                                                                              \\
\rowcolor[HTML]{EFEFEF} 
BLOCKCHAINED FEDERATED LEARNING                                                                                       & 10                                                                     & 2020.5           & 1.36            & 14\%                                                                                        & 7\%                                                                                          & 54\%                                                                             & 38\%                                                                                              \\
CRYPTOCURRENCY MARKET EFFIECIENCY                                                                                     & 5                                                                      & 2018.2           & 2.62            & 9\%                                                                                         & 15\%                                                                                         & 15\%                                                                             & 31\%                                                                                              \\
\rowcolor[HTML]{EFEFEF} 
BLOCKCHAIN APPLICATIONS IN CONSTRUCTION INDUSTRY                                                                      & 7                                                                      & 2020.4           & 1.67            & 11\%                                                                                        & 17\%                                                                                         & 47\%                                                                             & 28\%                                                                                              \\ \bottomrule
\end{tabular}
}
\label{tab:5-11}
\end{table}

\begin{table}[]
\centering
\caption{Top 10 ERAs Where USA Has the Highest Contribution in Blockchain Research}
\resizebox{\textwidth}{!}{%
\begin{tabular}{@{}lccccccc@{}}
\toprule
\textbf{Field}                                                                                          & \textbf{\begin{tabular}[c]{@{}c@{}}Size of \\ Foundation\end{tabular}} & \textbf{Recency} & \textbf{Growth} & \textbf{\begin{tabular}[c]{@{}c@{}}United \\ States \\ Contribution \\ to ERA\end{tabular}} & \textbf{\begin{tabular}[c]{@{}c@{}}United \\ Kingdom \\ Contribution \\ to ERA\end{tabular}} & \textbf{\begin{tabular}[c]{@{}c@{}}China \\ Contribution \\ to ERA\end{tabular}} & \textbf{\begin{tabular}[c]{@{}c@{}}Asia Pacific \\ Region \\ Contribution \\ to ERA\end{tabular}} \\ \midrule
\begin{tabular}[c]{@{}l@{}}BLOCKCHAIN-BASED BUSINESS PROCESS \\ MANAGEMENT (BPM) FRAMEWORK\end{tabular} & 2                                                                      & 2019.5           & 1.18            & 47\%                                                                                        & 3\%                                                                                          & 24\%                                                                             & 56\%                                                                                              \\
\rowcolor[HTML]{EFEFEF} 
\begin{tabular}[c]{@{}l@{}}BLOCKCHAIN-BASED PRIVACY-PRESERVING \\ AUTHENTICATION SYSTEM\end{tabular}    & 2                                                                      & 2020.5           & 0.75            & 42\%                                                                                        & 4\%                                                                                          & 58\%                                                                             & 42\%                                                                                              \\
BLOCKCHAIN-BASED HEALTHCARE                                                                             & 2                                                                      & 2022             & 1.50            & 33\%                                                                                        & 0\%                                                                                          & 0\%                                                                              & 78\%                                                                                              \\
\rowcolor[HTML]{EFEFEF} 
INDUSTRY 4.0 TECHNOLOGIES                                                                               & 3                                                                      & 2020.7           & 2.83            & 32\%                                                                                        & 3\%                                                                                          & 56\%                                                                             & 12\%                                                                                              \\
CRYPTOCURRENCY MARKETS                                                                                  & 13                                                                     & 2019.9           & 1.34            & 31\%                                                                                        & 19\%                                                                                         & 26\%                                                                             & 23\%                                                                                              \\
\rowcolor[HTML]{EFEFEF} 
FOOD INDUSTRY 4.0 TECHNOLOGIES                                                                          & 4                                                                      & 2022.3           & 1.14            & 28\%                                                                                        & 28\%                                                                                         & 28\%                                                                             & 52\%                                                                                              \\
BITCOIN CARBON FOOTPRINT                                                                                & 2                                                                      & 2020             & 0.74            & 28\%                                                                                        & 16\%                                                                                         & 32\%                                                                             & 20\%                                                                                              \\
\rowcolor[HTML]{EFEFEF} 
INTELLIGENT BLOCKCHAIN-BASED SUPPLY CHAINS                                                              & 2                                                                      & 2020             & 0.11            & 25\%                                                                                        & 25\%                                                                                         & 42\%                                                                             & 46\%                                                                                              \\
BLOCKCHAIN BASED HUMANITARIAN SUPPLY CHAIN                                                              & 2                                                                      & 2019.5           & 0.45            & 25\%                                                                                        & 27\%                                                                                         & 16\%                                                                             & 36\%                                                                                              \\
\rowcolor[HTML]{EFEFEF} 
BLOCKCHAIN TECHNOLOGIES                                                                                 & 2                                                                      & 2019             & 2.37            & 24\%                                                                                        & 11\%                                                                                         & 39\%                                                                             & 45\%                                                                                              \\ \bottomrule
\end{tabular}
}
\label{tab:5-12}
\end{table}
\begin{table}[]
\centering
\caption{Top 10 ERAs Where USA Has the Highest Contribution in Blockchain Research, Foundation Size >= 5}
\resizebox{\textwidth}{!}{%
\begin{tabular}{@{}lccccccc@{}}
\toprule
\textbf{Field}                                                                                          & \textbf{\begin{tabular}[c]{@{}c@{}}Size of \\ Foundation\end{tabular}} & \textbf{Recency} & \textbf{Growth} & \textbf{\begin{tabular}[c]{@{}c@{}}United \\ States \\ Contribution \\ to ERA\end{tabular}} & \textbf{\begin{tabular}[c]{@{}c@{}}United \\ Kingdom \\ Contribution \\ to ERA\end{tabular}} & \textbf{\begin{tabular}[c]{@{}c@{}}China \\ Contribution \\ to ERA\end{tabular}} & \textbf{\begin{tabular}[c]{@{}c@{}}Asia Pacific \\ Region \\ Contribution \\ to ERA\end{tabular}} \\ \midrule
CRYPTOCURRENCY MARKETS                                                                                  & 13                                                                     & 2019.9           & 1.34            & 31\%                                                                                        & 19\%                                                                                         & 26\%                                                                             & 23\%                                                                                              \\
\rowcolor[HTML]{EFEFEF} 
\begin{tabular}[c]{@{}l@{}}BLOCKCHAIN BASED SECURE DATA STORAGE AND \\ DATA SHARING\end{tabular}        & 6                                                                      & 2021             & 1.44            & 23\%                                                                                        & 11\%                                                                                         & 93\%                                                                             & 23\%                                                                                              \\
CENTRAL BANK DIGITAL CURRENCIES                                                                         & 5                                                                      & 2022             & 0.40            & 20\%                                                                                        & 23\%                                                                                         & 33\%                                                                             & 40\%                                                                                              \\
\rowcolor[HTML]{EFEFEF} 
BLOCKCHAIN AND FEDERATED LEARNING                                                                       & 6                                                                      & 2020.2           & 1.65            & 19\%                                                                                        & 5\%                                                                                          & 69\%                                                                             & 41\%                                                                                              \\
BLOCKCHAIN IN HEALTHCARE SECTOR                                                                         & 14                                                                     & 2018.6           & 3.32            & 17\%                                                                                        & 9\%                                                                                          & 19\%                                                                             & 48\%                                                                                              \\
\rowcolor[HTML]{EFEFEF} 
BLOCKCHAIN SUPPLY CHAIN TRACEABILITY                                                                    & 11                                                                     & 2020.1           & 0.75            & 16\%                                                                                        & 15\%                                                                                         & 70\%                                                                             & 19\%                                                                                              \\
\begin{tabular}[c]{@{}l@{}}BLOCKCHAIN-BASED INFORMATION SYSTEMS \\ MANAGEMENT AND SECURITY\end{tabular} & 14                                                                     & 2020.5           & 9.22            & 15\%                                                                                        & 11\%                                                                                         & 40\%                                                                             & 47\%                                                                                              \\
\rowcolor[HTML]{EFEFEF} 
BLOCKCHAINED FEDERATED LEARNING                                                                         & 10                                                                     & 2020.5           & 1.36            & 14\%                                                                                        & 7\%                                                                                          & 54\%                                                                             & 38\%                                                                                              \\
\begin{tabular}[c]{@{}l@{}}BLOCKCHAIN APPLICATIONS TO ENERGY TRADING \\ (PEER-TO-PEER)\end{tabular}     & 29                                                                     & 2019.2           & 2.56            & 12\%                                                                                        & 12\%                                                                                         & 29\%                                                                             & 28\%                                                                                              \\
\rowcolor[HTML]{EFEFEF} 
BLOCKCHAIN APPLICATIONS IN CONSTRUCTION INDUSTRY                                                        & 7                                                                      & 2020.4           & 1.67            & 11\%                                                                                        & 17\%                                                                                         & 47\%                                                                             & 28\%                                                                                              \\ \bottomrule
\end{tabular}
}
\label{tab:5-13}
\end{table}

\begin{table}[]
\centering
\caption{Top 10 ERAs Where UK Has the Highest Contribution in Blockchain Research, Foundation Size >= 5}
\resizebox{\textwidth}{!}{%
\begin{tabular}{@{}lccccccc@{}}
\toprule
\textbf{Field}                                                                                       & \textbf{\begin{tabular}[c]{@{}c@{}}Size of \\ Foundation\end{tabular}} & \textbf{Recency} & \textbf{Growth} & \textbf{\begin{tabular}[c]{@{}c@{}}United \\ States\\ Contribution \\ to ERA\end{tabular}} & \textbf{\begin{tabular}[c]{@{}c@{}}United \\ Kingdom \\ Contribution \\ to ERA\end{tabular}} & \textbf{\begin{tabular}[c]{@{}c@{}}China \\ Contribution \\ to ERA\end{tabular}} & \textbf{\begin{tabular}[c]{@{}c@{}}Asia Pacific \\ Region \\ Contribution \\ to ERA\end{tabular}} \\ \midrule
BLOCKCHAIN-DRIVEN SUPPLY CHAIN                                                                       & 3                                                                      & 2022             & 0.25            & 0\%                                                                                        & 58\%                                                                                         & 25\%                                                                             & 75\%                                                                                              \\
\rowcolor[HTML]{EFEFEF} 
BLOCKCHAIN-BASED TRUST MODELS                                                                        & 2                                                                      & 2020.5           & -0.25           & 22\%                                                                                       & 33\%                                                                                         & 44\%                                                                             & 89\%                                                                                              \\
DEFI ASSETS                                                                                          & 2                                                                      & 2022             & 1.60            & 5\%                                                                                        & 30\%                                                                                         & 60\%                                                                             & 50\%                                                                                              \\
\rowcolor[HTML]{EFEFEF} 
\begin{tabular}[c]{@{}l@{}}BLOCKCHAIN TECHNOLOGY APPLICATIONS \\ IN SUPPLY CHAIN CHAINS\end{tabular} & 2                                                                      & 2019             & 5.05            & 16\%                                                                                       & 30\%                                                                                         & 19\%                                                                             & 37\%                                                                                              \\
FOOD INDUSTRY 4.0 TECHNOLOGIES                                                                       & 4                                                                      & 2022.3           & 1.14            & 28\%                                                                                       & 28\%                                                                                         & 28\%                                                                             & 52\%                                                                                              \\
\rowcolor[HTML]{EFEFEF} 
BLOCKCHAIN BASED HUMANITARIAN SUPPLY CHAIN                                                           & 2                                                                      & 2019.5           & 0.45            & 25\%                                                                                       & 27\%                                                                                         & 16\%                                                                             & 36\%                                                                                              \\
INTELLIGENT BLOCKCHAIN-BASED SUPPLY CHAINS                                                           & 2                                                                      & 2020             & 0.11            & 25\%                                                                                       & 25\%                                                                                         & 42\%                                                                             & 46\%                                                                                              \\
\rowcolor[HTML]{EFEFEF} 
CENTRAL BANK DIGITAL CURRENCIES                                                                      & 5                                                                      & 2022             & 0.40            & 20\%                                                                                       & 23\%                                                                                         & 33\%                                                                             & 40\%                                                                                              \\
BLOCKCHAIN PRIVACY-PRESERVING                                                                        & 4                                                                      & 2021.8           & 0.43            & 8\%                                                                                        & 22\%                                                                                         & 51\%                                                                             & 59\%                                                                                              \\
\rowcolor[HTML]{EFEFEF} 
CRYPTOCURRENCY MARKETS                                                                               & 13                                                                     & 2019.9           & 1.34            & 31\%                                                                                       & 19\%                                                                                         & 26\%                                                                             & 23\%                                                                                              \\ \bottomrule
\end{tabular}
}
\label{tab:5-14}
\end{table}
\begin{table}[]
\centering
\caption{Top 10 ERAs Where UK Has the Highest Contribution in Blockchain Research, Foundation Size >= 5}
\resizebox{\textwidth}{!}{%
\begin{tabular}{@{}lccccccc@{}}
\toprule
\textbf{Field}                                                                                                        & \textbf{\begin{tabular}[c]{@{}c@{}}Size of \\ Foundation\end{tabular}} & \textbf{Recency} & \textbf{Growth} & \textbf{\begin{tabular}[c]{@{}c@{}}United \\ States \\ Contribution \\ to ERA\end{tabular}} & \textbf{\begin{tabular}[c]{@{}c@{}}United \\ Kingdom \\ Contribution \\ to ERA\end{tabular}} & \textbf{\begin{tabular}[c]{@{}c@{}}China \\ Contribution \\ to ERA\end{tabular}} & \textbf{\begin{tabular}[c]{@{}c@{}}Asia Pacific \\ Region \\ Contribution \\ to ERA\end{tabular}} \\ \midrule
CENTRAL BANK DIGITAL CURRENCIES                                                                                       & 5                                                                      & 2022             & 0.40            & 20\%                                                                                        & 23\%                                                                                         & 33\%                                                                             & 40\%                                                                                              \\
\rowcolor[HTML]{EFEFEF} 
CRYPTOCURRENCY MARKETS                                                                                                & 13                                                                     & 2019.9           & 1.34            & 31\%                                                                                        & 19\%                                                                                         & 26\%                                                                             & 23\%                                                                                              \\
\begin{tabular}[c]{@{}l@{}}BLOCKCHAIN APPLICATIONS IN CONSTRUCTION \\ INDUSTRY\end{tabular}                           & 7                                                                      & 2020.4           & 1.67            & 11\%                                                                                        & 17\%                                                                                         & 47\%                                                                             & 28\%                                                                                              \\
\rowcolor[HTML]{EFEFEF} 
BLOCKCHAIN SUPPLY CHAIN TRACEABILITY                                                                                  & 11                                                                     & 2020.1           & 0.75            & 16\%                                                                                        & 15\%                                                                                         & 70\%                                                                             & 19\%                                                                                              \\
CRYPTOCURRENCY MARKET EFFIECIENCY                                                                                     & 5                                                                      & 2018.2           & 2.62            & 9\%                                                                                         & 15\%                                                                                         & 15\%                                                                             & 31\%                                                                                              \\
\rowcolor[HTML]{EFEFEF} 
\begin{tabular}[c]{@{}l@{}}INTERRELATEDNESS OF CRYPTOCURRIENCIES, NFTS \\ CLEAN ENERGY AND GREEN MARKETS\end{tabular} & 23                                                                     & 2022.3           & 0.40            & 9\%                                                                                         & 15\%                                                                                         & 36\%                                                                             & 55\%                                                                                              \\
\begin{tabular}[c]{@{}l@{}}CRYPTOCURRENCIES AS A SAFE HAVEN FOR EQUITY \\ MARKETS\end{tabular}                        & 14                                                                     & 2019.4           & 2.67            & 9\%                                                                                         & 14\%                                                                                         & 19\%                                                                             & 38\%                                                                                              \\
\rowcolor[HTML]{EFEFEF} 
\begin{tabular}[c]{@{}l@{}}BLOCKCHAIN APPLICATIONS TO ENERGY TRADING \\ (PEER-TO-PEER)\end{tabular}                   & 29                                                                     & 2019.2           & 2.56            & 12\%                                                                                        & 12\%                                                                                         & 29\%                                                                             & 28\%                                                                                              \\
\begin{tabular}[c]{@{}l@{}}BLOCKCHAIN BASED SECURE DATA STORAGE AND \\ DATA SHARING\end{tabular}                      & 6                                                                      & 2021             & 1.44            & 23\%                                                                                        & 11\%                                                                                         & 93\%                                                                             & 23\%                                                                                              \\
\rowcolor[HTML]{EFEFEF} 
\begin{tabular}[c]{@{}l@{}}BLOCKCHAIN-BASED INFORMATION SYSTEMS \\ MANAGEMENT AND SECURITY\end{tabular}               & 14                                                                     & 2020.5           & 9.22            & 15\%                                                                                        & 11\%                                                                                         & 40\%                                                                             & 47\%                                                                                              \\ \bottomrule
\end{tabular}
}
\label{tab:5-15}
\end{table}

\smallskip \noindent \textbf{Top ERAs by Countries / Regions}
Tables ~\ref{tab:5-6} to ~\ref{tab:5-13} showcase the top 10 ERAs for China, APAC (excluding China), USA, and UK. 
As shown in Table~\ref{tab:5-6}, Among the top 10 ERAs where China has the highest contribution, China stands out by significantly surpassing other countries, with all of its top 10 ERAs having a contribution of over 90\% (or above 30\% for larger ERAs). Notably, China's contribution reached 100\% in the areas of \textit{Blockchain-Based Access Control Applications}, \textit{Blockchain and Green Innovation}, and \textit{Blockchain-Enabled Mobile-Edge Computing}.
However, it is worth mentioning that the majority of China's top 10 ERAs exhibit slow or declining growth.

For the USA, the highest contribution to a blockchain specific ERA was 47\% in \textit{Blockchain-based Business Process Management Framework}. However, the APAC region contributed the highest to this ERA with 56\% of publications in this ERA. The USA was dominant in the ERA which covers \textit{Cryptocurrency Markets} with 31\% contribution. 

For the UK, the highest contribution to a blockchain specific ERA was 58\% in \textit{Blockchain-driven Supply Chain} research. However, again the APAC region contributed the highest to this ERA with 75\% of the publications. The UK did not have the highest contribution to any of the blockchain specific ERAs. 

For APAC, the highest contribution to a blockchain specific ERA was 89\% in \textit{Blockchain-based Trust Models}. The APAC region was dominant in all of the Top 10 ERAs by largest contribution from APAC except for \textit{Risk Assessment of cryptocurrencies} where China has a higher share. The same trend also applies for larger ERAs. Many of the ERAs in the top 10 for APAC were related to blockchain applications to healthcare. 

\begin{mybox}
\textbf{Summary}: In many of the ERAs where China makes the highest contributions, its contributions are dominant, exceeding 90\%. As for other countries and regions, while they have their own ERAs where they make significant contributions, there are also numerous contributions from other countries or regions.
\end{mybox}


\section{The Analysis of Top Publications in Blockchain Research}\label{sec:top}

\begin{table}[]
\centering
\caption{Top 20 Publications in Blockchain Research}
\resizebox{\textwidth}{!}{%
\begin{tabular}{@{}lllcc@{}}
\toprule
\textbf{\begin{tabular}[c]{@{}l@{}}Citation \\ Velocity\end{tabular}} & \textbf{Article Title}                                                                                          & \textbf{Reference}                                                                                                      & \textbf{\begin{tabular}[c]{@{}c@{}}Publication \\ Year\end{tabular}} & \textbf{\begin{tabular}[c]{@{}c@{}}Highly \\ Cited Status\end{tabular}} \\ \midrule
29                                                                    & Hyperledger Fabric: A Distributed Operating System for Permissioned Blockchains                                 & \cite{androulaki2018hyperledger}                   & 2018                                                                 &                                                                         \\
\rowcolor[HTML]{EFEFEF} 
24.1                                                                  & Blockchains and Smart Contracts for the Internet of Things                                                      & \cite{christidis2016blockchains}                                                                                                                & 2016                                                                 & Y                                                                       \\
22.8                                                                  & Industry 4.0: state of the art and future trends                                                                & \cite{xu2018industry}                                                                               & 2018                                                                 & Y                                                                       \\
\rowcolor[HTML]{EFEFEF} 
21.6                                                                  & An Overview of Blockchain Technology: Architecture, Consensus, and Future Trends                                & \cite{zheng2017overview}                                                                           & 2017                                                                 &                                                                         \\
21.3                                                                  & Blockchain challenges and opportunities: a survey                                                               & \cite{zheng2018blockchain}                                                                          & 2018                                                                 & Y                                                                       \\
\rowcolor[HTML]{EFEFEF} 
21.1                                                                  & Blockchain technology and its relationships to sustainable supply chain management                              & \cite{saberi2019blockchain}                                                                              & 2019                                                                 & Y                                                                       \\
17.8                                                                  & IoT security: Review, blockchain solutions, and open challenges                                                 & \cite{khan2018iot}       & 2018                                                                 & Y                                                                       \\
\rowcolor[HTML]{EFEFEF} 
15.2                                                                  & Blockchain technology in the energy sector: A systematic review of challenges and opportunities                 & \cite{andoni2019blockchain}                                                                                    & 2019                                                                 & Y                                                                       \\
12.8                                                                  & A systematic literature review of blockchain-based applications: Current status, classification and open issues & \cite{casino2019systematic}                                                                                                 & 2019                                                                 & Y                                                                       \\
\rowcolor[HTML]{EFEFEF} 
12.7                                                                  & Designing microgrid energy markets A case study: The Brooklyn Microgrid                                         & \cite{mengelkamp2018designing}                           & 2018                                                                 & Y                                                                       \\
12.5                                                                  & 1 Blockchain's roles in meeting key supply chain management objectives                                          & \cite{kshetri20181}                                                                            & 2018                                                                 & Y                                                                       \\
\rowcolor[HTML]{EFEFEF} 
12.4                                                                  & Hawk: The Blockchain Model of Cryptography and Privacy-Preserving Smart Contracts                               & \cite{kosba2016hawk}                                                                           & 2016                                                                 &                                                                         \\
12.2                                                                  & Industry 4.0 and the current status as well as future prospects on logistics                                    & \cite{hofmann2017industry}                                                                                                     & 2017                                                                 & Y                                                                       \\
\rowcolor[HTML]{EFEFEF} 
11.9                                                                  & On blockchain and its integration with loT. Challenges and opportunities                                        & \cite{reyna2018blockchain}       & 2018                                                                 & Y                                                                       \\
11.8                                                                  & The impact of digital technology and Industry 4.0 on the ripple effect and supply chain risk analytics          & \cite{ivanov2019impact}                                                                               & 2019                                                                 & Y                                                                       \\
\rowcolor[HTML]{EFEFEF} 
11.8                                                                  & The Truth About Blockchain                                                                                      & \cite{iansiti2017truth}                           & 2017                                                                 & Y                                                                       \\
11.8                                                                  & Where Is Current Research on Blockchain Technology?-A Systematic Review                                         & \cite{yli2016current}                                                                                                                   & 2016                                                                 & Y                                                                       \\
\rowcolor[HTML]{EFEFEF} 
11.2                                                                  & Building dynamic capabilities for digital transformation: An ongoing process of strategic renewal               & \cite{warner2019building}                                                                                                   & 2019                                                                 & Y                                                                       \\
10.8                                                                  & A Blockchain-Based Privacy-Preserving Intelligent Charging Station Selection for Electric Vehicles              & \cite{danish2020blockchain} & 2020                                                                 &                                                                         \\
\rowcolor[HTML]{EFEFEF} 
10.8                                                                  & Decentralizing Privacy: Using Blockchain to Protect Personal Data                                               & \cite{zyskind2015decentralizing}                                                                             & 2015                                                                 &                                                                         \\ \bottomrule
\end{tabular}
}
\label{tab:6-1}
\end{table}

Many ways exist to estimate the quality of research. Influence of research is a commonly used estimator and it is usually measured by citations. This is commonly referred to as “citation impact”. In research like blockchain technologies that have grown from almost nothing 10 years ago (see previous chapters), it is important to measure citation impact in the context of this rapid growth. To achieve this, Clarivate used \textbf{citation velocity}. This measure captures the rate of citation accumulation each month, starting from the month it is published until the data was extracted for this report. Because citations accumulate at different rates between fields, papers were ranked by considering the year of publication and the field of publication. Clarivate analysts have therefore identified the most consistently-cited publications by selecting the top 10\% in terms of citation velocity.

Out of the total papers in blockchain research, citation velocity can be calculated for 27,574 (67\% of cited papers) and after ranking these papers, 2,564 papers are in the top 10\%.
Due to the page limitation, we show top 20 publicaitons based on citation velocity in Table~\ref{tab:6-1}.

The research paper on Hyperledger Fabric~\cite{androulaki2018hyperledger} has achieved the highest citation velocity, reaching a value of 29. Hyperledger Fabric, developed under the Hyperledger project by the Linux Foundation, offers a modular architecture that facilitates the establishment of private and permissioned blockchain networks. Consequently, this technology has gained extensive adoption within both the research and industry communities.

Among these top 20 papers, there are many that belong to the category of review articles~\cite{zheng2017overview, zheng2018blockchain, casino2019systematic}. These papers cover both the research on the development trends and challenges of blockchain technology itself, as well as the exploration of blockchain applications in other fields, such as IoT (Internet of Things)~\cite{khan2018iot, reyna2018blockchain} and Supply Chain~\cite{kshetri20181, ivanov2019impact}.

There are two papers that focus on privacy~\cite{kosba2016hawk, zyskind2015decentralizing}. Privacy and the inherent characteristics of public blockchains (data transparency, immutability, and traceability) are contradictory. However, in practical applications, privacy is of great importance to users. We also notice that two papers~\cite{zheng2017overview, zheng2018blockchain} are from the same team.  

\begin{mybox}
\textbf{Summary}: The paper on Hyperledger Fabric ranks first based on the citation velocity. Many of the top 20 papers are survey papers related to Blockchain technology
\end{mybox}

\section{The Analysis of Key Research Fronts in Blockchain Research}\label{sec:key}

In this section, we select five representative key Research Fronts identified in Section~\ref{sec:era}. These Research Fronts include:
\begin{itemize}
    \item INTERRELATEDNESS OF CRYPTOCURRIENCIES, NFTS CLEAN ENERGY AND GREEN MARKETS
    \item BLOCKCHAIN BASED SECURE DATA STORAGE AND DATA SHARING
    \item BLOCKCHAIN PRIVACY \& SECURITY
    \item BLOCKCHAIN AND IOT INTEGRATION AND SECURITY
    \item BLOCKCHAIN AND FEDERATED LEARNING
\end{itemize}
These research fronts combine with blockchain from different perspectives, including cryptocurrencies with clean energy and green markets, data storage and sharing, privacy and security, IoT integration, and federated learning. Among these research directions, four of them rank among the top 10 youngest ERAs in blockchain research, with a foundation size of 5 or more, except for \textit{Blockchain and IoT Integration and Security}. However, this particular research front is among the top 10 ERAs with the highest growth rate.  
Note that due to the multitude of research fronts in blockchain research, it is not feasible to provide individual analysis for all of them. Therefore, in this paper, we have chosen these five research fronts as examples to illustrate our points.
We aggregated the data of the five selected Research Fronts, given the small size of the research output. 



\subsection{National Affiliations Analysis}

\begin{table}[]
\centering
\caption{Top 10 Countries By Number of Publications for Selected RFs}
\resizebox{\textwidth}{!}{%
\begin{tabular}{@{}lccccc@{}}
\toprule
\textbf{Country / Region} & \textbf{\begin{tabular}[c]{@{}c@{}}Number of \\ Publications\end{tabular}} & \textbf{\begin{tabular}[c]{@{}c@{}}Category Normalized \\ Citation Impact\end{tabular}} & \textbf{\begin{tabular}[c]{@{}c@{}}Citation \\ Impact\end{tabular}} & \textbf{\begin{tabular}[c]{@{}c@{}}\% of Documents in \\ Top 10\%\end{tabular}} & \textbf{\begin{tabular}[c]{@{}c@{}}\% of International \\ Collaboration\end{tabular}} \\ \midrule
CHINA MAINLAND            & 39                                                                         & 17.48                                                                                   & 123.03                                                              & 94.87                                                                           & 87.18                                                                                 \\
AUSTRALIA                 & 30                                                                         & 19.99                                                                                   & 148.63                                                              & 100.00                                                                          & 90.00                                                                                 \\
UNITED ARAB EMIRATES      & 16                                                                         & 28.66                                                                                   & 111.00                                                              & 100.00                                                                          & 100.00                                                                                \\
IRELAND                   & 11                                                                         & 30.00                                                                                   & 132.55                                                              & 100.00                                                                          & 81.82                                                                                 \\
MALAYSIA                  & 10                                                                         & 30.23                                                                                   & 39.50                                                               & 100.00                                                                          & 100.00                                                                                \\
RUSSIA                    & 10                                                                         & 21.46                                                                                   & 51.60                                                               & 100.00                                                                          & 100.00                                                                                \\
PAKISTAN                  & 9                                                                          & 28.97                                                                                   & 171.00                                                              & 100.00                                                                          & 100.00                                                                                \\
USA                       & 8                                                                          & 16.96                                                                                   & 118.75                                                              & 100.00                                                                          & 100.00                                                                                \\
UNITED KINGDOM            & 6                                                                          & 29.91                                                                                   & 61.50                                                               & 100.00                                                                          & 100.00                                                                                \\
VIETNAM                   & 5                                                                          & 34.45                                                                                   & 57.80                                                               & 100.00                                                                          & 100.00                                                                                \\ \bottomrule
\end{tabular}
}
\label{tab:7-1}
\end{table}
\begin{table}[]
\centering
\caption{Top 10 Countries By Category Normalized Citation Impact for Selected RFs (Number of Publications > 2)}
\resizebox{\textwidth}{!}{%
\begin{tabular}{@{}lccccc@{}}
\toprule
\textbf{Country / Region} & \textbf{\begin{tabular}[c]{@{}c@{}}Number of \\ Publications\end{tabular}} & \textbf{\begin{tabular}[c]{@{}c@{}}Category Normalized \\ Citation Impact\end{tabular}} & \textbf{\begin{tabular}[c]{@{}c@{}}Citation \\ Impact\end{tabular}} & \textbf{\begin{tabular}[c]{@{}c@{}}\% of Documents in \\ Top 10\%\end{tabular}} & \textbf{\begin{tabular}[c]{@{}c@{}}\% of International \\ Collaboration\end{tabular}} \\ \midrule
FRANCE                    & 4                                                                          & 35.39                                                                                   & 78.00                                                               & 100.00                                                                          & 50.00                                                                                 \\
VIETNAM                   & 5                                                                          & 34.45                                                                                   & 57.80                                                               & 100.00                                                                          & 100.00                                                                                \\
MALAYSIA                  & 10                                                                         & 30.23                                                                                   & 39.50                                                               & 100.00                                                                          & 100.00                                                                                \\
IRELAND                   & 11                                                                         & 30.00                                                                                   & 132.55                                                              & 100.00                                                                          & 81.82                                                                                 \\
UNITED KINGDOM            & 6                                                                          & 29.91                                                                                   & 61.50                                                               & 100.00                                                                          & 100.00                                                                                \\
PAKISTAN                  & 9                                                                          & 28.97                                                                                   & 171.00                                                              & 100.00                                                                          & 100.00                                                                                \\
UNITED ARAB EMIRATES      & 16                                                                         & 28.66                                                                                   & 111.00                                                              & 100.00                                                                          & 100.00                                                                                \\
RUSSIA                    & 10                                                                         & 21.46                                                                                   & 51.60                                                               & 100.00                                                                          & 100.00                                                                                \\
AUSTRALIA                 & 30                                                                         & 19.99                                                                                   & 148.63                                                              & 100.00                                                                          & 90.00                                                                                 \\
CHINA MAINLAND            & 39                                                                         & 17.48                                                                                   & 123.03                                                              & 94.87                                                                           & 87.18                                                                                 \\ \bottomrule
\end{tabular}
}
\label{tab:7-2}
\end{table}

\begin{figure}
  \centering
  \subfloat[Number of Publications]{\includegraphics[width=0.45\textwidth]{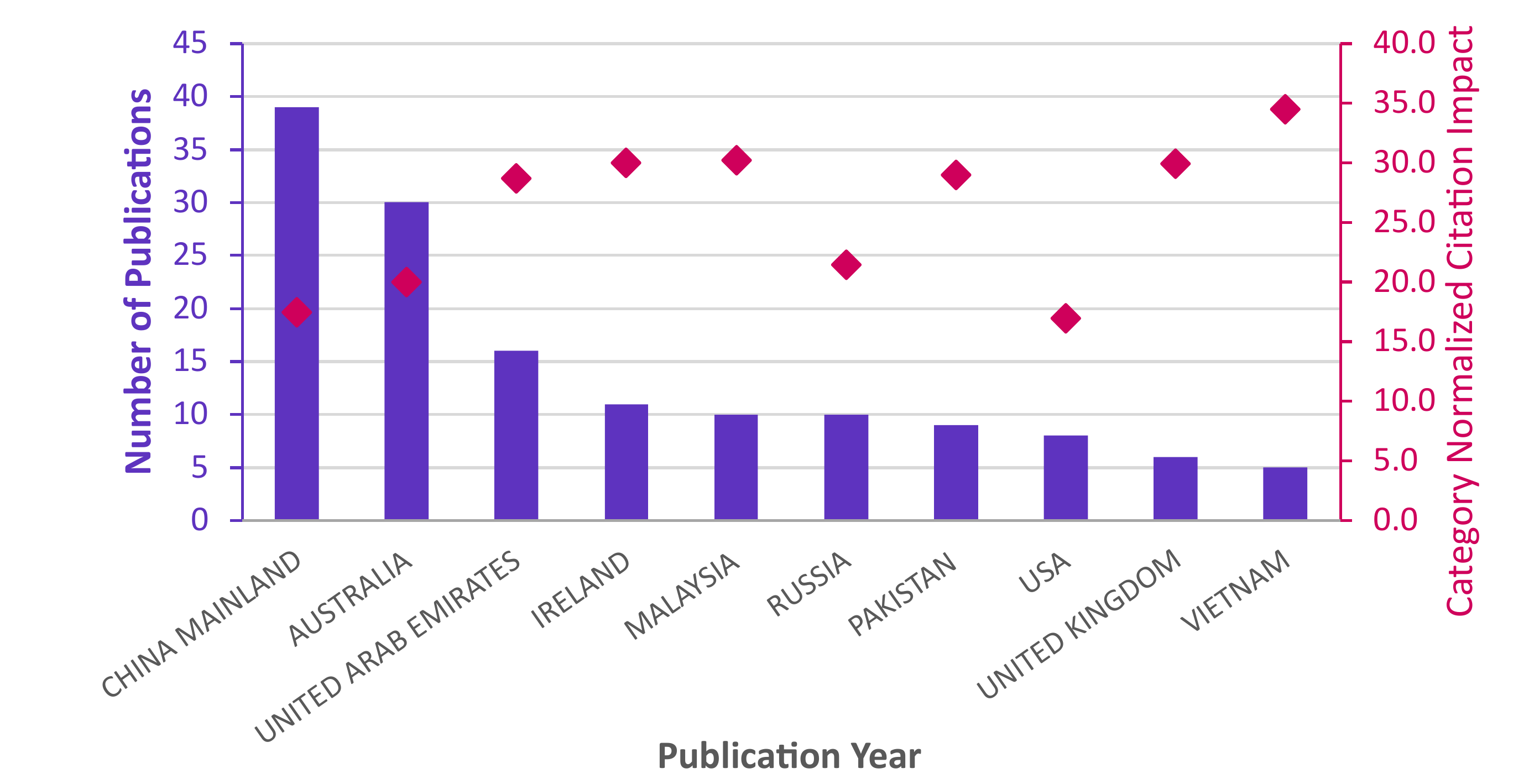}}
  \hfill
  \subfloat[Category Normalized Citation Impact]{\includegraphics[width=0.45\textwidth]{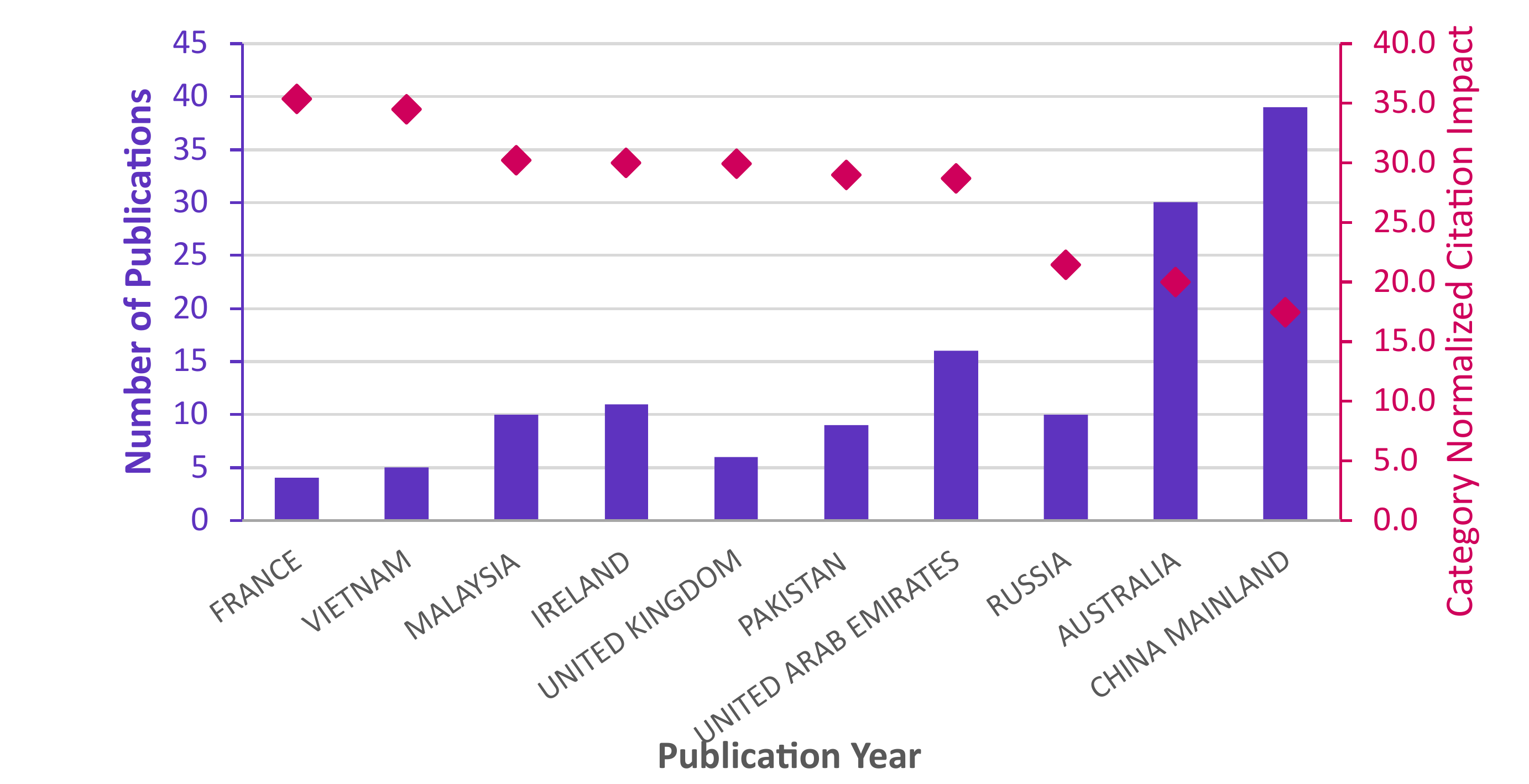}}
  \caption{Top 10 Countries/Regions for Selected RFs}
  \label{fig:7-country}
\end{figure}



Table~\ref{tab:7-1} and~\ref{tab:7-2} present the top 10 counties/regions by number of publications and category normalized citation impact for the five selected Research Fronts, respectively. 
Figure~\ref{fig:7-country} provides the visualization of the top 10 countries/regions by number of publications and category normalized citation impact for the selected RFs. 

According to Table~\ref{tab:7-1}, China leads the ranking of countries in terms of the number of publications, with 39 core papers across the 5 research fronts, and its citation impact is 17 times higher than the world average of 1. Despite having one of the lowest international collaboration rates among the top 10 countries, China still maintains a relatively high rate at 87\%. Australia follows China with a total of 30 core papers and a slightly higher citation impact of 20. Among all the countries in the top 10 based on publication output, Vietnam has the highest citation impact (34.45), closely followed by Malaysia (30.23) and Ireland (30).
France has the highest category normalized citation impact (35.39) but only has four publications (see Table~\ref{tab:7-2}). 


\subsection{Institutional Affiliation Analysis}

\begin{table}[]
\centering
\caption{Top 10 Institutions By Number of Publications for Selected RFs}
\resizebox{\textwidth}{!}{%
\begin{tabular}{@{}lccccc@{}}
\toprule
\textbf{Instituion}                               & \textbf{\begin{tabular}[c]{@{}c@{}}Number of \\ Publications\end{tabular}} & \textbf{\begin{tabular}[c]{@{}c@{}}Category Normalized \\ Citation Impact\end{tabular}} & \textbf{\begin{tabular}[c]{@{}c@{}}Citation \\ Impact\end{tabular}} & \textbf{\begin{tabular}[c]{@{}c@{}}\% of Documents in \\ Top 10\%\end{tabular}} & \textbf{\begin{tabular}[c]{@{}c@{}}\% of International \\ Collaboration\end{tabular}} \\ \midrule
Swinburne University of Technology                & 22                                                                         & 19.83                                                                                   & 158.41                                                              & 100.00                                                                          & 86.36                                                                                 \\
Taiyuan University of Science \& Technology       & 15                                                                         & 16.34                                                                                   & 184.53                                                              & 100.00                                                                          & 100.00                                                                                \\
United Arab Emirates University                   & 13                                                                         & 30.31                                                                                   & 44.77                                                               & 100.00                                                                          & 100.00                                                                                \\
South Ural State University                       & 9                                                                          & 21.78                                                                                   & 50.78                                                               & 100.00                                                                          & 100.00                                                                                \\
Chinese Academy of Sciences                       & 7                                                                          & 17.24                                                                                   & 172.57                                                              & 100.00                                                                          & 100.00                                                                                \\
Institute of Automation, CAS                      & 6                                                                          & 17.90                                                                                   & 182.67                                                              & 100.00                                                                          & 100.00                                                                                \\
Sunway University                                 & 5                                                                          & 44.89                                                                                   & 28.20                                                               & 100.00                                                                          & 100.00                                                                                \\
Beijing Wuzi University                           & 5                                                                          & 20.50                                                                                   & 236.80                                                              & 100.00                                                                          & 100.00                                                                                \\
University of Nottingham Malaysia                 & 5                                                                          & 15.56                                                                                   & 50.80                                                               & 100.00                                                                          & 100.00                                                                                \\
Trinity College Dublin                            & 4                                                                          & 30.02                                                                                   & 65.50                                                               & 100.00                                                                          & 100.00                                                                                \\
Ho Chi Minh City University Economics             & 4                                                                          & 29.78                                                                                   & 64.75                                                               & 100.00                                                                          & 100.00                                                                                \\
University of Wollongong                          & 4                                                                          & 28.49                                                                                   & 62.50                                                               & 100.00                                                                          & 100.00                                                                                \\
Nanjing University of Aeronautics \& Astronautics & 4                                                                          & 15.60                                                                                   & 80.25                                                               & 100.00                                                                          & 100.00                                                                                \\
International Business Machines (IBM)             & 4                                                                          & 15.60                                                                                   & 80.25                                                               & 100.00                                                                          & 100.00                                                                                \\ \bottomrule
\end{tabular}
}
\label{tab:7-3}
\end{table}

\begin{table}[]
\centering
\caption{Top 10 Institutions By Category Normalized Citation Impact for Selected RFs (Number of Publications > 2)}
\resizebox{\textwidth}{!}{%
\begin{tabular}{@{}lccccc@{}}
\toprule
\textbf{Instituion}                        & \textbf{\begin{tabular}[c]{@{}c@{}}Number of \\ Publications\end{tabular}} & \textbf{\begin{tabular}[c]{@{}c@{}}Category Normalized \\ Citation Impact\end{tabular}} & \textbf{\begin{tabular}[c]{@{}c@{}}Citation \\ Impact\end{tabular}} & \textbf{\begin{tabular}[c]{@{}c@{}}\% of Documents in \\ Top 10\%\end{tabular}} & \textbf{\begin{tabular}[c]{@{}c@{}}\% of International \\ Collaboration\end{tabular}} \\ \midrule
Sunway University                          & 5                                                                          & 44.89                                                                                   & 28.20                                                               & 100.00                                                                          & 100.00                                                                                \\
Jiangxi University of Finance \& Economics & 3                                                                          & 33.52                                                                                   & 66.67                                                               & 100.00                                                                          & 100.00                                                                                \\
United Arab Emirates University            & 13                                                                         & 30.31                                                                                   & 44.77                                                               & 100.00                                                                          & 100.00                                                                                \\
Trinity College Dublin                     & 4                                                                          & 30.02                                                                                   & 65.50                                                               & 100.00                                                                          & 100.00                                                                                \\
Ho Chi Minh City University Economics      & 4                                                                          & 29.78                                                                                   & 64.75                                                               & 100.00                                                                          & 100.00                                                                                \\
University of Wollongong                   & 4                                                                          & 28.49                                                                                   & 62.50                                                               & 100.00                                                                          & 100.00                                                                                \\
South Ural State University                & 9                                                                          & 21.78                                                                                   & 50.78                                                               & 100.00                                                                          & 100.00                                                                                \\
Beijing Wuzi University                    & 5                                                                          & 20.50                                                                                   & 236.80                                                              & 100.00                                                                          & 100.00                                                                                \\
Swinburne University of Technology         & 22                                                                         & 19.83                                                                                   & 158.41                                                              & 100.00                                                                          & 86.36                                                                                 \\
Institute of Automation, CAS               & 6                                                                          & 17.90                                                                                   & 182.67                                                              & 100.00                                                                          & 100.00                                                                                \\ \bottomrule
\end{tabular}
}
\label{tab:7-4}
\end{table}

\begin{figure}
  \centering
  \subfloat[Number of Publications]{\includegraphics[width=0.45\textwidth]{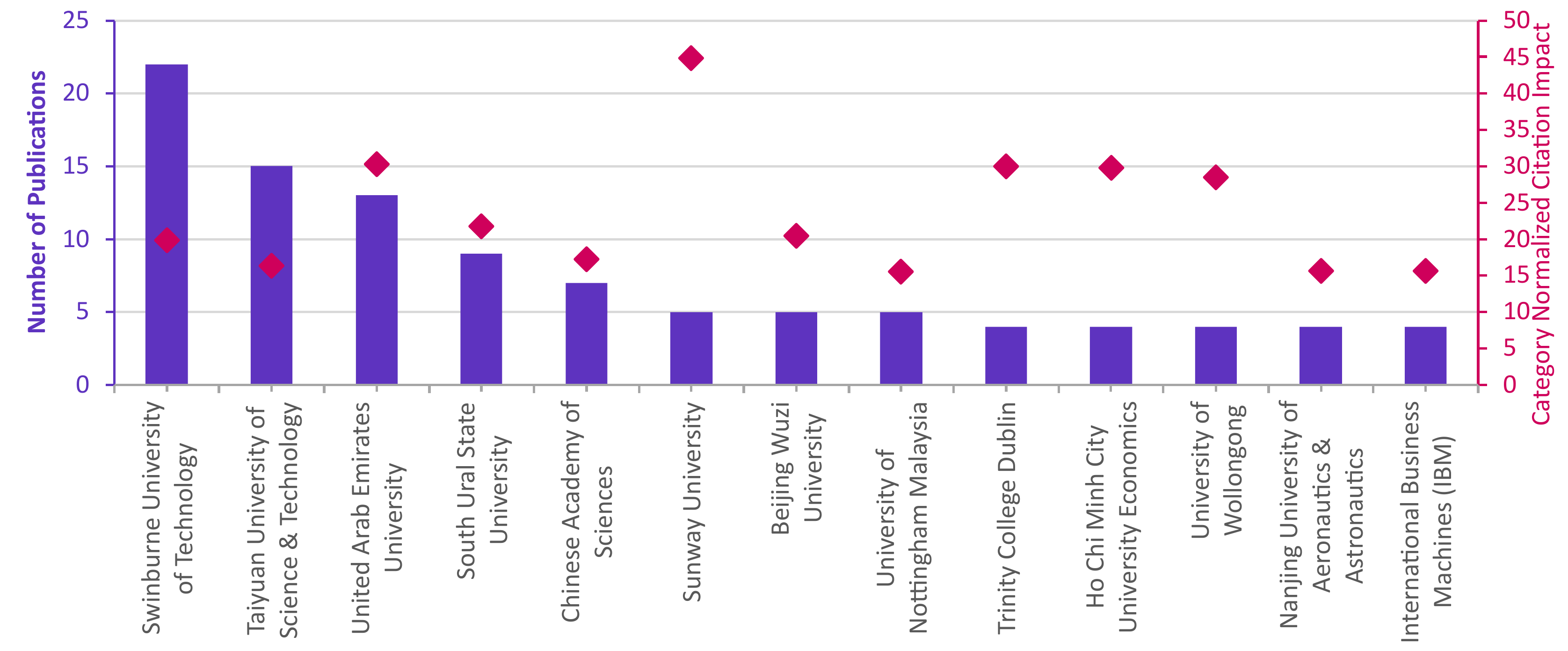}}
  \hfill
  \subfloat[Category Normalized Citation Impact]{\includegraphics[width=0.45\textwidth]{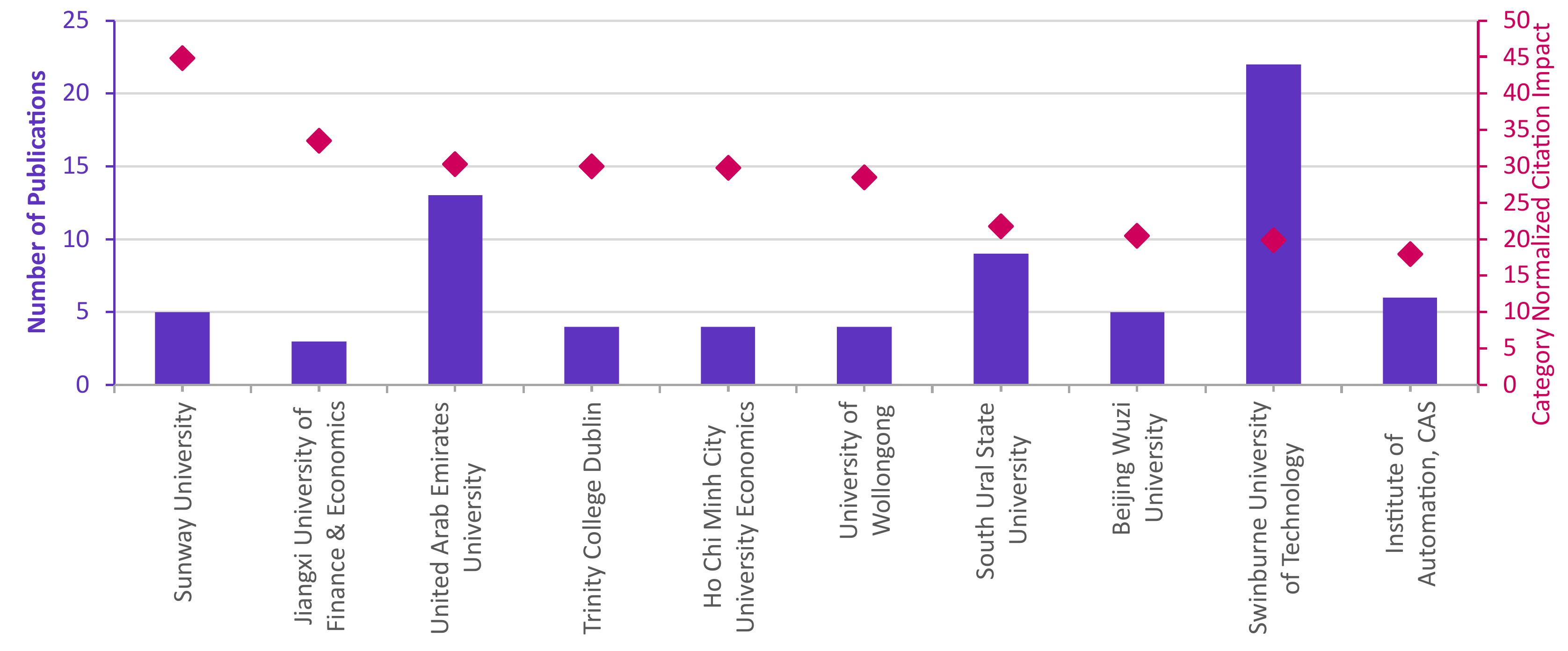}}
  \caption{Top 10 Institutions for Selected RFs}
  \label{fig:7-institution}
\end{figure}




We identify and rank the top institutions within the Research Front as determined by publication output and scholarly impact.
Table~\ref{tab:7-3} and~\ref{tab:7-4} present the top 10 institutions by number of publications and category normalized citation impact for the selected RFs, respectively. Figure~\ref{fig:7-institution} provides the visualization of the top 10 institutions by number of publications and category normalized citation impact for the selected RFs. 

For the top 10 institutions by publication, Taiyuan University of Science and Technology had the largest number of core papers (15) and a citation impact of 16 times the world average. Of the top 10 by publication count, Sunway University had the highest citation impact of more than 44 times the world average. Notably, there is one private sector institution (IBM) on the list with 4 core papers and a citation impact of 15.60. These papers also appear to be the same papers as that for Nanjing University of Aeronautics \& Astronautics. 

For those institutions that had at least 3 core papers and ranked by category normalized citation impact, Sunway University had the highest citation impact (44.89) as noted previously. Followed by Jiangxi University of Finance and Economics (33.52) which is the only university in this ranking that was not also in the Top 10 by publication output. 

\subsection{Journals and Conferences Analysis}

\begin{table}[]
\centering
\caption{Top 10 Journals/Conferences By Category Normalized Citation Impact for Selected RFs (Number of Publications > 5)}
\resizebox{\textwidth}{!}{%
\begin{tabular}{@{}lccccc@{}}
\toprule
\textbf{Journal/Conference}                                              & \textbf{\begin{tabular}[c]{@{}c@{}}Number of \\ Publications\end{tabular}} & \textbf{\begin{tabular}[c]{@{}c@{}}Category Normalized \\ Citation Impact\end{tabular}} & \textbf{\begin{tabular}[c]{@{}c@{}}Citation \\ Impact\end{tabular}} & \textbf{\begin{tabular}[c]{@{}c@{}}\% of Documents in \\ Top 10\%\end{tabular}} & \textbf{\begin{tabular}[c]{@{}c@{}}\% of International \\ Collaboration\end{tabular}} \\ \midrule
ACM COMPUTING SURVEYS                                                    & 13                                                                         & 21.73                                                                                   & 48.69                                                               & 84.62                                                                           & 53.85                                                                                 \\
IEEE COMMUNICATIONS SURVEYS AND TUTORIALS                                & 10                                                                         & 19.94                                                                                   & 141.30                                                              & 70.00                                                                           & 90.00                                                                                 \\
ENERGY ECONOMICS                                                         & 20                                                                         & 15.68                                                                                   & 17.20                                                               & 80.00                                                                           & 90.00                                                                                 \\
TECHNOLOGICAL FORECASTING AND SOCIAL CHANGE                              & 6                                                                          & 13.39                                                                                   & 37.50                                                               & 83.33                                                                           & 83.33                                                                                 \\
FINANCE RESEARCH LETTERS                                                 & 29                                                                         & 12.57                                                                                   & 31.14                                                               & 82.76                                                                           & 58.62                                                                                 \\
INTERNATIONAL REVIEW OF FINANCIAL ANALYSIS                               & 12                                                                         & 10.49                                                                                   & 20.42                                                               & 83.33                                                                           & 66.67                                                                                 \\
IEEE TRANSACTIONS ON INDUSTRIAL INFORMATICS                              & 8                                                                          & 8.09                                                                                    & 74.63                                                               & 75.00                                                                           & 62.50                                                                                 \\
INTERNATIONAL REVIEW OF ECONOMICS \& FINANCE                             & 12                                                                         & 7.96                                                                                    & 5.25                                                                & 75.00                                                                           & 91.67                                                                                 \\
FUTURE GENERATION COMPUTER SYSTEMS-THE INTERNATIONAL JOURNAL OF ESCIENCE & 22                                                                         & 7.65                                                                                    & 134.50                                                              & 68.18                                                                           & 63.64                                                                                 \\
IEEE INTERNET OF THINGS JOURNAL                                          & 45                                                                         & 7.15                                                                                    & 63.20                                                               & 57.78                                                                           & 73.33                                                                                 \\ \bottomrule
\end{tabular}
}
\label{tab:7-5}
\end{table}

\begin{figure}
    \centering
    \includegraphics[width=0.5\textwidth]{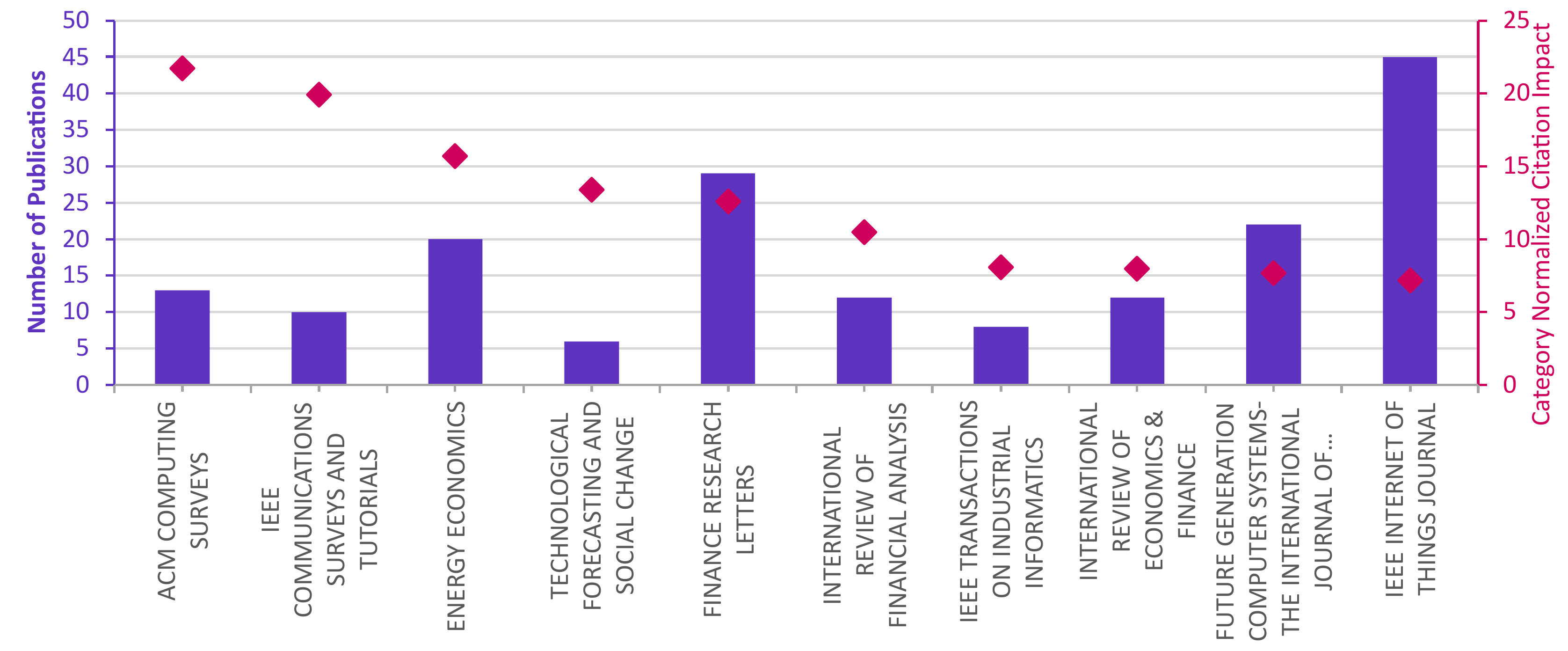}
    \caption{Top 10 Journals/Conferences by Category Normalized Citation Impact for Selected RFs}
    \label{fig:7-5}
\end{figure}


We identify the journals and conferences that have featured the most/highest impact publications or presentations falling within the client-selected ERAs.
Table~\ref{tab:7-5} presents the top 10 journals/conferences by category normalized citation impact for the selected RFs. Figure~\ref{fig:7-5} provides the visualization of the top 10 journals/conferences by category normalized citation impact for the selected RFs.

The Journal with the highest impact publications was \textit{ACM Computing Surveys} with a citation impact of more than 20 times higher than the world average of 1. This is followed by \textit{IEEE Communications and Surveys and Tutorials} with a citation impact of 20. The journal \textit{IEEE Internet of Things} had the largest number of publications (45) of the journals in the top 10 by impact.


\section{Conclusion}\label{sec:conclusion}

In this paper, we analyzed 41,497 blockchain research publications from 2008 to 2023 using Clarivate databases. Through bibliometric and citation analyses, we gained valuable insights, providing an extensive overview of the blockchain research landscape, including country, institution, authorship, and subject categories.
We also identified emerging research areas (ERA) within blockchain using co-citation clustering, considering factors like recency, growth, and contributions from different regions. Additionally, we determined influential publications based on citation velocity and conducted a detailed analysis of five representative research fronts.
Our analysis offers a fine-grained examination of specific areas in blockchain research, enhancing understanding of evolving trends, emerging applications, and future research directions. This comprehensive analysis serves as a valuable resource for researchers and practitioners interested in exploring the advancements and potential of blockchain technology.


\printcredits

\bibliographystyle{model1-num-names}

\bibliography{reference}



\end{document}